\UseRawInputEncoding
\documentclass[showpacs,twocolumn,superscriptaddress,nofootinbib]{revtex4-1}
\pdfoutput=1
\usepackage{amsfonts,amssymb,amsmath}
\usepackage[usenames,dvipsnames]{xcolor}
\usepackage[unicode, colorlinks=true, 
	linkcolor=blue, 
	urlcolor=blue, 
	citecolor=blue]{hyperref}
\usepackage[all]{hypcap}
\usepackage[utf8]{inputenc}
\usepackage{graphicx}
\usepackage{xspace}
\usepackage{tikz}
\DeclareUnicodeCharacter{2212}{\textendash}
\usepackage{float}
\usepackage{csquotes}
\usetikzlibrary{shapes,arrows}
\usetikzlibrary{matrix}
\usetikzlibrary{shapes.multipart}
\usetikzlibrary{er,positioning}
\usepackage[normalem]{ulem} 
\usepackage{url}
\usepackage{color,soul}
\usepackage[caption=false]{subfig}
\usepackage[T1]{fontenc}
\usepackage{lmodern}
\usepackage{microtype}
\usepackage{cleveref}

\crefname{equation}{Eqn.}{Eqns.}
\crefname{figure}{Fig.}{Figs.}
\crefname{section}{Sec.}{Secs.}
\usepackage{orcidlink}
\usepackage{enumitem}

\newcommand{\GKU}{\affiliation{Department of Physics, Gurukula Kangri (Deemed to be University), Haridwar - 249 404,  Uttarakhand, India}}

\newcommand{\HNB}{\affiliation{Department of Physics,Hemvati Nandan Bahuguna Garhwal University, Srinagar Garhwal – 246174, Uttarakhand, India}}
\newcommand{\HEMS}{\affiliation{Center for Space Research, North-West University, Mafikeng 2745, South Africa.}}
\newcommand{\SAN}{\affiliation{Institute for Theoretical Physics and Cosmology, Zheijiang University of Technology, Hangzhou 310023, China}}
\newcommand{\SANJ}{\affiliation{Akfa University,  Milliy Bog Street 264, Tashkent 111221, Uzbekistan}}
\newcommand{\SANJA}{\affiliation{Ulugh Beg Astronomical Institute, Astronomy St. 33, Tashkent 100052, Uzbekistan}}
\newcommand{\SANJAR}{\affiliation{National University of Uzbekistan, Tashkent 100174, Uzbekistan}}
\newcommand{\SANJARS}{\affiliation{National Research University TIIAME, Kori Niyoziy 39, Tashkent 100000, Uzbekistan}}

\makeatletter
\newsavebox\myboxA
\newsavebox\myboxB
\newlength\mylenA

\newcommand*\xoverline[2][0.75]{%
	\sbox{\myboxA}{$\m@th#2$}%
	\setbox\myboxB\null
	\ht\myboxB=\ht\myboxA%
	\dp\myboxB=\dp\myboxA%
	\wd\myboxB=#1\wd\myboxA
	\sbox\myboxB{$\m@th\overline{\copy\myboxB}$}
	\setlength\mylenA{\the\wd\myboxA}
	\addtolength\mylenA{-\the\wd\myboxB}%
	\ifdim\wd\myboxB<\wd\myboxA%
	\rlap{\hskip 0.5\mylenA\usebox\myboxB}{\usebox\myboxA}%
	\else
	\hskip -0.5\mylenA\rlap{\usebox\myboxA}{\hskip 0.5\mylenA\usebox\myboxB}%
	\fi}
\makeatother

\begin{document}

\title{
Chaos motion and Periastron precession of spinning test particles moving in the vicinage of a Schwarzschild black hole surrounded by a quintessence matter field
}

\author{Shobhit Giri \orcidlink{0000-0002-7444-0389}}
\email{shobhit6794@gmail.com}
\GKU
\author{Pankaj Sheoran \orcidlink{0000-0001-8283-8744}}
\email{hukmipankaj@gmail.com}
\GKU
\author{Hemwati Nandan\orcidlink{0000-0002-1183-4727}}
\email{hnandan@associates.iucaa.in}
\HNB
\GKU
\HEMS
\author{Sanjar Shaymatov
\orcidlink{0000-0002-5229-7657}}
\email{sanjar@astrin.uz}
\SAN
\SANJ
\SANJA
\SANJAR
\SANJARS

\date{\today}

\begin{abstract}
In the present work, our main objective is to investigate the orbits of spinning test particles around a Schwarzschild black hole under the influence of a quintessence matter field (SQBH). We begin with the dynamics of the spinning test particles around SQBH which is governed by the Mathisson-Papapetrou-Dixon (MPD) equations under the pole-dipole approximation, where the gravitational field and the higher multipoles of the particle are neglected. Depending on the types of saddle points, 
the effective potential are classified and the possibility of chaotic orbits is discussed.  
The inner most stable circular orbits (ISCOs) of the spinning particle around SQBH are addressed, as are the effects of the parameters $S$ (particles' spin) and $\epsilon$ (equation of state parameter). 
Later, Periastron precession is investigated up to the first-order spin correction for a spinning particle moving in nearly circular orbits around SQBH. It is noted that the addition of particle's spin revamps the results obtained for the non-spinning particles and also articulates the some interesting observational properties of the SQBH. Additionally, we discuss the ramifications of employing first-order spin corrections for analysing ISCOs, as well as compare our results to the Schwarzschild black hole to ensure that they are consistent in the limit when equation of state parameter $\epsilon=-1/3$ and normalization factor $\alpha \to 0$.

\end{abstract}

\maketitle

\section{Introduction}

Black holes (BHs) are one of Einstein's most enigmatic predictions from general relativity (GR). They are strange in a way that they are discovered in a region of spacetime where gravity is so strong that nothing, not even light, can escape its gravitational field. This feature of BHs, along with the observational limitations in the past, makes them difficult to detect. However, in this era of advanced laser interferometers (such as the Laser Interferometer Gravitational Wave Observatory (LIGO) \cite{Harry:2010zz}, Virgo \cite{Accadia:2011zzc}, and KAGRA \cite{Somiya:2011np}) and high resolution telescopes (such as the Event Horizon Telescope (EHT) \cite{2015IAUGA..2257792J,2016ApJ...829...11C,2017NatAs...1..646D}), BHs are no longer just theoretical objects, as their existence is almost confirmed by the EHT, which provided the first ever direct image of supermassive BHs hosted by the elliptical galaxy M87 \cite{EventHorizonTelescope:2019dse,EventHorizonTelescope:2019ths,EventHorizonTelescope:2019ggy} and Sgr A \cite{EventHorizonTelescope:2022xnr}.

From the astrophysical point of view it is particularly significant to understand the qualitative aspects of the existing fields and to bring out their effects on geodesics of massive test particles around a black hole. 
Hence, these existing fields would play an important role in altering the geodesics of massive particles, thereby strongly affecting observable properties~\cite{Herrera05,Bini12,Shaymatov21pdu,Shaymatov21d,Narzilloev20b}. With this in mind, in a realistic astrophysical scenario, it is increasingly important to consider the effect arising from quintessential field in the environment surrounding a black hole as well as at large scales. Relying on the recent astronomical observations, it is believed that the accelerated expansion of the universe can be explained by the vacuum energy referred to as the cosmological constant $\Lambda$ with repulsive effect in Einstein field equations~\cite{Cruz05,Stuchlik11,Grenon10,Rezzolla03a,Arraut15,Faraoni15,Shaymatov18a,Rayimbaev-Shaymatov21a}. Later on, the quintessential matter field was proposed to explain the repulsive behavior of the dark energy~\cite{Peebles03,Wetterich88,Caldwell09}, and thus alternatives have come into play. With this motivation, Kiselev derived general solution that represents a black hole surrounded by the quintessence matter field given by the equation of state $p = \omega_{q}\rho$. Interestingly, it turns out that $\omega_q$ takes the values in the range $(-1;-1/3)$ \cite{Kiselev2003aa}. It does however takes the values in the range $(-1;-2/3)$ according to the model proposed by Ref.~\cite{Hellerman2001JHEP}. The point to be noted here is that the vacuum energy field with cosmological constant $\Lambda$ can be explained by $\omega_{q} = −1$, and similarly the case with $\omega_{q} = −1/3$ reflects another matter field. This is what Kiselev proposed a black hole solution with the existing matter fields represented by the equation of state.

The motion of massive spinning and non-spinning test particles around a compact object (such as a BH) is of critical importance from the view point of astrophysics \cite{hartl2003dynamics,han2008chaos}. 
However, in this paper, we are solely interested in the motion of spinning particles in a curved spacetime around a BH. The main motivation for studying the motion of spinning particles is that we are in the era of powerful laser interferometry and have already begun receiving data on gravitational waves released from the coalescence of two compact objects, such as neutron star - neutron star \cite{LIGOScientific:2017vwq}, neutron-star-BH binaries \cite{LIGOScientific:2021qlt}, and BH-BH binaries \cite{LIGOScientific:2016sjg}. Now, in order to compare these massive data sets of gravitational waves emitted by the systems mentioned above, we need to create more precise theoretical predictions. Because the comparison of gravitational wave data with the theoretical predictions helps us not only to determine a number of astrophysical parameters of compact objects (i.e., their masses, spins and their orbital information), but also to obtain a deeper understanding of fundamental physics \cite{Cutler:1992tc}. 

With this motivation in hand, in recent times many studies have been performed to investigate the relativistic spin effect of a spinning particle moving around a BH in various interesting scenarios such as scattering \cite{PhysRevD.96.084044,Maybee:2019jus}, particle acceleration \cite{Zhang:2016btg,PhysRevD.93.084025,Zaslavskii:2016dfh,2016CQGra..33j5014A,An:2017hby,PhysRevD.97.064024,PhysRevD.98.044006,PhysRevD.98.064027,PhysRevD.99.064022,PhysRevD.102.064046,Zhang:2020cpu,Liu:2019wvp} and others \cite{1976NCimB..34..365T,1979MNRAS.189..621A,Plyatsko:2005bh,Chicone:2005jj,Han:2008zzf,PhysRevD.81.084024,PhysRevD.82.084013,PhysRevLett.108.051104,PhysRevD.88.084005,PhysRevD.93.044015,PhysRevD.96.064051,PhysRevD.97.084056,Antoniou:2019lit,PhysRevD.99.104059,Nucamendi:2019qsn,2020IJMPD..2950121L,PhysRevD.102.024021,Shahzadi:2021upd,Atamurotov:2021imh,Bonocore:2020xuj,PhysRevD.105.104059}. Although, the work on spinning particle moving in a curved spacetime is first started with the works of Mathisson \cite{Mathisson:1937zz} and Papapetrou \cite{Papapetrou:1951pa} (hereafter called as MP), they first derived the equation of motion for the spinning particle in the context of GR, and extended by Tulczyjew \cite{Tulczyjew:1959} and Taub \cite{1964JMP.....5..112T}. Whereas the theory of spinning particles is reformulated by Dixon \cite{Dixon:1970zza,Dixon:1970zz,dixon1964covariant}. The spinning particle in MP theory is an extended entity with a finite size that is significantly less than the characteristic length of spacetime, and so has a dipole moment in addition to the monopole that describes point particles. Another rationale to investigate the spinning particle in a curved spacetime comes from the MP equations of motion, which tell us that a path taken by a spinning particle is non-geodesic due to the coupling between the particle's spin and gravity.

In fact, the motion of spinning test particle deviates from geodesics due to gravitational interaction between the curvature of the spacetime and the spin of the particle as a consequence of spin-orbit coupling, thus corresponding to non-geodesic motion \cite{verhaaren2010chaotic}. The spin of test particles approximates the motion to higher degree and the orbits of spinning particles around central compact massive objects are calculated based on the equations which were first derived by MP \cite{Mathisson:1937zz,papapetrou1951spinning} and then reformulated by Dixon \cite{suzuki1997chaos,kyrian2007spinning,harms2016spinning}. The equations of motion obtained under the approximation of Mathisson-Papapetrou-Dixon are popularly known as MPD equations. These equations describe the motion of spinning particles in the pole-dipole approximation where the multipole moments of the particle higher than mass monopole and spin dipole are ignored \cite{0264-9381-28-19-195025,suzuki1998innermost,hojman1977spinning}. These equations generalize geodesic motion to a spinning particle and are written in terms of the momentum ($p^\mu $) and antisymmetric spin tensor ($S^{\mu\nu}$) of the particle.

It would therefore be quite interesting to investigate the motion of spinning test particles around a Schwarzschild BH spacetime surrounded by quintessence matter field (SQBH) \cite{uniyal2015geodesic}. Our current interest in the idea of quintessence, the dark energy in the universe is dominated by the potential of a scalar field that can be parametrized by providing the equation of state parameter ($\epsilon<0$) for the quintessence fluid. The existence of dark energy cannot be neglected to have an impact on our universe at any scale instead of the effect of this kind of energy is much negligible.  In this context, our approach to studying the possibility of chaotic orbits of spinning particles even in the presence of the most intriguing dark energy candidate ``quintessence'' by using effective potential analysis as an indicator of chaos, in conjunction with the study of ISCO and Periastron precession, is significant from an astrophysical standpoint.

The paper is organized as follows.  To make the study self-contained, \cref{sec:overiew_spin} discusses a brief description of the dynamics of massive spinning test particles in a curved spacetime. The motion of massive spinning test particle around SQBH is investigated in \cref{sec:SQBH_metric}.  In \cref{sec:effective_pot}, the effective potential for SQBH is derived followed by the classification of possibile chaotic orbits. We analyze the properties of ISCOs for a spinning particle moving in the vicinity of SQBH in \cref{sec:ISCO_Parameters}. Further, small-spin corrections and Periastron precession of spinning particles around SQBH are discussed in \cref{sec:ssc_and_peri}.  We finally give concluding remarks in \cref{sec:concluding_remarks}.
In addition, as supplemental material for the study of spinning particle dynamics in a spherically symmetric spacetime, we present the explicit form of MPD \cref{eq:MPD1,eq:MPD2}, and components of four-velocity $u^{\mu}$ in the equatorial plane ($\theta=\pi/2$) in \cref{sec:MPDeqns}.
Throughout the paper, we use natural system of units $G = c = 1$ and choose the signature of the metric as ($-,+,+,+$).
\section{A brief overview of the dynamics of massive Spinning Test Particles in a curved spacetime}\label{sec:overiew_spin}

 In this section, we briefly review the basic equations of motion for a massive spinning test particle moving in the background of a massive compact object. To describe the motion of a massive spinning test particle in a curved spacetime, the set of MPD equations in tensorial form read as
\begin{align}
u^{\mu} &= \frac{d x^{\mu}}{d \tau}, \label{eq:MPD0}\\
\frac{D p^\mu}{D \tau} &= - \frac{1}{2} u^{\pi} S^{\rho \sigma} R^{\mu}_{\pi \rho \sigma}, \label{eq:MPD1}\\
\frac{D S^{\mu \nu}}{D \tau} &= p^\mu u^\nu- p^\nu u^\mu, \label{eq:MPD2}
\end{align}
where $\tau$, $u^{\mu}$, $p^\mu$ and $S^{\mu \nu}$ are the particle's proper time, the four-velocity, four-momentum of the particle and the spin tensor, respectively. The term $R^{\mu}_{\pi \rho \sigma}$ is the Riemann curvature tensor of the spacetime and $D/D\tau$ denotes the covariant derivative with respect to proper time. An interaction term between the curvature of the spacetime  $R^{\mu}_{\pi \rho \sigma}$ and the spin of the particle $S^{\mu \nu}$ as shown in \cref{eq:MPD1} is referred to as spin-orbit interaction, through which we can define the dynamics of a spinning particle. These \crefrange{eq:MPD0}{eq:MPD2} describe a spinning particle in the pole-dipole approximation, where the particle is characterized as mass monopole and spin dipole (i.e., the multipole moments of the particle are neglected). Consequently, this is the generalization of geodesic equation to spinning particles around a central gravitating objects. It is important to note here that throughout this work we are considering only the spin-orbit coupling up to the first order and hence, discard the newly found quadratic spin corrections to the MPD eqquations in \cite{Deriglazov:2017jub,Deriglazov:2018vwa} for the shake of simplicity.

The set of MPD \crefrange{eq:MPD0}{eq:MPD2} does not form a closed system. Hence, we need an extra constraint condition popularly known as spin supplementary condition. In this work, we work with the Tulczyjew-Dixon spin supplemental condition (TDSSC),
\begin{align}
p_{\mu} S^{\mu \nu}=0. \label{eq:SSS}
\end{align}
The TDSSC is commonly applied in order to find an acceptable trajectory of the spinning particle's centre of mass \cite{suzuki1999signature}. 
Using \cref{eq:SSS}, one can immediatley find a relation between $u^{\mu}$ and $p^{\mu}$ \cite{suzuki1998innermost}, which reads as
\begin{align}
u^{\mu}=\frac{\mathcal{N}}{m}\left(p^{\mu}+\frac{2S^{\mu\nu}p^{\lambda}R_{\nu\lambda\rho\sigma}S^{\rho\sigma}}{\mathcal{Y}}\right),\label{eq:u_mu}
\end{align}
where
\begin{align}
    \mathcal{Y}=4m^{2}+R_{\alpha\beta\gamma\delta}S^{\alpha\beta}S^{\gamma\delta},\label{eq:Y}
\end{align}
and $\mathcal{N}$ is a normalization constant, which can be fixed by choosing a specific choice of proper time $\tau$. 
As we are choosing $\tau$ as the proper time, then the four-velocity is normalized i.e., $u_{\mu}u^{\nu}=-1$. 
Further, $v^{\mu}\equiv P^{\mu}/m$ is defined as the unit vector parallel to the four-momentum $P^{\mu}$
and $m$ is taken as the mass of spinning test particle defined by 
\begin{align}
m^{2}= - p_{\nu} p^{\nu}.\label{eq:conservation_4p}
\end{align} 
The magnitude of spin $S$ is however determined by relation\\
\begin{equation}
S^{2}= \frac{1}{2}S_{\mu \nu}S^{\mu \nu}.
\label{eq:conservation_spin}
\end{equation} 
Both $m$ and $S$ are constants of motion for a test particle in addition to conserved quantities that are coming from the symmetry of the spacetime. 

In particular, if the spacetime possesses some symmetry described by a Killing vector $\xi^{\mu}$ corresponding to which a constant of motion associates. The constant of motion which is given by the standard expression for geodesics 
and a contribution due to a coupling with the spin tensor, can be expressed as \cite{hartl2003survey}, 
\begin{equation}
C=\xi^{\mu} p_{\mu}-\frac{1}{2}\xi_{\mu;\nu}S^{\mu \nu},
\label{conservation_eq}
\end{equation}
where, $\xi_{\mu;\nu}$ is the covariant derivative of the Killing vector.

Now, as the mass $m$ and spin $S$ both are conserved quantities, and remain invariant under the reparametrization of the proper time parameter $\tau$, we can fix $\tau$ so that it satisfies the relation
\begin{align}
    p^{\mu}u_{\mu}=-m.
\end{align}
Using this relation and the \cref{eq:MPD2}, the four-momentum  can be defined as:
\begin{equation}
p^{\mu}=-m u^{\mu}+u_{\nu}\frac{D S^{\mu \nu}}{d\tau},
\end{equation}

The magnitude of spin $S$ quantifies the size of the spin which plays a crucial role in determining the behavior of spinning test particle systems.
In order to understand the physical constraints on spin ($S$), we measure distances and time in terms of mass of central object, $M$  and momentum in terms of mass of orbiting particle, $m$. In these units, spin $S$ is measured in units of $m M$. Since the Papapetrou-Dixon equations are valid in the test particle approximation that hold for $m\ll M$ \cite{verhaaren2010chaotic}. Thus, for all physically realistic systems, an important physical constraint is that the spin parameter $S$ must satisfy $S\ll 1$. However, for astrophysical systems such as intermediate mass ratio inspirals, $S$ can be even more.
\begin{figure}
\begin{tabular}{c c}
\includegraphics[scale=0.28]{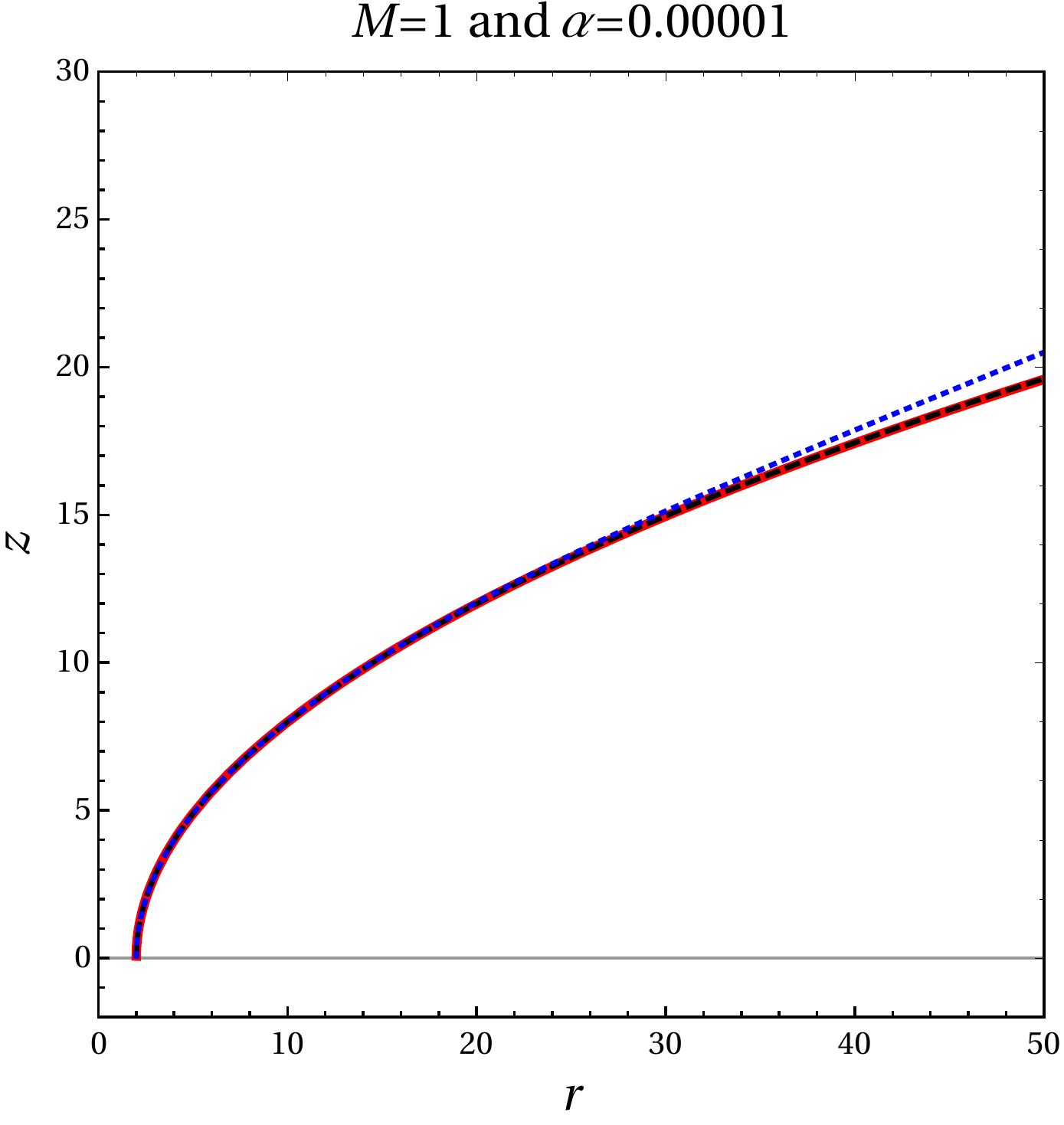}
&\includegraphics[scale=0.28]{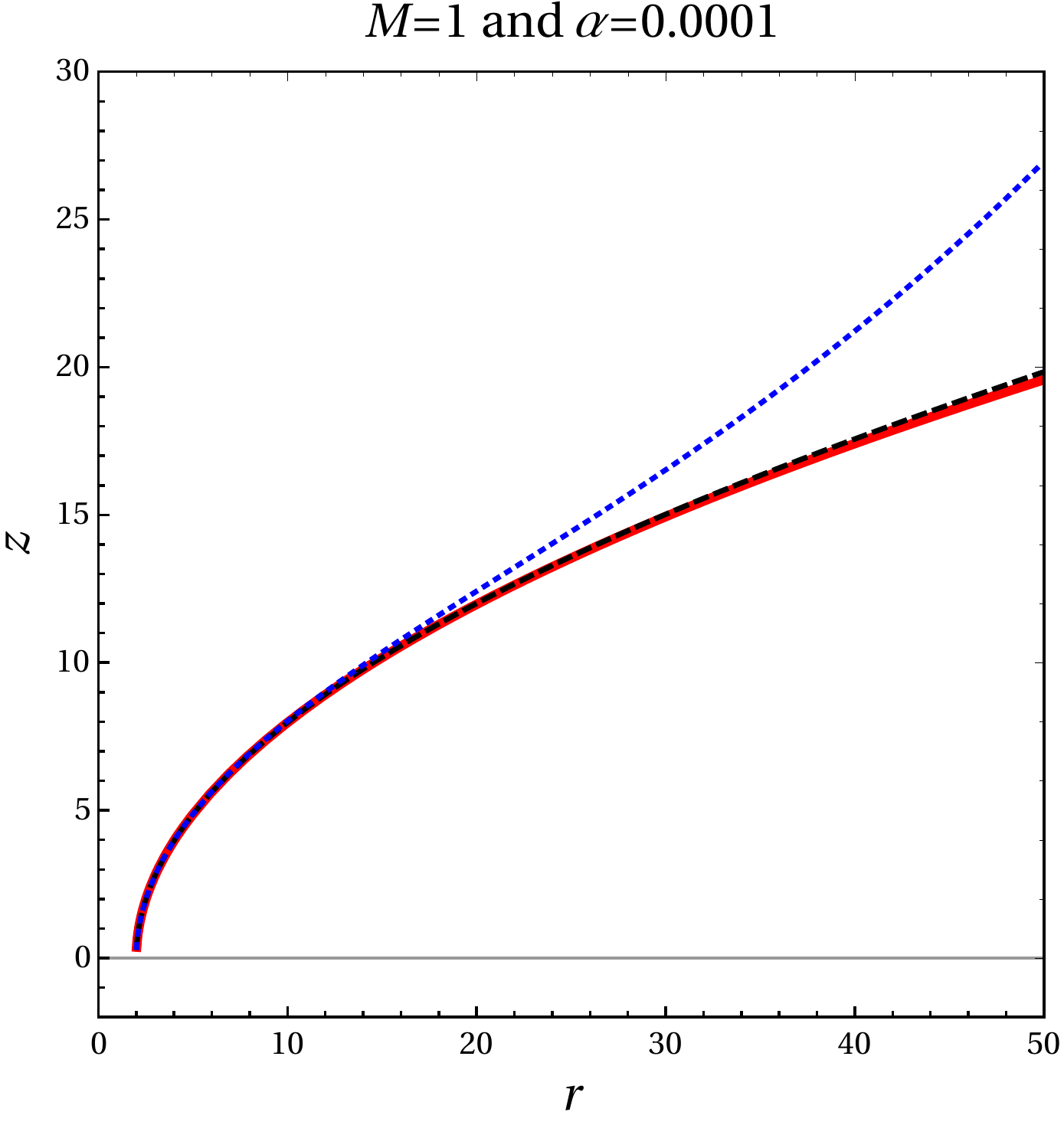}\\
\includegraphics[scale=0.28]{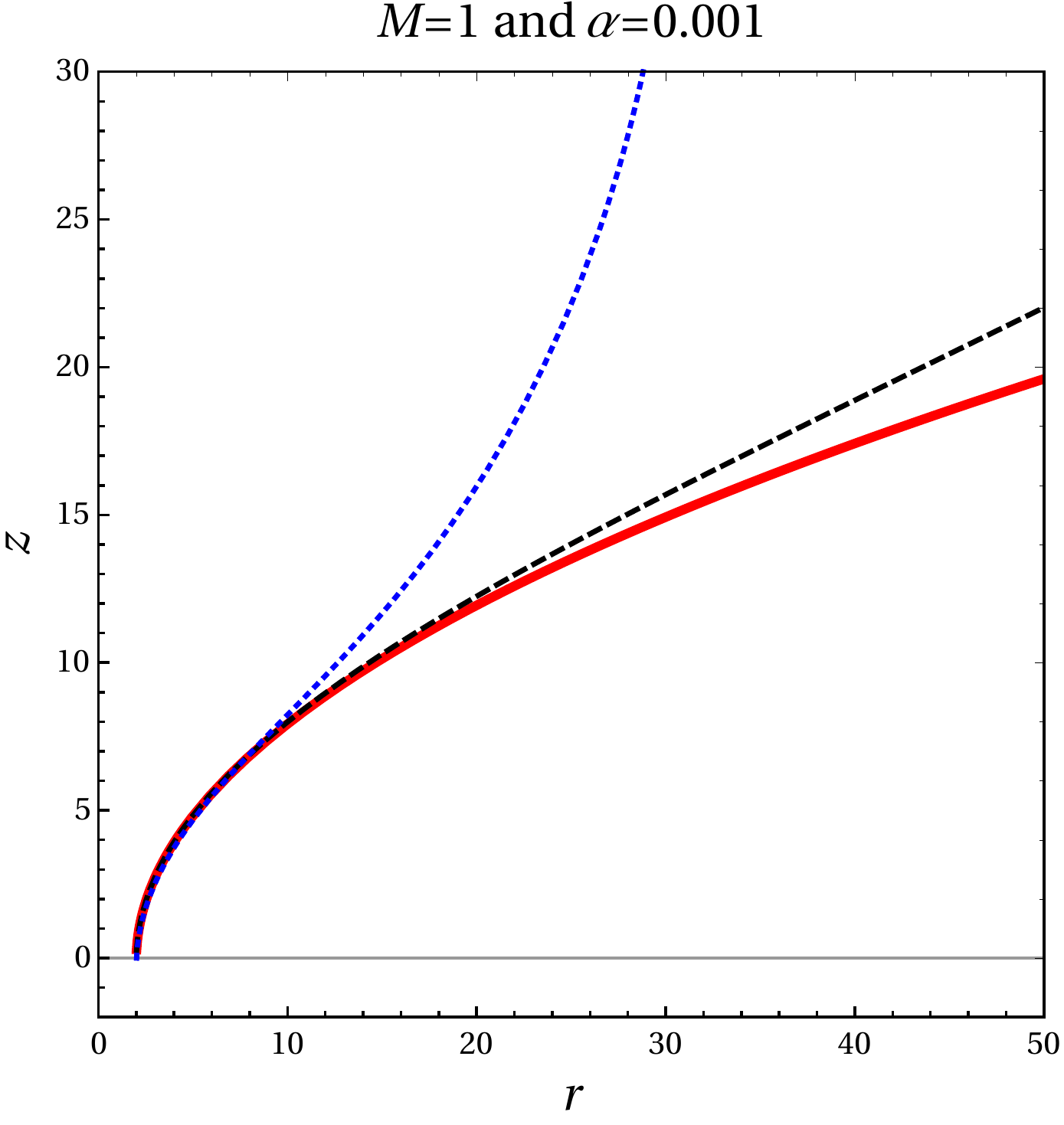}
&\includegraphics[scale=0.28]{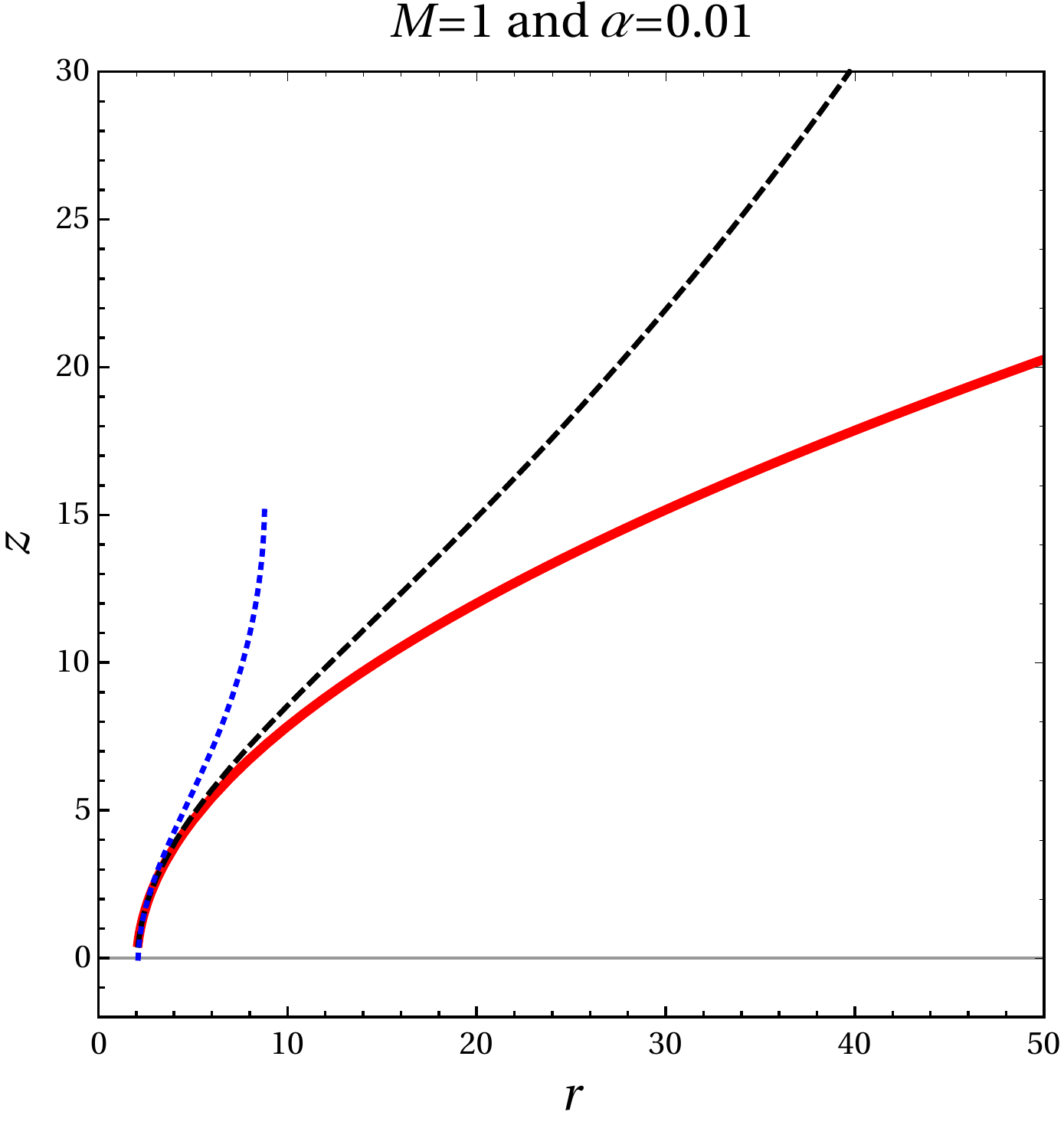}
\end{tabular}
\caption{The variation of $z$ as a function of $r$ for different combinations of parameters $\alpha$ and $\epsilon$. Here, the red (solid), black (dashed) and blue (dotted) curves respectively correspond to $\epsilon=-1/3$, $-2/3$ and $-1$. The mass parameter $M$ is set to unity.}\label{fig:2D_Embedded}
\end{figure}
\begin{figure*}[ht]
\begin{tabular}{c c}
\includegraphics[scale=0.5]{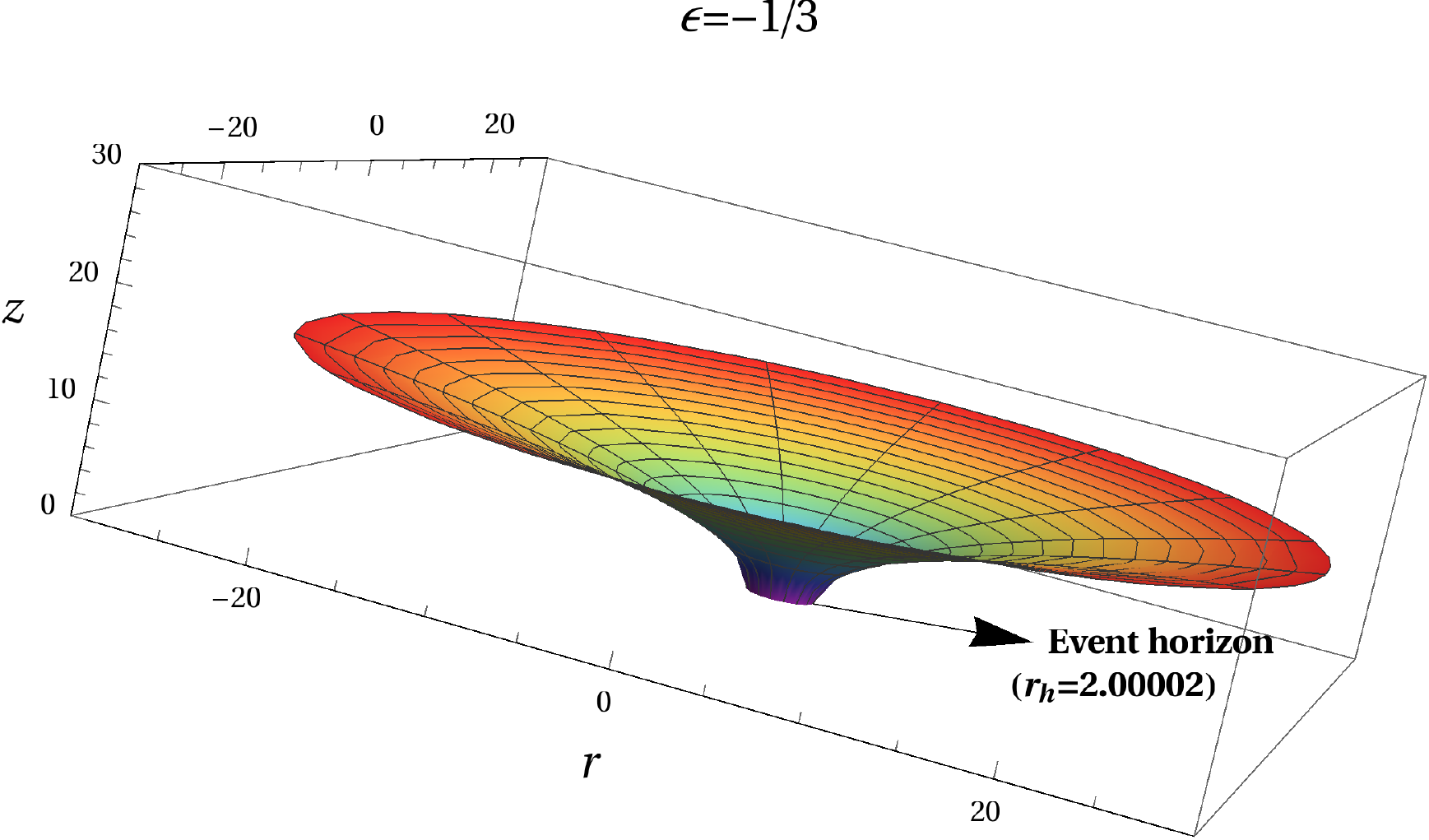}
&\includegraphics[scale=0.5]{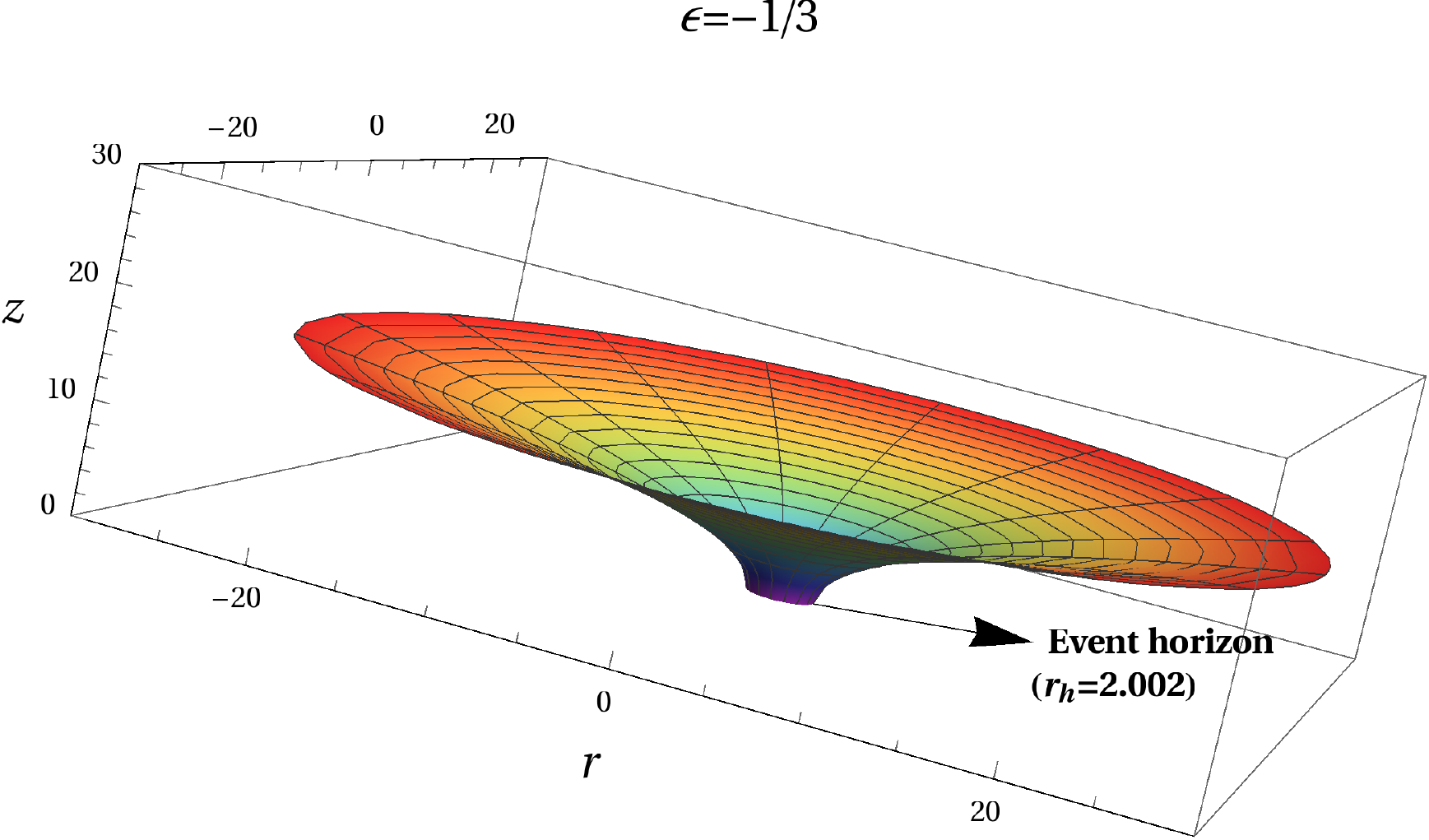}\\
\includegraphics[scale=0.5]{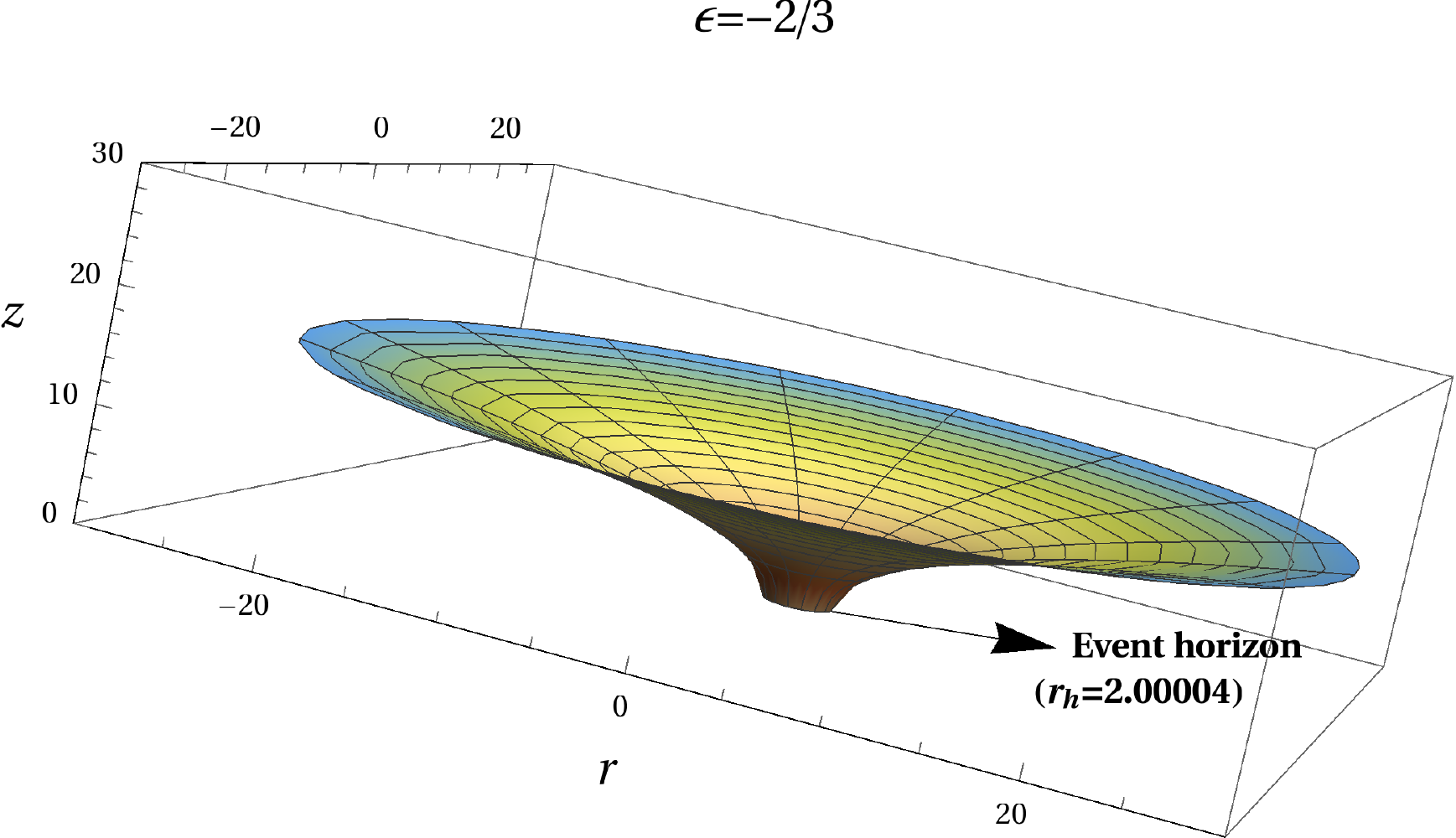}
&\includegraphics[scale=0.5]{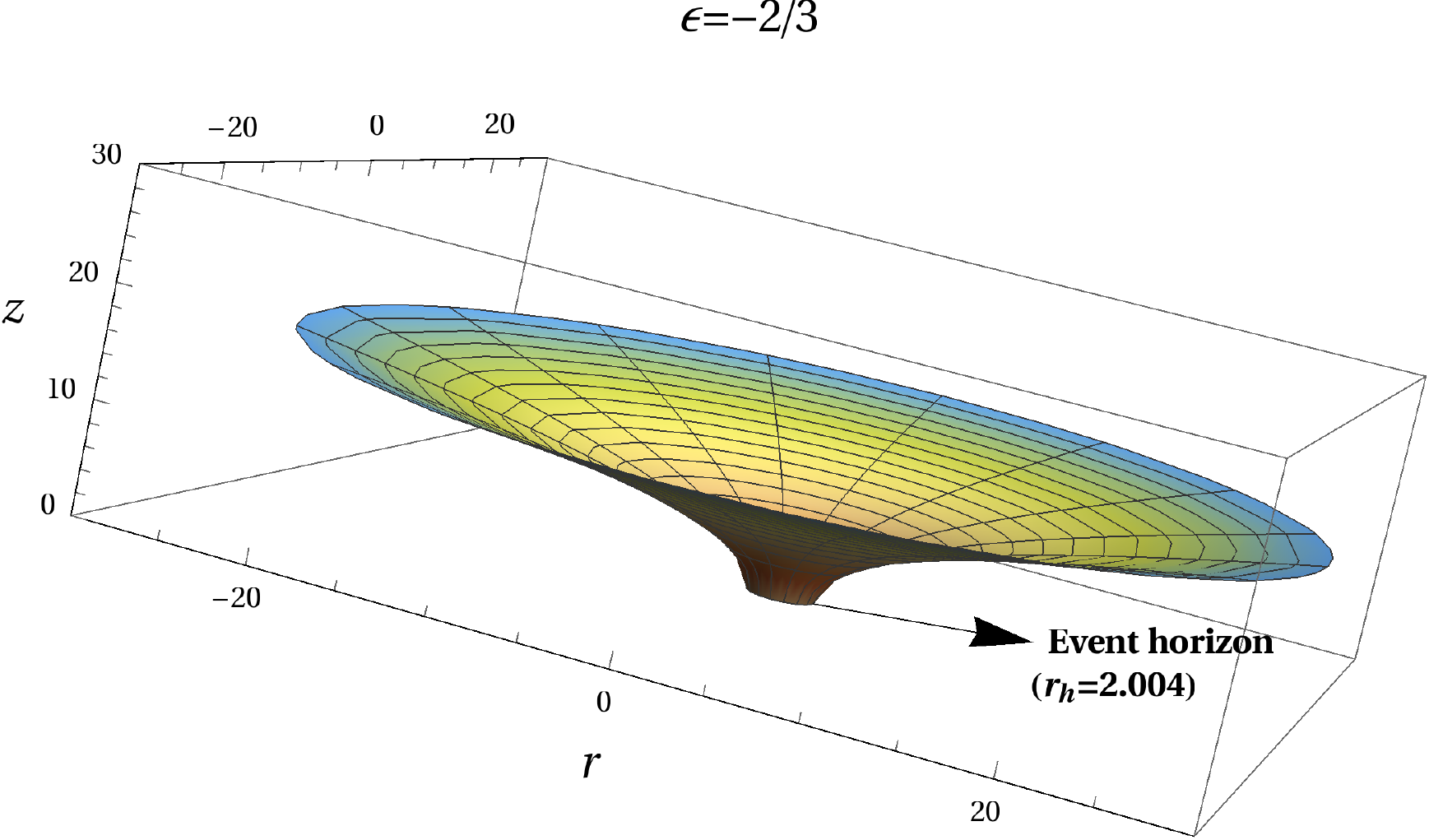}\\
\includegraphics[scale=0.5]{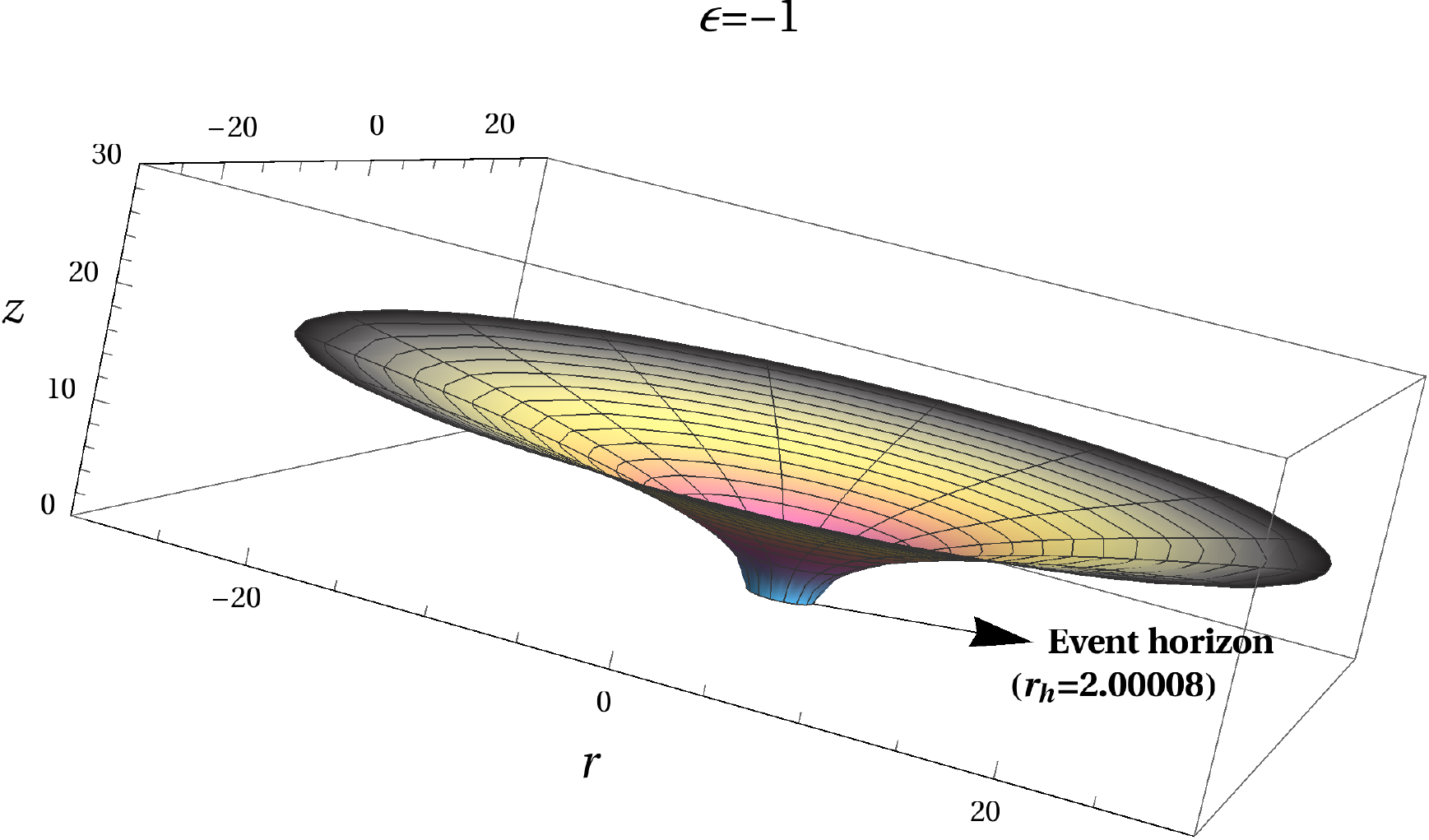}
&\includegraphics[scale=0.5]{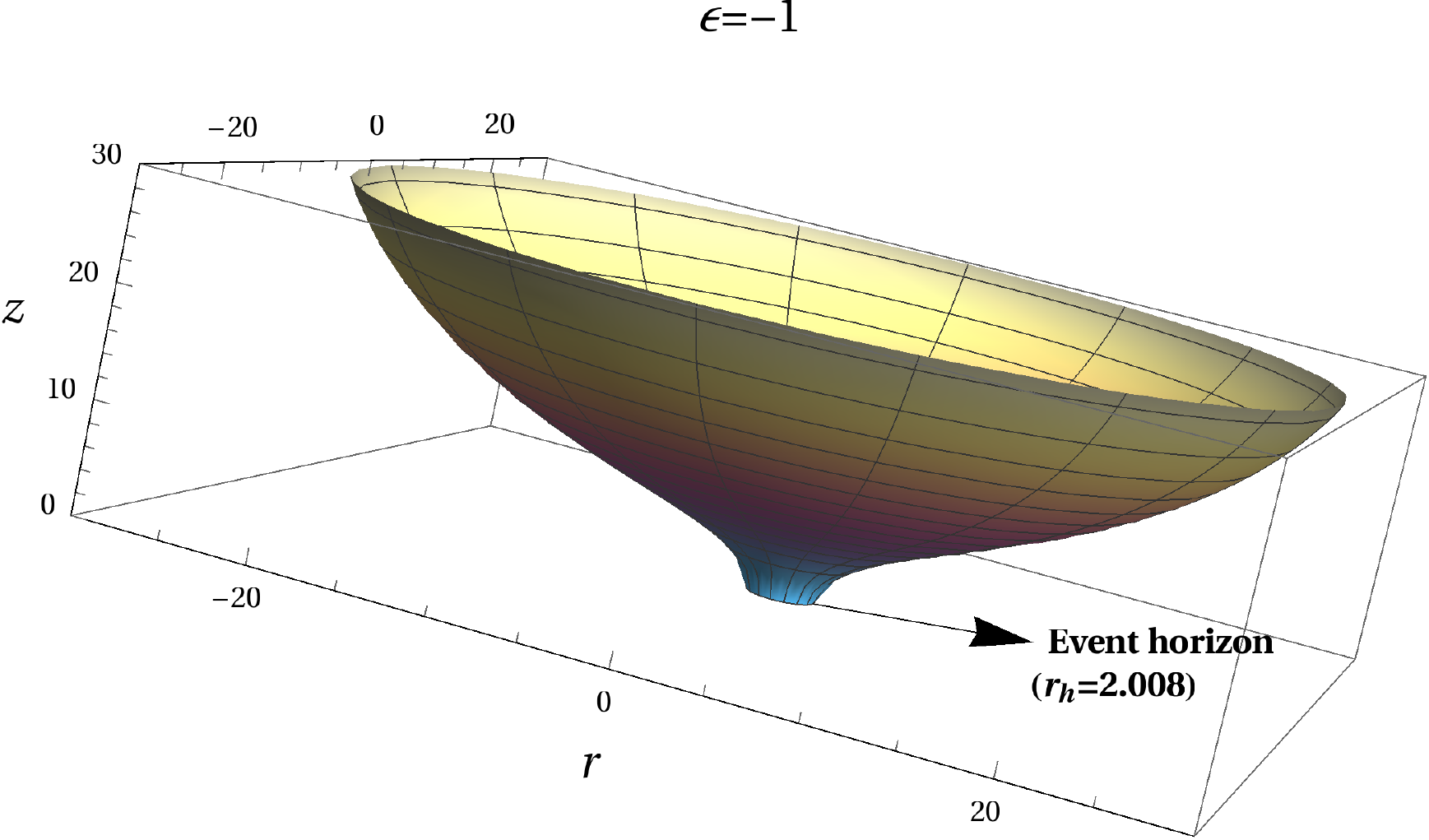}
\end{tabular}
\caption{$3D$ embedded diagrams of the SQBH in cylindrical coordinates ($r, \phi, z$). The normalization factor $\alpha$ is set to $0.00001$ for the \textit{left column} whereas it is set to $0.001$ for the \textit{right column}. Note that the mass parameter $M$ is fixed to unity.}\label{fig:3D_Embedded}
\end{figure*}
\section{Massive Spinning test Particle Motion in Background of SQBH}\label{sec:SQBH_metric}
\subsection{Embedded geometry of SQBH in cylindrical coordinates}
To study the motion of spinning test particle in field of a static and spherically symmetric SQBH. The line element of this type of spacetime in general is given by
\begin{align}
ds^{2} =  - f\left(r\right) dt^2+f^{-1}\left(r\right) dr^{2} + r^{2} d\theta^{2} + r^{2} \text{sin}^{2}\theta d\phi^{2},\label{eq:SQBH_metric}
\end{align}
where the metric function is given by
\begin{equation}
f(r)=  \left(1-\frac{2M}{r} - \frac{\alpha}{r^{3 \epsilon+1}}\right),
\end{equation} 
with $M$, $\alpha$ and $\epsilon$ being the mass, normalization factor and equation of state parameter $\left(-1\leq\epsilon\leq-1/3\right)$ of the SQBH, respectively. The spatial curvature of the SQBH for $r>r_{h}$ can be easily visualized by embedding the line element in a three-dimensional space at a constant time equatorial slice ($t=constant, \theta=\pi/2$) and allowing the position of a moving particle to be described solely with the help of remaining coordinates (i.e., $r$ and $\phi$). Now, with these limitations on the coordinates $t$ and $\theta$, the metric given by \cref{eq:SQBH_metric} takes the form
\begin{align}
ds^{2} = \left(1-\frac{2M}{r} - \frac{\alpha}{r^{3 \epsilon+1}}\right)^{-1} dr^{2} +  r^{2} d\phi^{2}.\label{eq:SQBH_metric1}
\end{align}
To embed the line element described by the \cref{eq:SQBH_metric1}, let us choose a Euclidean metric that in cylindrical coordinates $({r,\phi,z})$ reads as 
\begin{align}
    ds^{2}=dz^{2}+dr^{2}+r^{2}d\phi^{2}.\label{eq:Euclidean_metric}
\end{align}
The Euclidean metric \cref{eq:Euclidean_metric} can be furhter expressed as if the surface is described by the function $z=z(r)$
\begin{align}
    ds^{2}=\left[1+\left(\frac{dz}{dr}\right)^{2}\right]dr^{2}+r^{2}d\phi^{2},\label{eq:Euclidean_metric_1}
\end{align}
comparing \cref{eq:Euclidean_metric_1} with the SQBH metric (\ref{eq:SQBH_metric1}), leads to $dz/dr$ as
\begin{align}
    \frac{dz}{dr}
    =\pm \sqrt{\frac{2 M r^{3\epsilon}+\alpha}{r^{3\epsilon+1}-2 M r^{3\epsilon}-\alpha}}.\label{dzbydr}
\end{align}
Since, the above \cref{dzbydr} is difficult to analyze analytically, we numerically solve it to find the slope that represents the curvature of the spacetime (see, for example, \cref{fig:2D_Embedded,fig:3D_Embedded}). \cref{fig:2D_Embedded} shows the variation $z$ as a function of $r$ for different combinations of parameters $\alpha$ and $\epsilon$. Also, \cref{fig:2D_Embedded} shows that the spatial geometry of SQBH in cylindrical coordinate system is highly sensitive to the parameters $\alpha$ and $\epsilon$. The spatial geometry of SQBH gets separated more from one another when the parameter $\alpha$ grows for the respective equation of state parameter (i.e., -1/3 (red), -2/3 (black, dashed) and -1 (blue, dotted)).
The complete three-dimemnsional view of SQBH obtained from the solution of \cref{dzbydr} is presented in \cref{fig:3D_Embedded}. Here, we consider only two values of parameter $\alpha=0.00001$ and $0.001$ for various combinations of equation of state parameter $\epsilon=-1/3, -2/3\; \text{and} -1$, respectively.
\subsection{Conserved quantities}
Since the SQBH spacetime is static and spherically symmetric, it corresponds to two Killing vector fields,  i.e., $\xi^t$ and $\xi^\phi$.
Thus, the conservation \cref{conservation_eq} provides the conserved quantities associated with these two Killing vector fields as
\begin{align}
E &= -C_{t}= -p_{t}-\left[\frac{M}{r^{2}}+\frac{\alpha\left(3\epsilon+1\right)}{2r^{3\epsilon+2}}\right] S^{tr},\label{eq:conserved_E}
\\
J_{z} &=C_{\phi} = p_{\phi}-r \text{sin}^{2}\theta S^{\phi r}+r^{2} \text{sin}\theta\; \text{cos}\theta S^{\theta \phi}.\label{eq:conserved_Jz}
\end{align}
Here, $E$ and $J_{z}$ are defined as the conserved energy and z-component of the total angular momentum associated with the spinning particle, respectively. Additionally, there are two more conserved quantities for a spherically symmetry spacetime as mentioned in \cite{suzuki1997chaos}, namely $J_{x}$ and $J_{y}$ associated with the $x$ and $y$ components of the total angular momentum. These two conserved components of total angular momentum for SQBH are obtained as
\begin{align}
J_{x} \sin \theta &= -p_{\phi}\sin\phi-p_{\phi} \cos\theta \cos\phi - r \sin\phi S^{r\theta} \nonumber
\\ & +r^{2} \cos\phi \sin^{2}\theta S^{\theta \phi} + r \sin\theta \cos\theta \cos\phi S^{\phi r},\label{eq:conserved_Jx}
\\
J_{y} \sin \theta&= p_{\theta} \cos\phi- p_{\phi} \cos\theta \sin\phi+ r \cos\phi S^{r \theta} \nonumber\\
& + r^{2} \sin^{2}\theta \sin\phi S^{\theta \phi}
+ r \sin\theta \cos\theta \sin\phi S^{\phi r}.
\label{eq:conserved_Jy}
\end{align}
Now, without the loss of generality for a spherically symmetric spacetime, one can always choose that the total angular momentum $J$ pointed in the z-direction only:
\begin{align}
    J_{x}=0,\;\;\;J_{y}=0\;\;\;\text{and}\;\;\;J_{z}=J,
    \label{eq:Jz=J}
\end{align}
where, $J>0$. Then, using \cref{eq:conserved_E,eq:conserved_Jz,eq:conserved_Jx,eq:conserved_Jy} along with \cref{eq:Jz=J}, the $S^{r \theta},S^{\theta \phi}\;\text{and}\;S^{\phi r}$ components of spin tensor are obtained as
\begin{align}
S^{r \theta}&= \frac{-p_{\theta}}{r},\label{eq:srtheta}\\
S^{\theta \phi}&= \frac{J \text{cos}\theta}{ r^{2}~\text{sin} \theta},\label{eq:sthetaphi}\\
S^{\phi r}&=\frac{\left( p_{\phi}-J \text{sin}^{2}\theta \right)}{r \text{sin}^{2}\theta}.\label{eq:sphir}
\end{align}
The remaining spin components $S^{ti} (i= r,\theta,\phi )$ are derived from \cref{eq:SSS} together with \cref{eq:srtheta,eq:sthetaphi,eq:sphir} as follows
\begin{align}
S^{tr}&= \frac{- \left(p_{\theta}^{2} \text{sin}^{2}\theta +p_{\phi}^{2}-J p_{\phi}\text{sin}^{2}\theta \right)}{ p_{t}r\text{sin}^{2}\theta},\label{eq:str}
\\
S^{t\theta}&= \frac{\left(p_{r} p_{\theta} r \text{sin}\theta+ J p_{\phi} \text{cos}\theta\right)}{ p_{t}r^{2} \text{sin}\theta} ,\label{eq:sttheta}
\\
S^{t \phi}&= \frac{-\left(J p_{r} r\text{sin}^{2}\theta- p_{r} p_{\theta}r+ J p_{\phi} \text{cos}\theta \text{sin}\theta\right)}{p_{t} r^{2} \text{sin}^{2}\theta}.\label{eq:stphi}
\end{align}
Now, substituting $S^{tr}$ in \cref{eq:conserved_E}, we obtain the expression of energy as
\begin{align}
E=-p_{t}+ \frac{1}{r p_{t}}\left(\frac{M}{r^{2}}+\frac{\alpha\left(3\epsilon+1\right)}{2r^{3\epsilon+2}}\right) \left(p_{\theta}^{2}+\frac{p_{\phi}^{2}}{\text{sin}^{2}\theta}
-J p_{\phi}\right),\label{eq:conserved_E_wo_Str}
\end{align}
which is conserved energy of spinning particle.
 \section{Effective Potential for SQBH}\label{sec:effective_pot}
\begin{figure}
\includegraphics[scale=0.6]{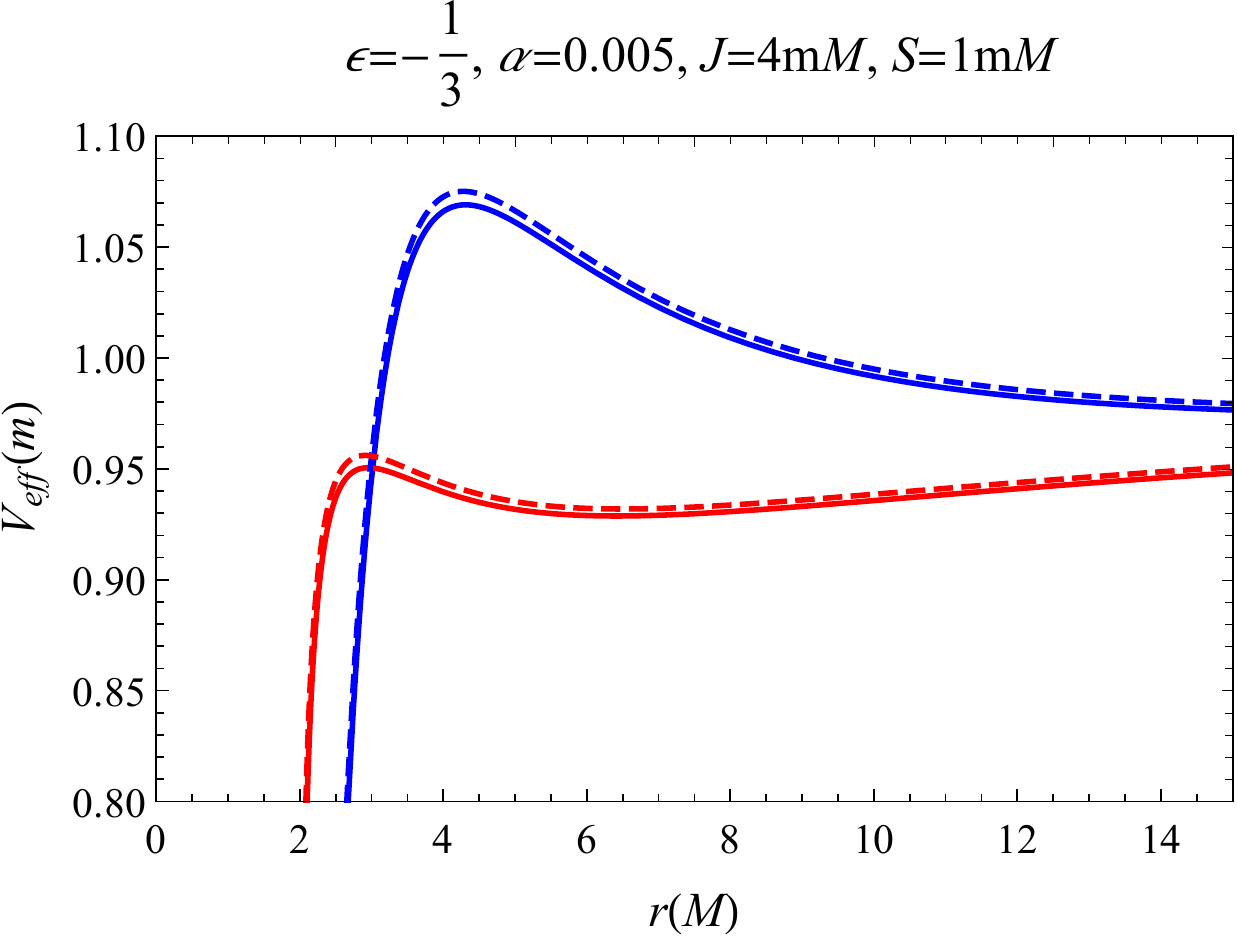}      
\includegraphics[scale=0.6]{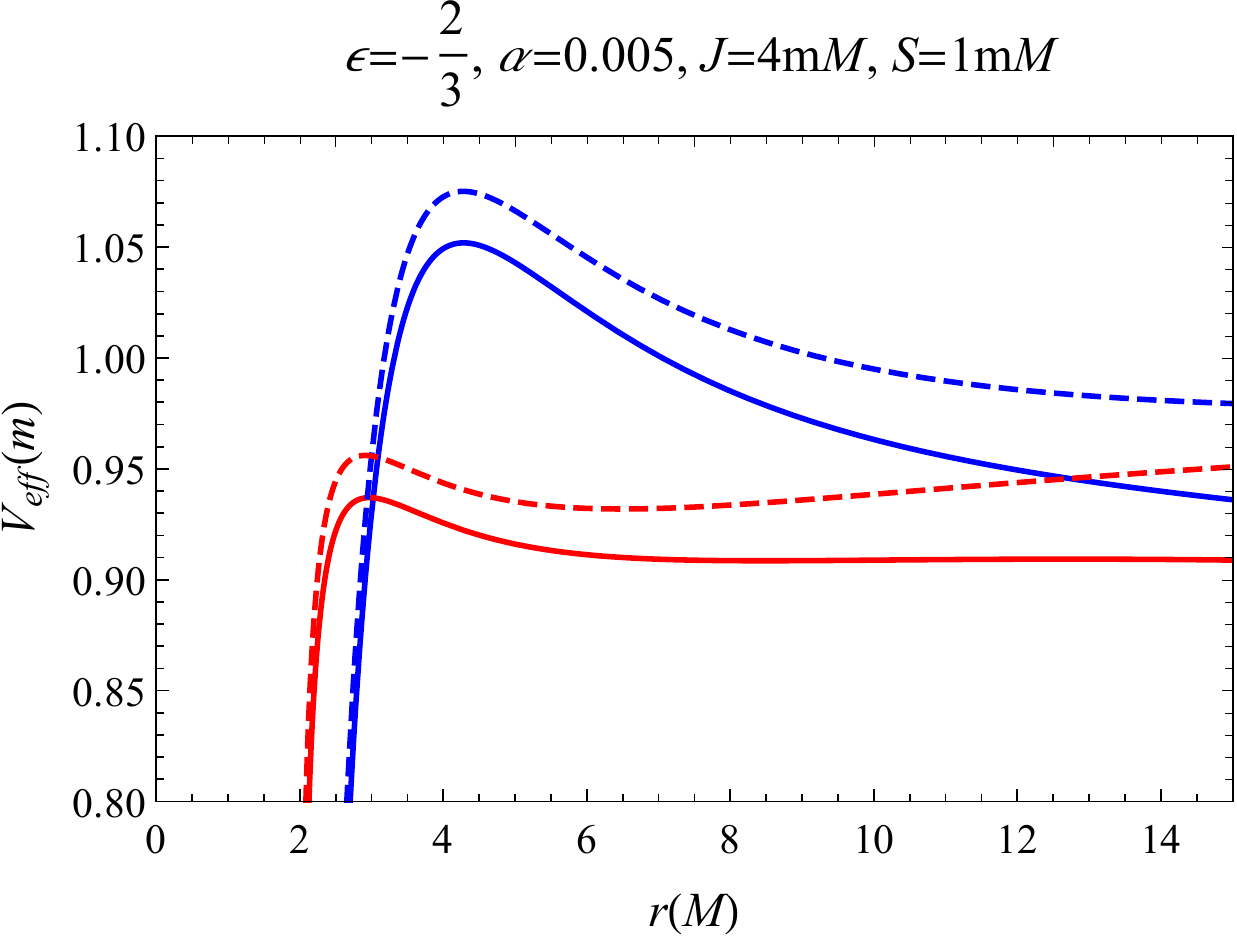}
\includegraphics[scale=0.6]{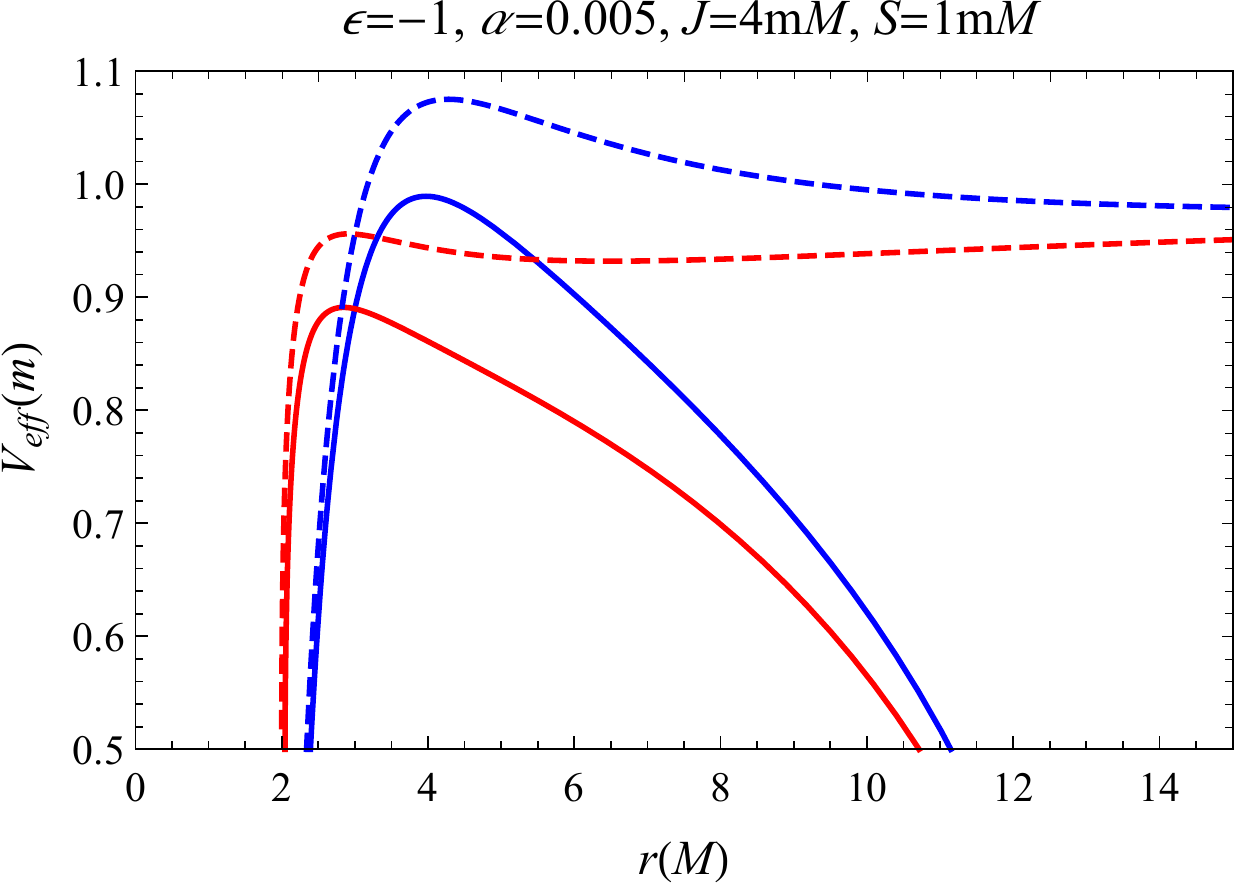}
\caption{A comparison of the effective potentials $V_{eff_\pm}$ 
for the SQBH (solid lines) for $\alpha=0.005$ and different $\epsilon= -1/3, -2/3, -1$ (top to bottom) with $J=4 mM$ and $S=1 mM$ and the Schwarzschild BH (dashed lines). Here, the blue and red curves represent $V_{eff_+}$ and $V_{eff_-}$, respectively.
}\label{fig:Veff_comparison}
\end{figure}
\begin{figure}
\begin{tabular}{c}
\includegraphics[scale=0.6]{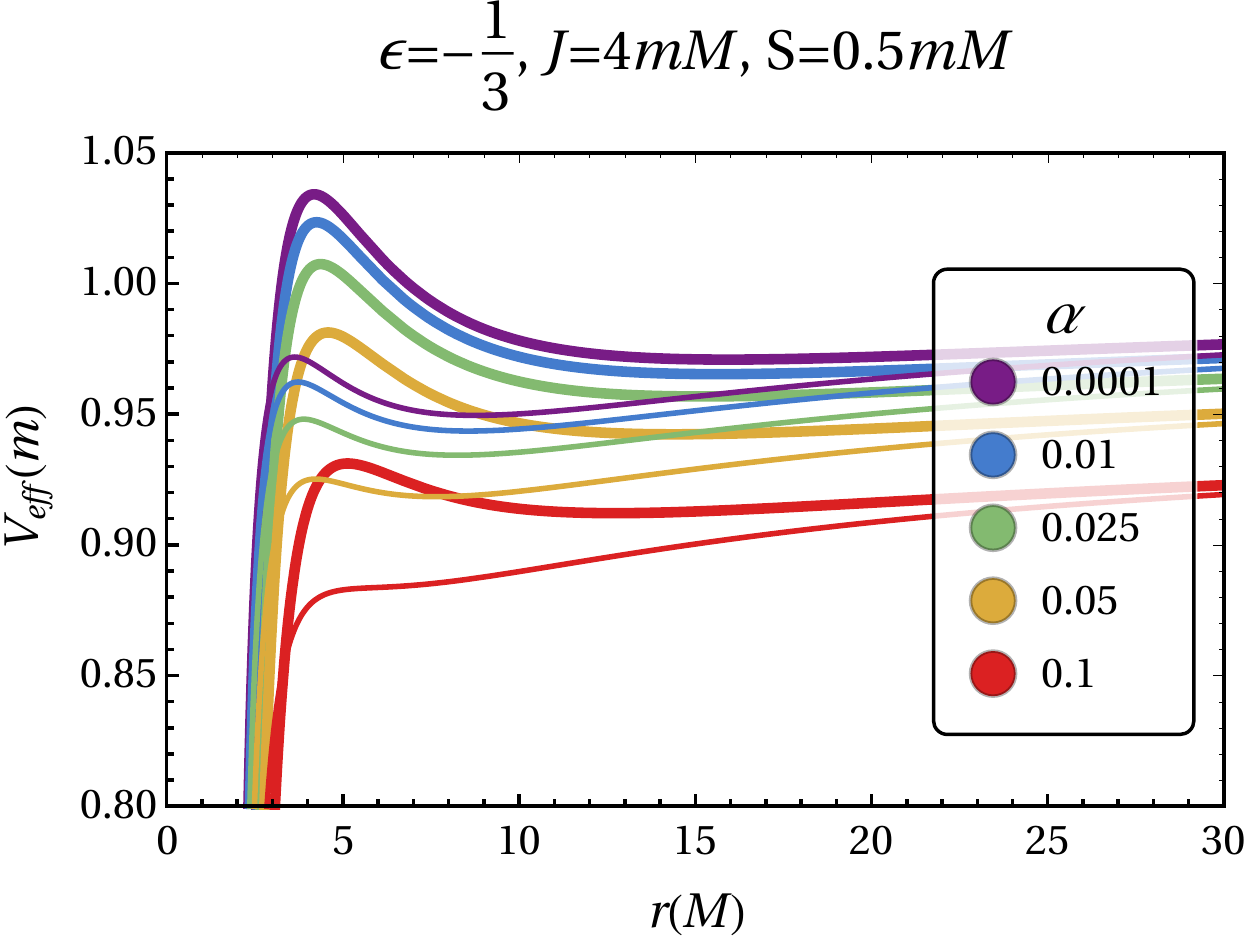}
\\
\includegraphics[scale=0.6]{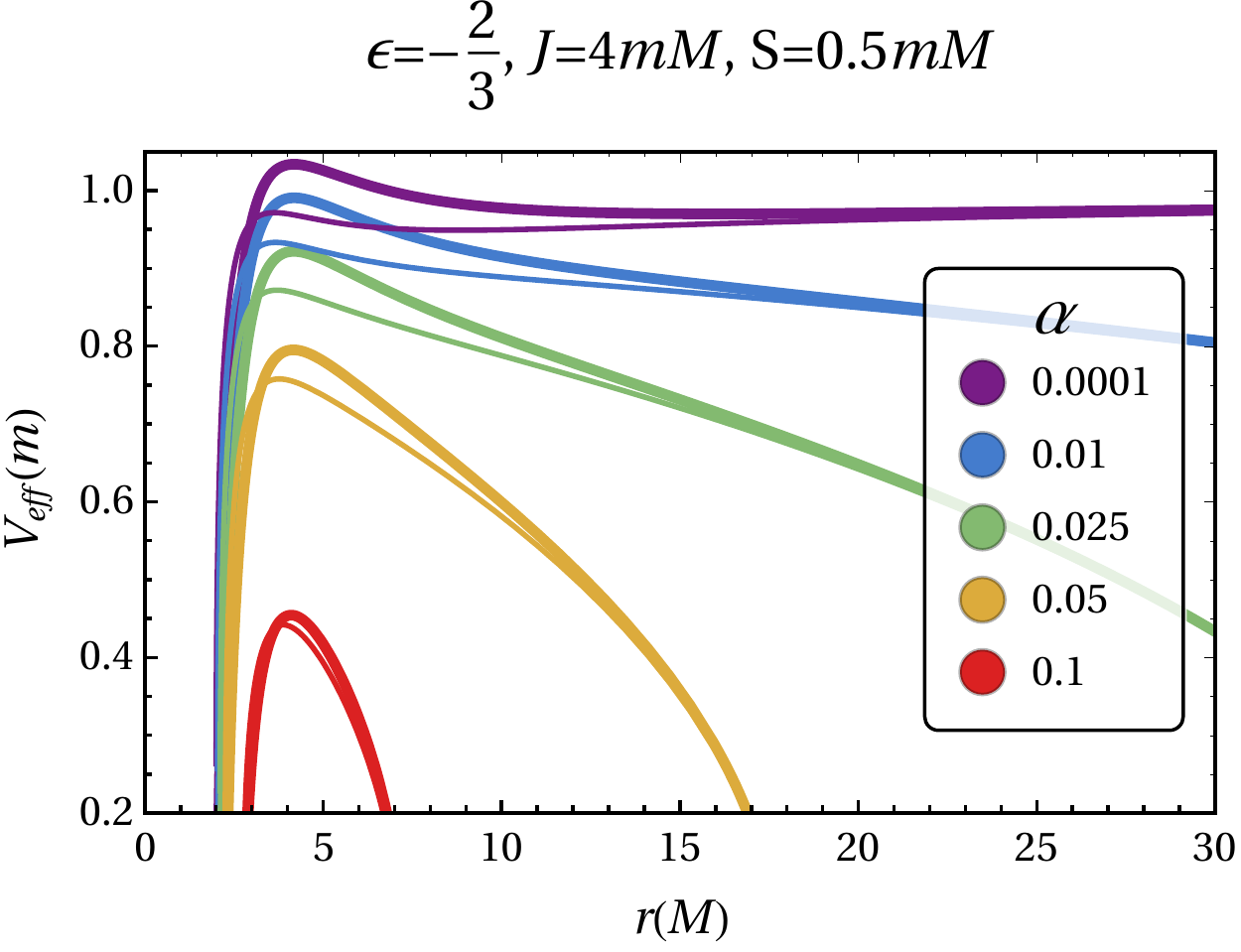}
\\
\includegraphics[scale=0.6]{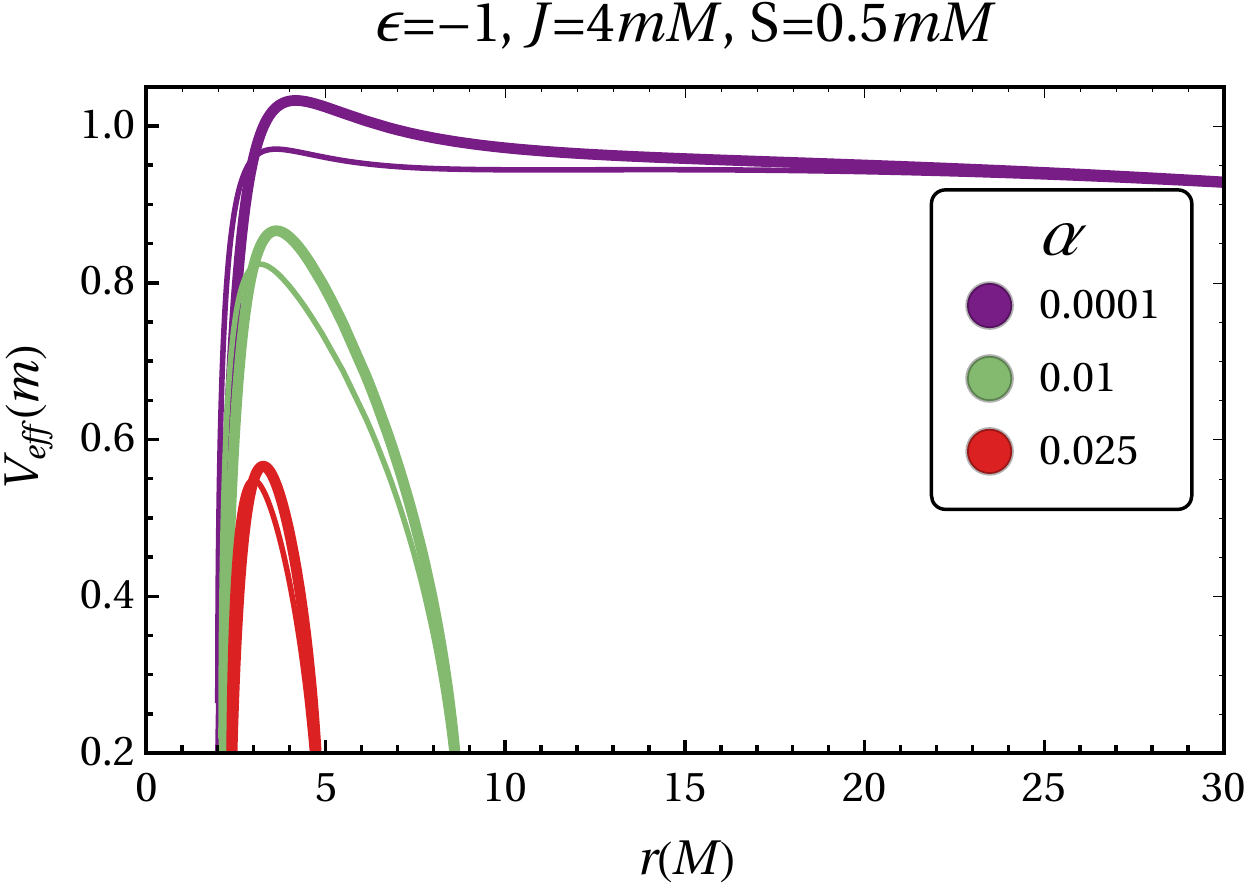}
\\
\end{tabular}
\caption{The variation of effective potential $V_{eff_\pm}$ 
with radial distance $r$ on the equatorial plane of the SQBH for different values of $\alpha$. In each panel, we fix the parameters $\epsilon =-1/3\;\text{top}$, \, $ -2/3\;\text{middle}$,\, $\text{and}\;-1\;\text{bottom}$,  $J=4mM$ and $S=0.5mM$.
Here, thick and thin lines correspond to $V_{eff_+}$ and $V_{eff_-}$, respectively.}\label{fig:effective_pot_for_dif_alpha}
\end{figure}
\begin{figure}
\begin{tabular}{c c}
\includegraphics[scale=0.85]{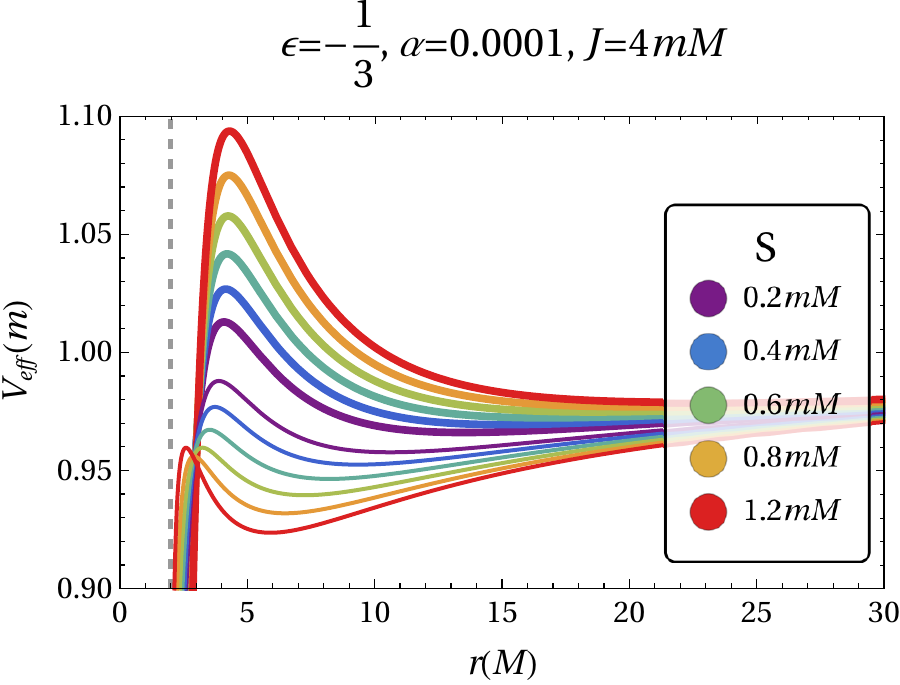}\\
\includegraphics[scale=0.85]{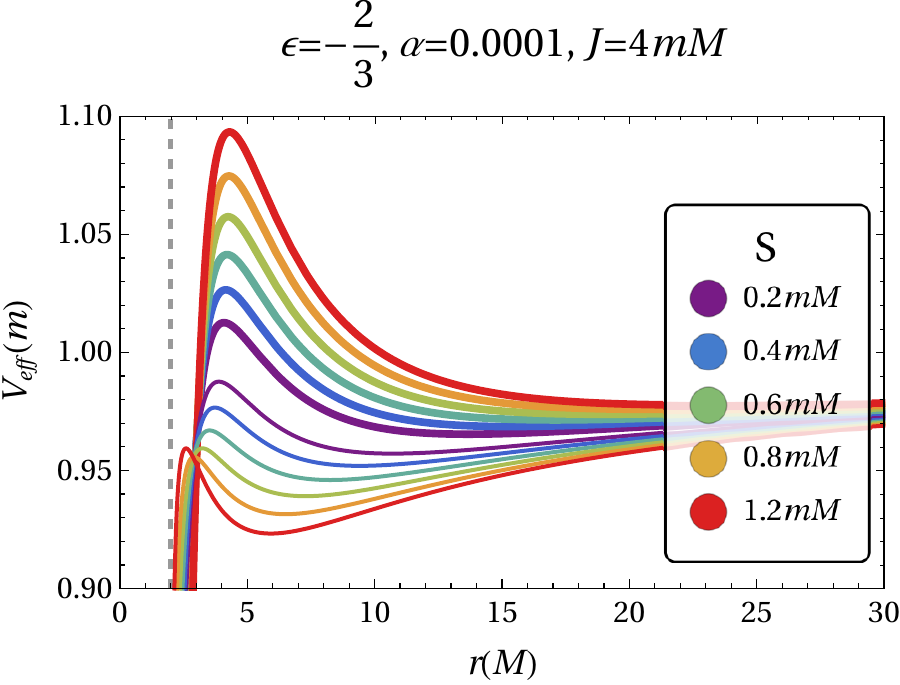}\\
\includegraphics[scale=0.85]{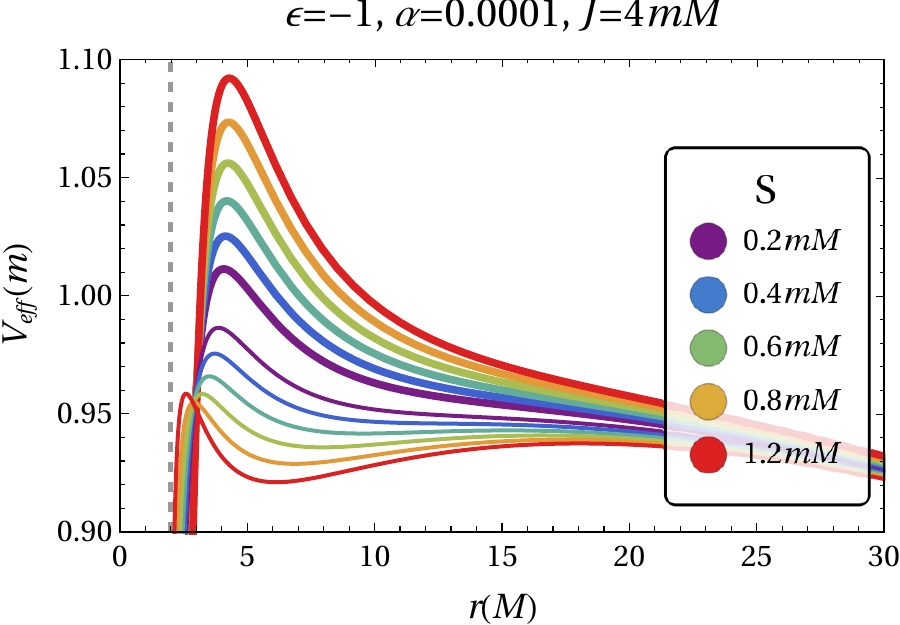}\\
\end{tabular}
\caption{Plots show the variation of effective potential $V_{eff_\pm}$ 
with radial distance $r$ on the equatorial plane of the SQBH. Here, thick, thin and dashed lines correspond to $V_{eff_+}$, $V_{eff_-}$ and the event horizon, respectively. The parameter $\epsilon$ is taken to be as $-1/3$ for {top}, $-2/3$ for {middle} and $-1$ for {bottom}, whereas we fixed $J=4mM$ and $\alpha=0.0001$ for  various combinations of the parameter $S$ in each panel. }\label{fig:effective_pot_for_dif_S}
\end{figure}
Since, the effective potential analysis have shown a good indications for a chaotic behaviour in spinning particles \cite{suzuki1997chaos}. In order to investigate such behaviour, we consider a spinning test particle that neither escapes to infinity nor falls into a SQBH. Therefore, to define the above mentioned spinning particle one has to choose its energy $E$, total angular momentum $J$ and spin $S$, accordingly. However, the choice of $E, J\; \text{and}\; S$ would be easy if we have an expression for the effective potential of a spinning particle. Contrary to the non-spinning particle case,  the case of particle with a spin is rather complicated because there exists an extra dynamical variable (i.e., direction of the spin), in addition to spin-orbits coupling. Hence, the spin-orbit coupling cannot be taken as a potential force for the case of spinning particle which in turn indicates that it is not possible to find an effective potential $V_{eff}$ in the $r-\theta$ plane only. Hence, in that situation it is sufficient only to find the boundary of a region in the $r-\theta$ plane where the particle can move. For such a curve the respective components of four-momentum vanish, i.e., $p^{r}=0=p^{\theta}$. This automatically vanishes $S^{r\theta}$ components of spin tensor (see; \cref{eq:srtheta}), implies that the spin direction is constrained now and it lies only in the meridian plane. Now, using the condition $p^{r}=p^{\theta}=0 $ with the conservation of four-momentum \cref{eq:conservation_4p}, one can define  
\begin{align}
&\;\;\;\; p_{t}= -m f(r)^{1/2} \cosh Z,\label{eq:pt}\\
\hspace{-6cm}
\text{and}\;\;\;\;\;\;\;\;\;\;\;\;&\nonumber\\
&\;\;\;p_{\phi}= m r \sin \theta \sinh Z ,\label{eq:p_phi}
\end{align}
where, $Z$ is an unknown function of $r\;\text{and}\;\theta$. It can be determined by inserting \cref{eq:pt,eq:p_phi} into the conservation of spin \cref{eq:conservation_spin} with relationships \cref{eq:srtheta,eq:sthetaphi,eq:sphir,eq:str,eq:sttheta,eq:stphi}. 

\begin{align}
&\left(m^{2} r^{2}-S^{2}f(r) \right) \sinh^{2}Z- 2J m r \sin\theta \sinh Z\nonumber\\ 
&+\left(J^{2}-S^{2}\right) f(r)+ \left(\frac{2M}{r}+\frac{\alpha}{r^{3\epsilon+1}}\right) J^{2} \sin^{2}\theta=0.\label{eq:quad_sinhz}
\end{align}
The above \cref{eq:quad_sinhz} is quadratic in $\sinh{Z}$, whose solution gives
\begin{align}
\sinh Z_{\pm}&= \frac{J m r \sin \theta}{m^{2}r^{2}- S^{2}f(r)}\pm \left[\frac{m^{2}J^{2}r^{2}\sin^{2}\theta}{\left(m^{2}r^{2}- S^{2}f(r)\right)^{2}}\right.\nonumber\\
&- \left. \frac{\left(J^{2}-S^{2}\right) f(r)+\left(\frac{2M}{r}+\frac{\alpha}{r^{3\epsilon+1}}\right)J^{2}\sin^{2}\theta}{m^{2}r^{2}- S^{2}f(r)}\right]^{1/2}.\label{eq:sinhz}
\end{align}
Here, the subscript $\pm$ refers to the direction of the particle's spin. It is negative $``-"$, when the total angular momentum $J$ and $z-$ component of the particle's spin both are parallel to each other, whereas it is positive $``+"$, when they are anti-parallel to each other.
Finally, from \cref{eq:conserved_E_wo_Str,eq:pt,eq:p_phi,eq:sinhz}, and using the relation $E= V_{\pm}(r,\theta, J, S)$, we obtain the expression for the potential of the spinning particle moving on the boundary of a region where $p^{r}=0=p^{\theta}$ in $r-\theta$ plane as
\begin{align}
   V_{\pm}&\equiv V_{\pm}(r,\theta, J, S)= m \left[f(r)^{1/2}\cosh Z_{\pm}+\left(\frac{M}{r}\right.\right.\nonumber\\
&\left.\left. +\frac{\alpha\left(3\epsilon+1\right)}{2r^{3\epsilon+1}}\right) \frac{\sinh Z_{\pm}}{f(r)^{1/2} \cosh Z_{\pm}} \left(\frac{J \sin \theta}{m r}-\sinh Z_{\pm}\right)\right]. \label{eq:v_pm}
\end{align}
The above \cref{eq:v_pm} reduces to the effective potential for Schwarzschild case \cite{suzuki1997chaos} in the prescribed limit $\alpha=0$. On the other hand, in the case when the conditions $\alpha\neq0, S=0\; \text{and\;} \theta=0$ are taken into account it then reduces to the potential for non-spinning test particle in SQBH spacetime. The form of the potential derived in \cref{eq:v_pm} is referred to as the $``$effective potential$"$ $V_{eff}$ of a spinning particle moving in the vicinity of a SQBH subjected to the condition $V^{2}_{\pm}<E^{2}$. 

In \cref{fig:Veff_comparison}, we show the variation of $V_{eff_\pm}$ on the equatorial plane, $\theta=\pi/2$,  as the function of radial distance $r$ for 
keeping normalization constant fixed (i.e., $\alpha=0.005$) as a consequence of increasing the equation of state parameter $\epsilon$ from top ($\epsilon=-1/3$) to bottom ($\epsilon=-1$) in steps of $-1/3$. Also, we present the comparison of $V_{eff_\pm}$ of SQBH with the Schwarzschild BH in \cref{fig:Veff_comparison}. From this figure, it is easy to see that $V_{eff_\pm}$ for SQBH at larger values of $r$ deviates significantly from the Schwarzschild BH case and this deviation increases as the parameter $\epsilon$ increases from -1/3 to -1 for keeping parameters $\alpha, J$ and $S$ fixed. Additionally, from these figures, one can conclude that the particle for the case of $V_{-}$ can move closer to the event horizon ($\approx 2M$) of SQBH than that for $V_{+}$ case. This is similar to that observed for the Schwarzschild BH in \cite{suzuki1997chaos}. In \cref{fig:effective_pot_for_dif_alpha,fig:effective_pot_for_dif_S}, we plot $V_{eff_\pm}$ for different combinations of $\alpha$ and $S$, respectively and increase the parameter $\epsilon$ from top ($-1/3$) to bottom ($-1$). From \cref{fig:effective_pot_for_dif_alpha}, one can easily see that as the parameter $\alpha$ increases the maximum of $V_{eff_\pm}$ decreases for each value of equation of state parameter $\epsilon$. One can also observe here that when the parameter $\alpha$ rises, the closest distance which a spinning particle can reach moves further away from the event horizon, for all the values of $\epsilon$. Contrary to this, we interestingly observe from \cref{fig:effective_pot_for_dif_S} that when the particle's spin increases while keeping all other parameters (i.e., $\epsilon, \alpha\;\text{and}\; J$) fixed, it can move closer to the event horizon.

\subsection{Classification of the Effective Potential and possibility of chaotic orbits}
    To investigate the chaotic behavior of the spinning particle around the SQBH, we consider the cases where the gravitational field will be strong and account for large spin-orbit interactions. In order then to satisfy these conditions, the spinning particle needs to move closer to the event horizon and its natural to choose $(V_{eff})_{-}$ instead of $(V_{eff})_{+}$. As a result, in this subsection, we just consider $(V_eff)_{-}$, and for the sake of simplicity, we omit the subscript ($-$) from here on.

\begin{figure*}
\begin{tabular}{c c}
\includegraphics[scale=0.75]{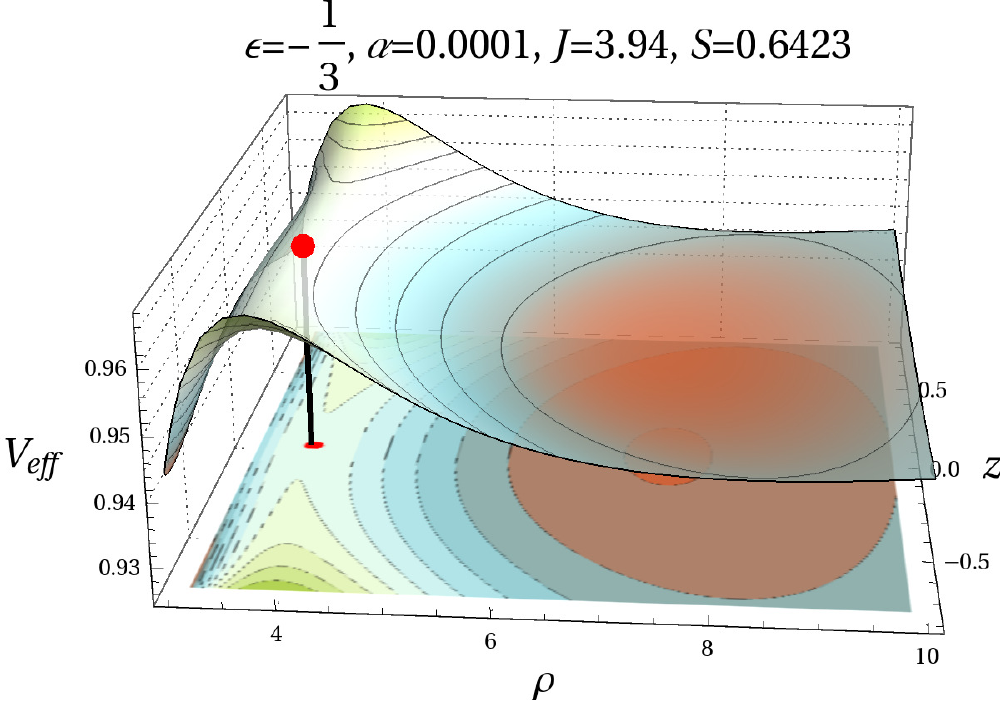}{(a)}\hspace{-0.2cm}
\includegraphics[scale=0.55]{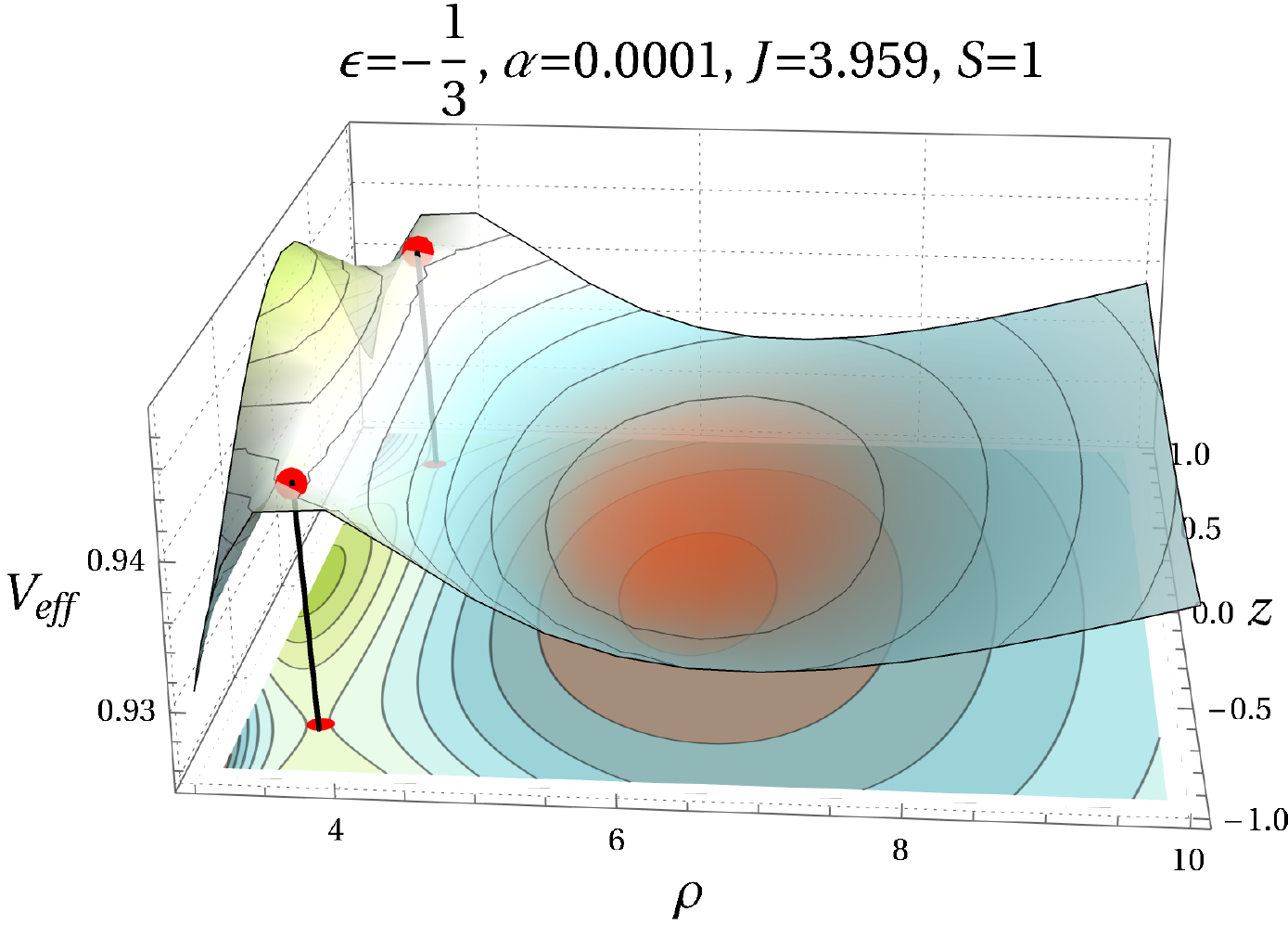}{(b)}\\
\includegraphics[scale=0.75]{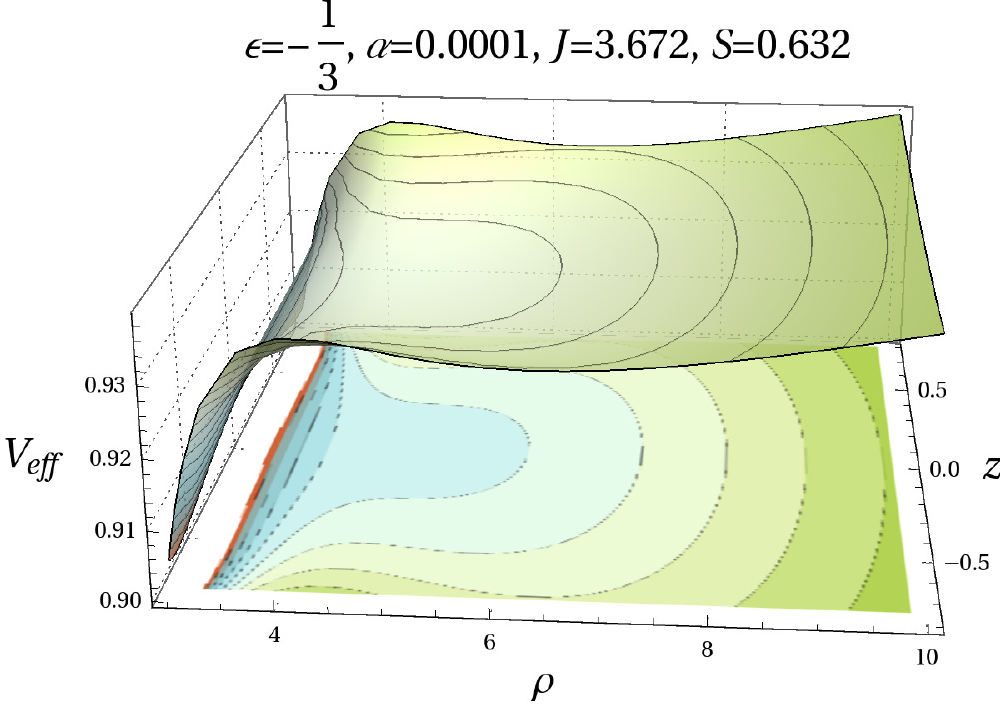}{(c)}\hspace{0.2cm}
\includegraphics[scale=0.75]{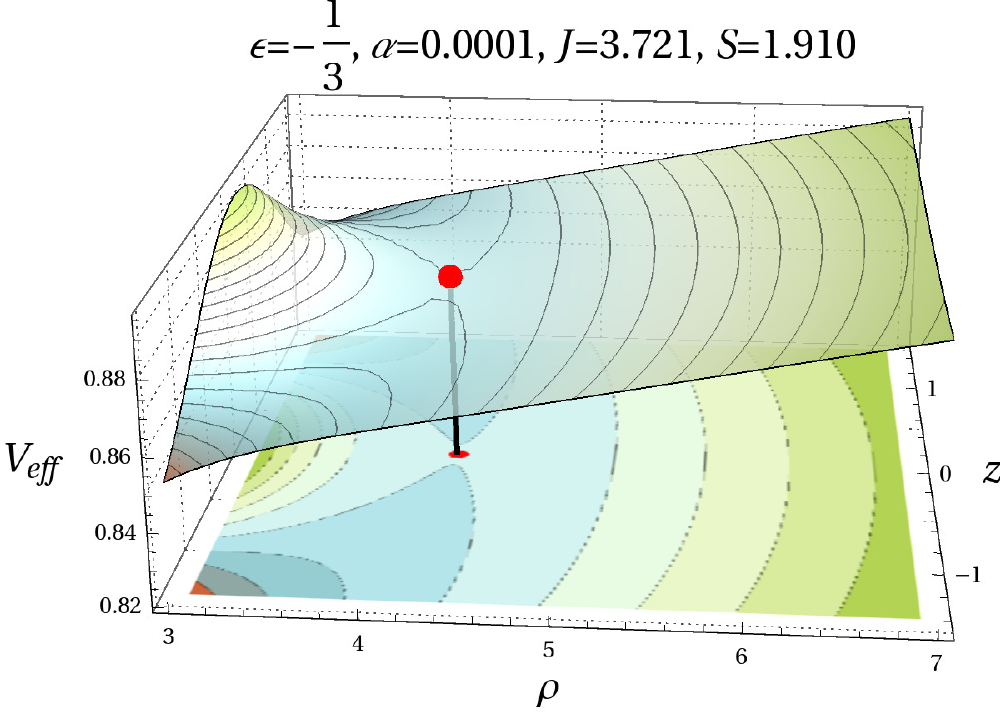}{(d)}
\end{tabular}
\caption{Classification of the effective potential $V_{eff}$ 
in cylindrical polar coordinate system for $ \epsilon =-1/3$ and $ \alpha =0.0001$ depending on various combinations of $J$ and $S$. Top row, left panel (a): For $J=3.94$ and $S=0.6423$, one saddle point (marked by a red dot) and one minimum exist on the $\theta=\pi/2$ plane, i.e., z=0. Top row, right panel (b): For $J=3.959$ and  $S=1$, two saddle points appear on both side perpendicular to the $\theta=\pi/2$ plane along side with a minimum. Bottom row, left panel (c): For $J=3.672$ and $S=0.632$, no saddle point and no minimum appear. Bottom row, right panel (d): For $J=3.721, S=1.910$, there exists a saddle point on the $\theta=\pi/2$ plane, which is locally minimal in radial direction and maximal in angular direction (d). Here, we have defined $J\to J/mM$ and $S\to S/mM$ as well as 
the cylindrical coordinates by $\rho\to \rho/M$ and $z\to z/M$ 
}\label{fig:3D_Veff}
\end{figure*}

\begin{figure}
\begin{tabular}{c c}
\includegraphics[scale=0.45]{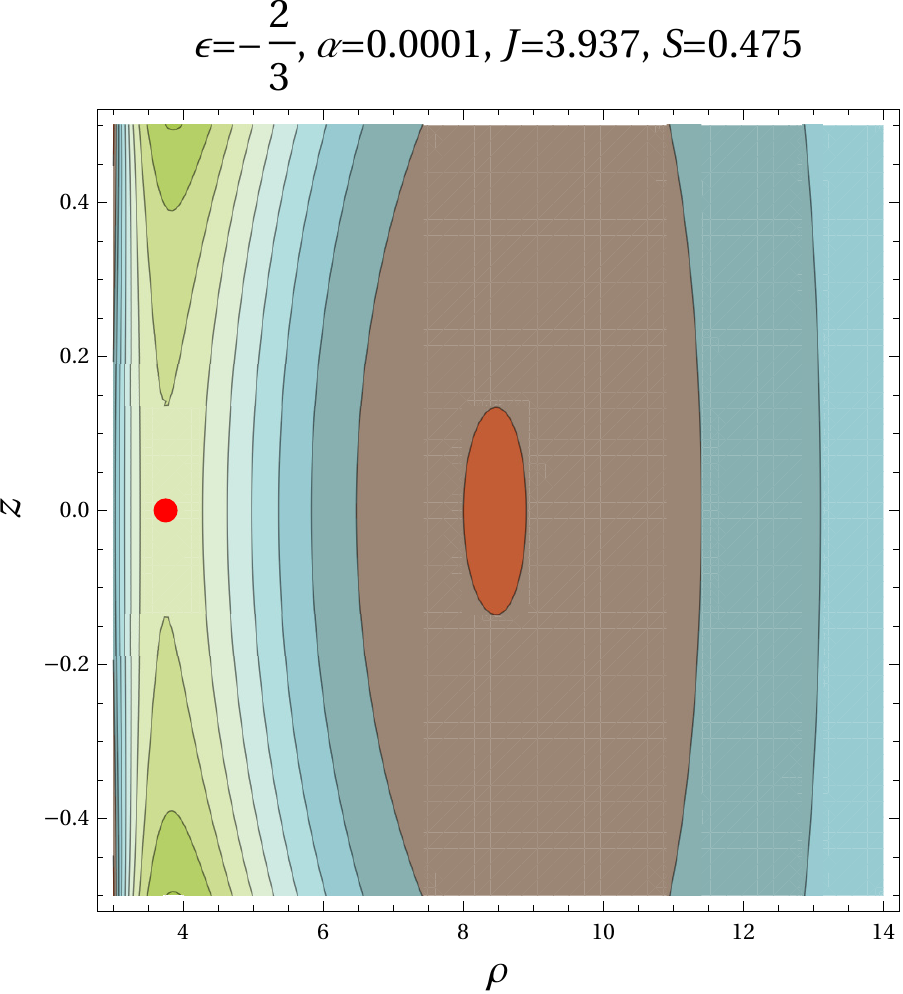}\hspace{0.2cm}
\includegraphics[scale=0.45]{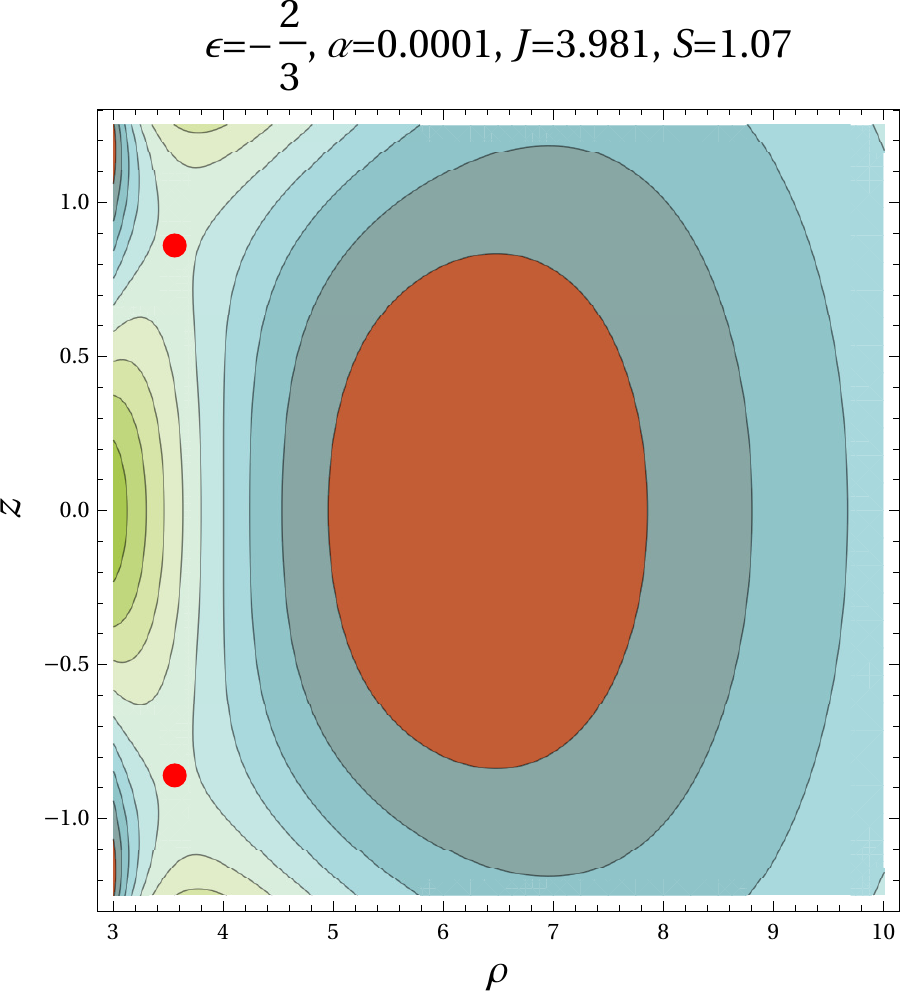}\\
\includegraphics[scale=0.45]{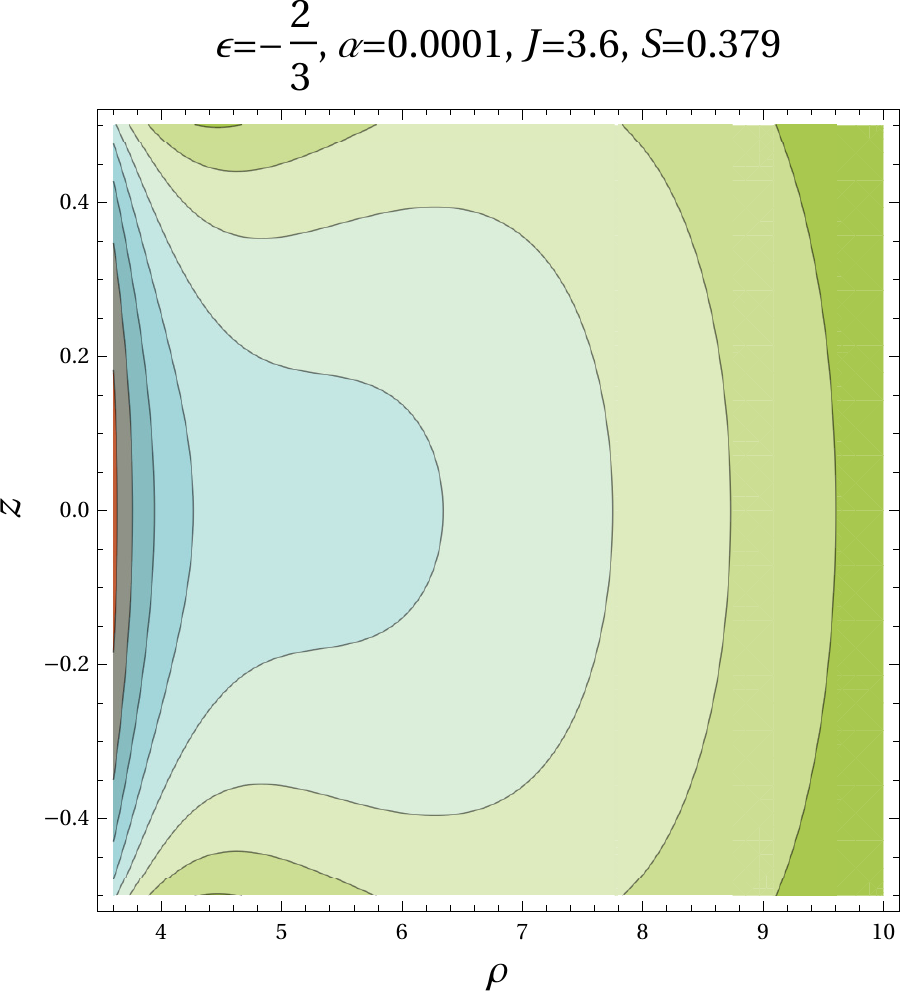}\hspace{0.2cm}
\includegraphics[scale=0.45]{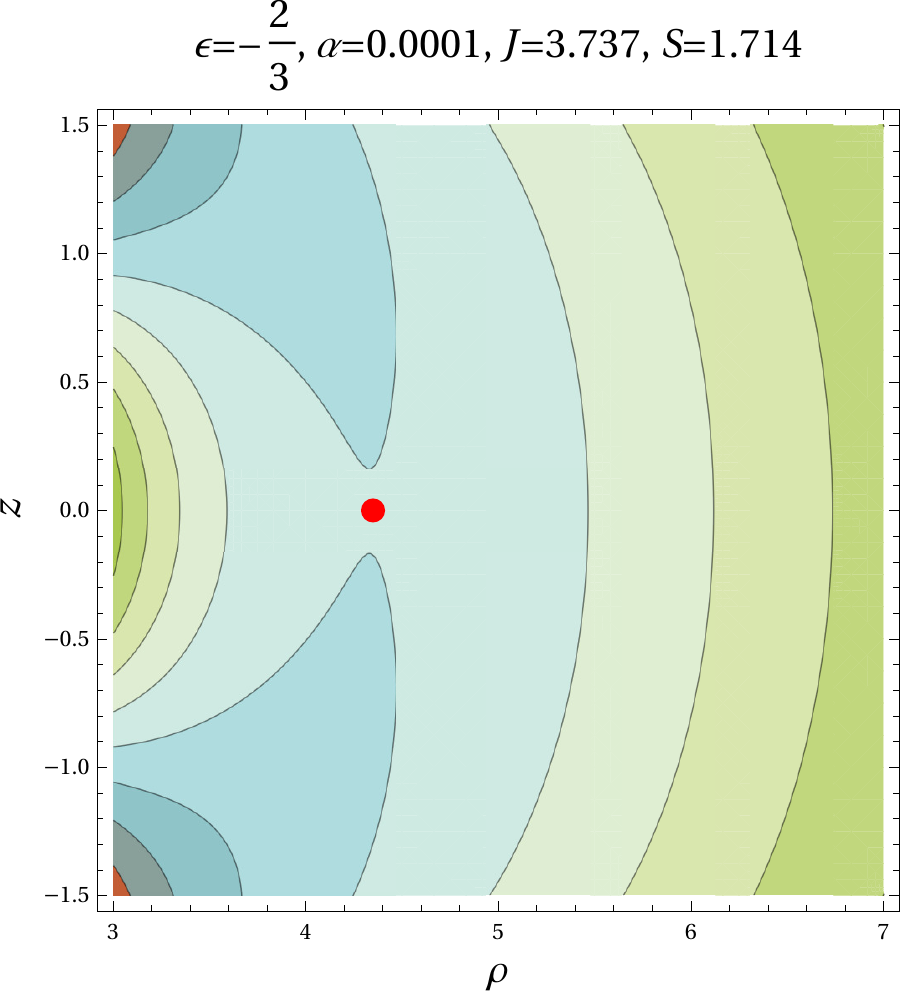}
\end{tabular}
\caption{Classification of the effective potential $V_{eff}$ for $\epsilon =-2/3$ and $ \alpha =0.0001$ depending on various values of $J$ and $S$. Here, we have also defined $J\to J/mM$ and $S\to S/mM$ as well as 
the cylindrical coordinates by $\rho\to \rho/M$ and $z\to z/M$.
}\label{fig:2D_Veff1}
\end{figure}
\begin{figure}
\begin{tabular}{c c}
\includegraphics[scale=0.45]{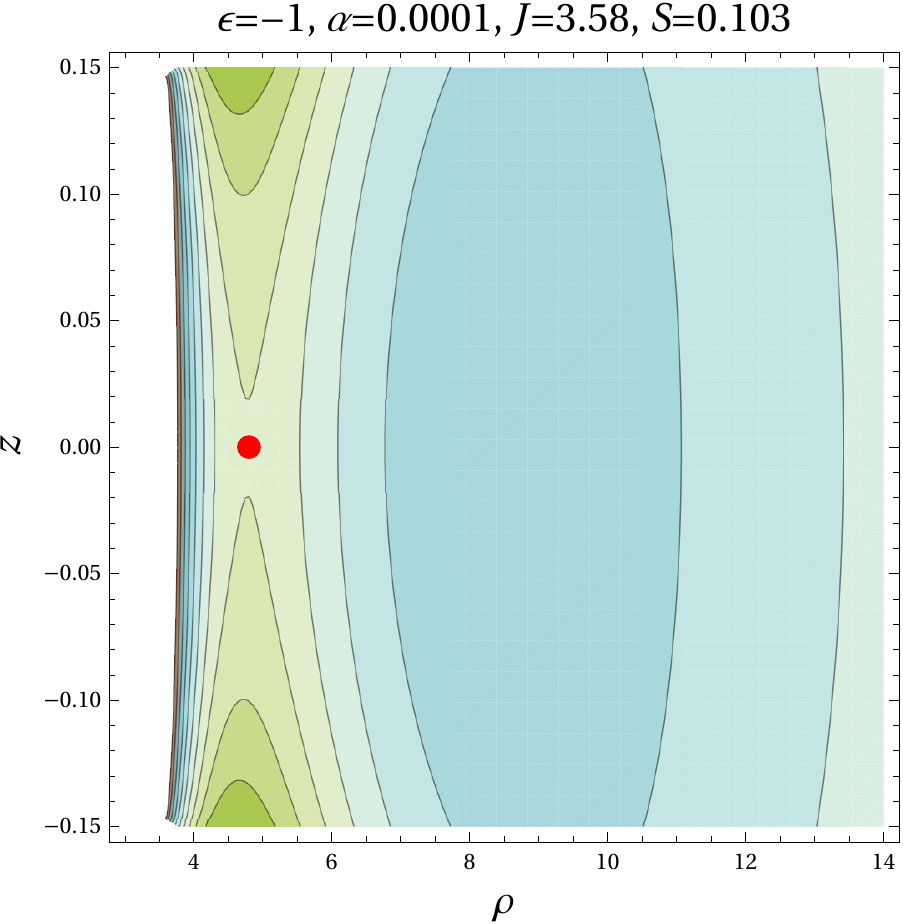}
\includegraphics[scale=0.45]{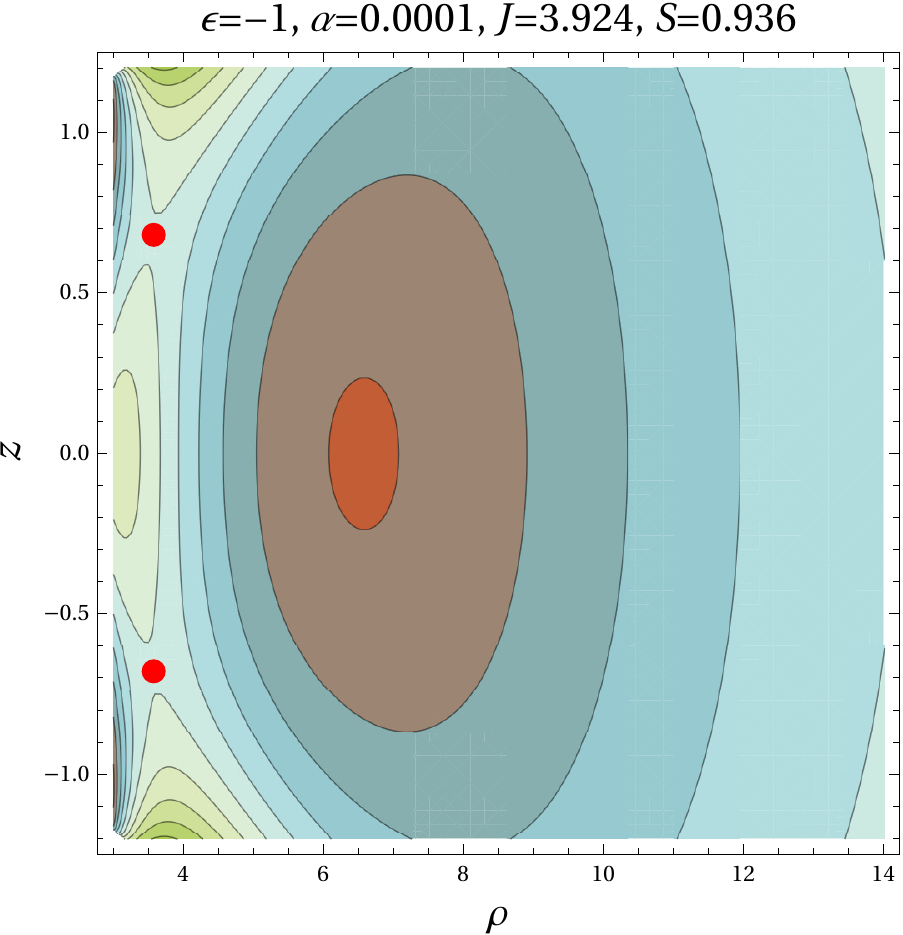}\\
\includegraphics[scale=0.45]{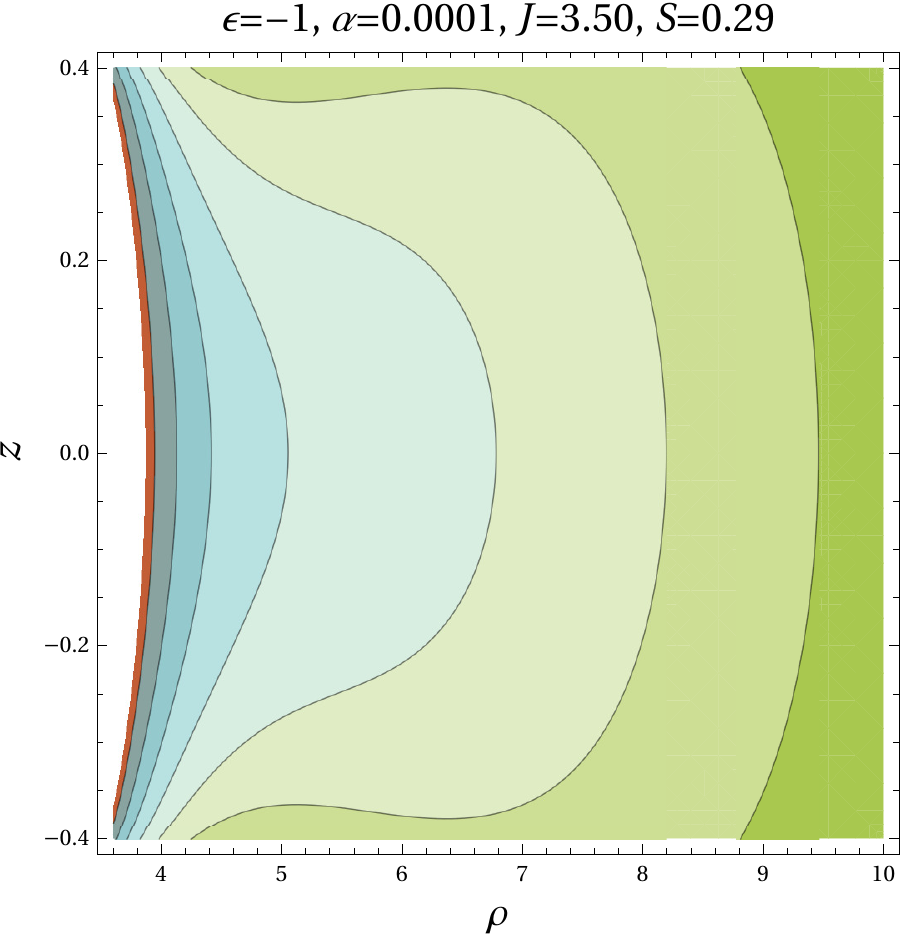}
\includegraphics[scale=0.45]{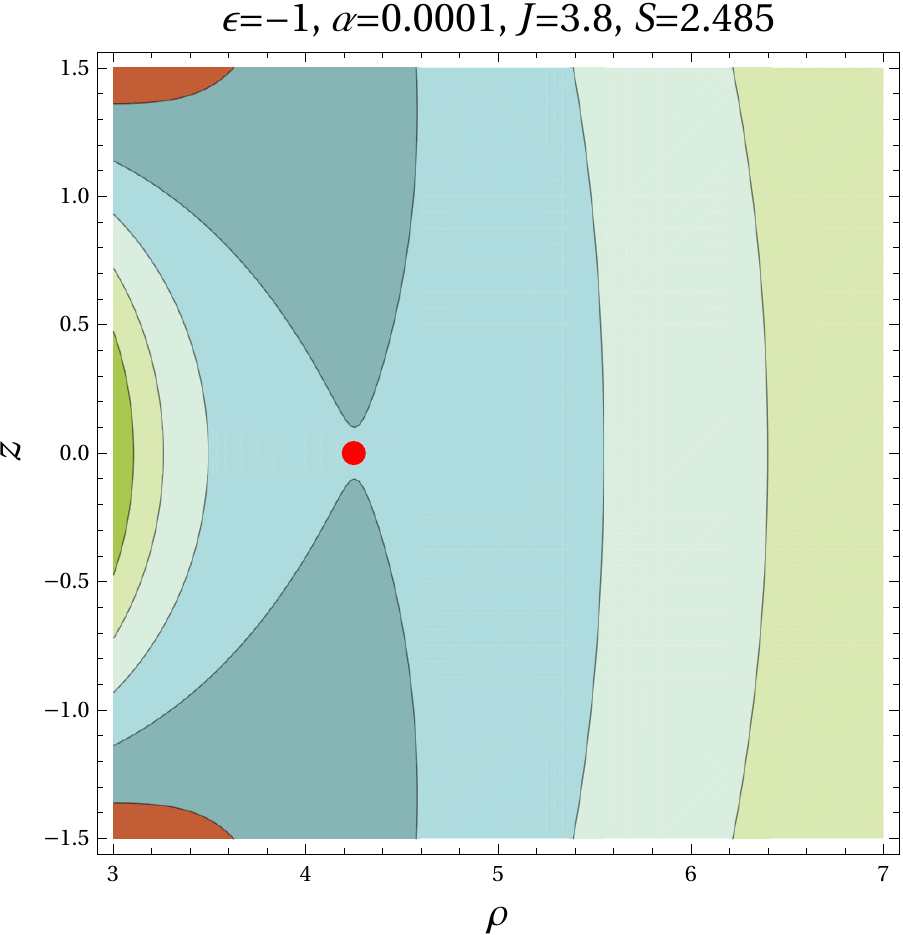}
\end{tabular}
\caption{Classification of the effective potential $V_{eff}$ for $\epsilon =-1$ and $ \alpha = 0.0001$ depending on various values of $J$ and $S$. 
}\label{fig:2D_Veff2}
\end{figure}
Here, we analyze the contour maps of $V_{eff}$ for the different combinations of the parameters $J, S$ and $\epsilon$ as it allows to identify the region in the $r-\theta$ plane where a spinning particle with $E$ can move. As we vary the parameters $J$ and $S$ for each $\epsilon$, we observe four distinct forms of $V_{eff}$ (see;  \cref{fig:3D_Veff,fig:2D_Veff1,fig:2D_Veff2}), i.e., 
    \begin{itemize}
        \item $V_{eff}$ that has only one saddle point and one minimal point on the equatorial plane $\theta=\pi/2$ (i.e., $z=0$ in cylindrical coordinate system), is named as type ($\mathcal{A}$) effective potential (for example). This type of potential arises when $S<1<<J$. As the spin effect is minimal, the potential has a comparable form to that of a non-spinning (geodesic) particle. Hence, in this situation, the orbit of the spinning particle around the SQBH never becomes chaotic.
        \item $V_{eff}$ that has two saddle points off $\theta=\pi/2$ plane (i.e., $z\neq0$) and one minimal point at $\theta=\pi/2$ (i.e. $z=0$), is named as type ($\mathcal{B}$) effective potential. In this case, the two saddle points are found symmetrically on the both sides of $z=0$. When we decrease the parameter $S$ from a higher value (more than $1.2$) till it is near to unity but close to it while maintaining the parameter $J$ approximately constant, we get this type of $V_{eff}$ from type ($\mathcal{A}$) effective potential. This is due to the fact that spin-orbit interactions are repulsive in nature as well as angle dependent.
        \item $V_{eff}$ that has no minimal (bound region) and no saddle point is named as type ($\mathcal{C}$) effective potential. This type of $V_{eff}$ occurs when parameter $J$ is comparatively smaller than the type ($\mathcal{A}$) and ($\mathcal{B}$) effective potentials while the parameter $S<1$. As this potential has no bound region, the spinning particle will pass through the event horizon and eventually fall into the SQBH. This form of potential occurs when the centrifugal force is insufficient to balance the gravity of SQBH.
        \item $V_{eff}$ that has a saddle point on the equatorial plane $\theta=\pi/2$ (i.e., at $z=0$) and no bound region but has a minimal in the radial direction and maximal in the $\theta$ direction, we name it as a type ($\mathcal{D}$) effective potential. This type of $V_{eff}$ also occurs when parameter $J$ is comparatively smaller than the type ($\mathcal{A}$) and ($\mathcal{B}$) effective potentials, similar to type ($\mathcal{C}$) $V_{eff}$. However, unlike type ($\mathcal{C}$), the parameter $S$ for this type of $V_{eff}$ must be greater than $1$. 
       \end{itemize}
$V_{eff}$ for ($\mathcal{C}$) and ($\mathcal{D}$) arises when the total angular momentum $J$ is small enough so that the centrifugal force becomes insufficient to balance the gravitational force. Further, for the type ($\mathcal{D}$) $V_{eff}$, a test particle will progressively deviate from the $\theta=\pi/2$ plane and eventually plunge into the SQBH. This might happen because a spin-orbit interaction creates a potential barrier on the $\theta=\pi/2$ plane (see; plot (d) of \cref{fig:3D_Veff}). For a detailed discussion on classification of $V_{eff}$, one can refer to Suzuki and Maeda \cite{suzuki1997chaos}.
    
It is worth to note here that these four different types of $V_{eff}$ are observed for all three cases of equation of state parameter $\epsilon$, i.e., $-1/3, -2/3\;\text{and}\;-1$. Therefore, for the first case only (i.e., $\epsilon=-1/3$), we show the $3D$ plot of $V_{eff}$ in cylindrical coordinate system as well as its projection in $z-\rho$ plane for a better visualization in \cref{fig:3D_Veff}. However, for the remaining two cases of equation of state parameter (i.e., $\epsilon=-2/3\;\text{and}\;-1$) in \cref{fig:2D_Veff1,fig:2D_Veff2}, we only show $2D$ projection plots.
    
\begin{figure*}[t!]
\begin{tabular}{c c c}
\includegraphics[width=5cm, height=4cm]{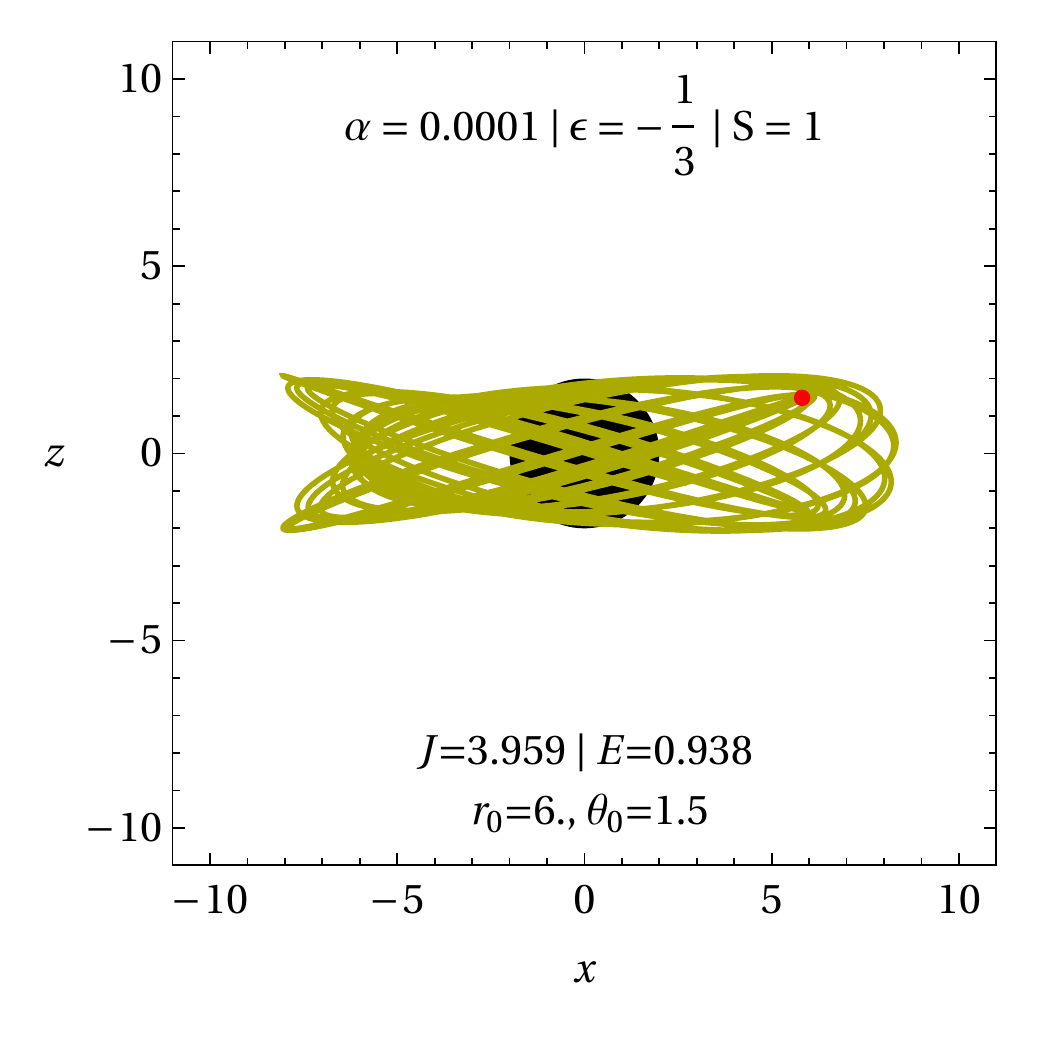}
\includegraphics[width=5cm, height=4cm]{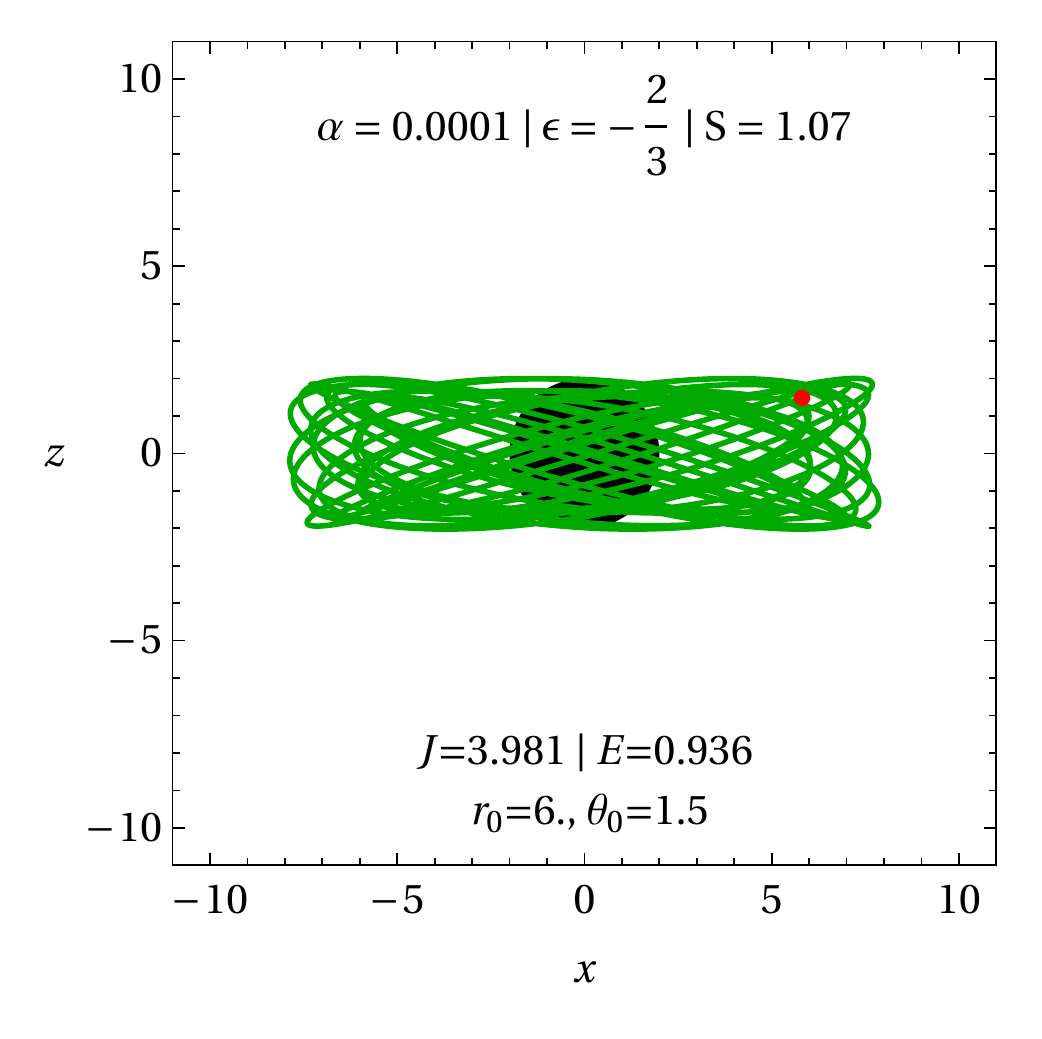}
\includegraphics[width=5cm, height=4cm]{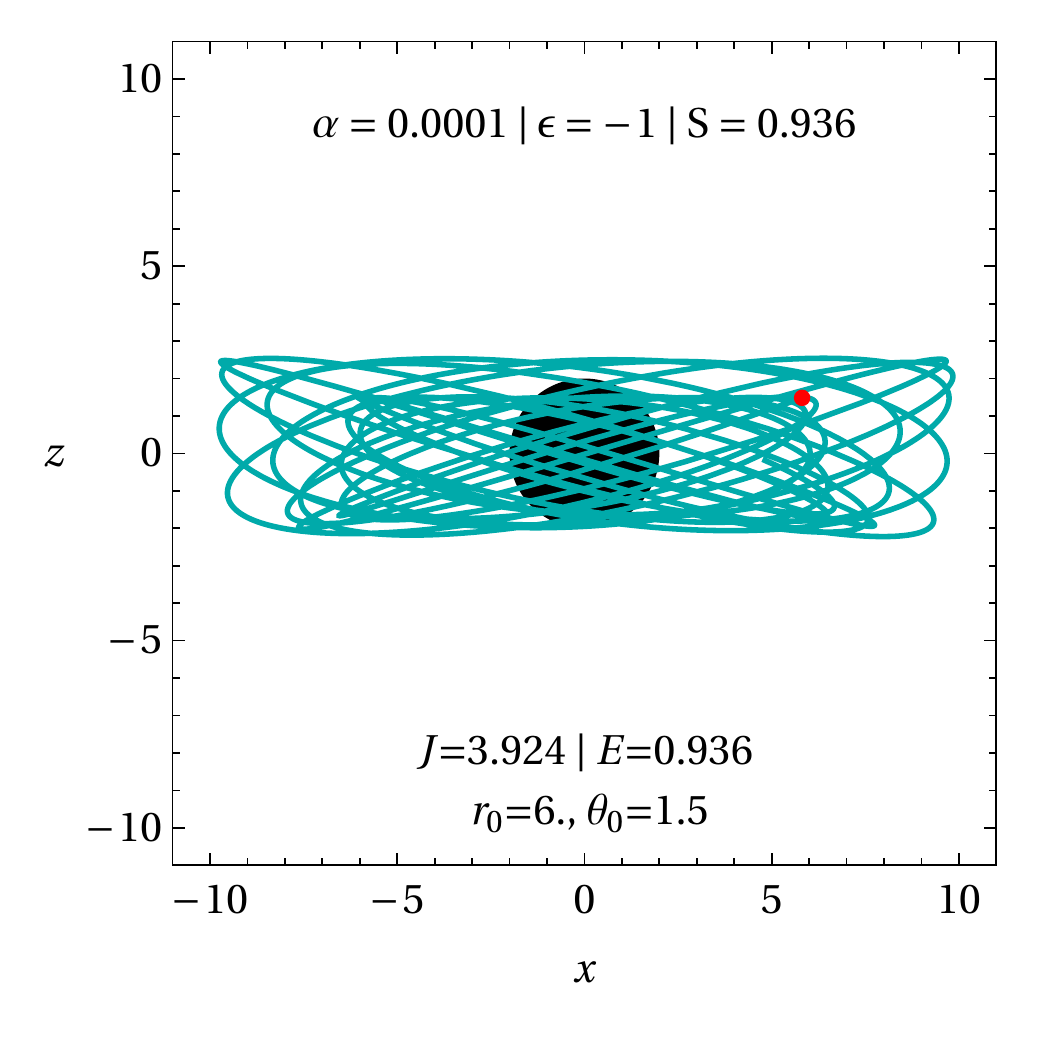}\\
\includegraphics[width=5cm, height=4cm]{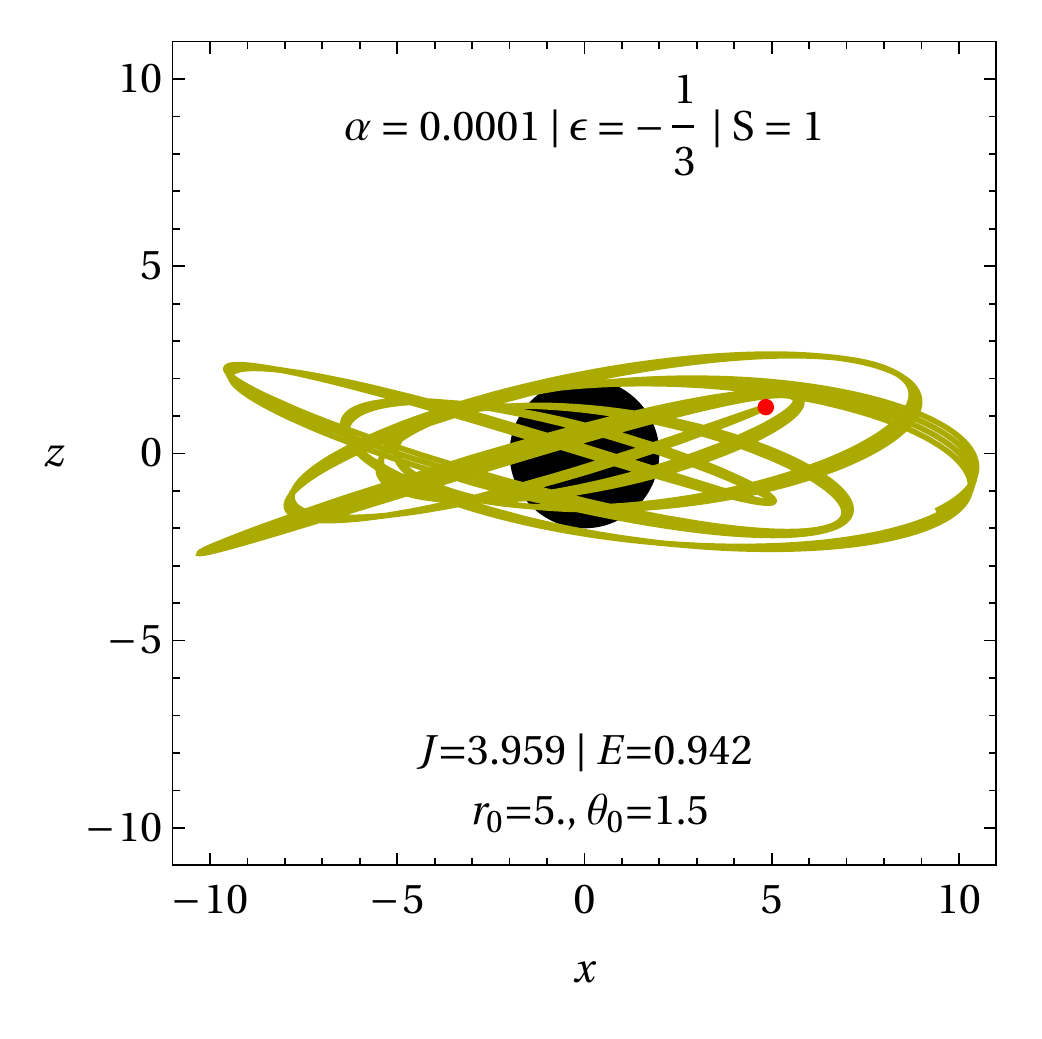}
\includegraphics[width=5cm, height=4cm]{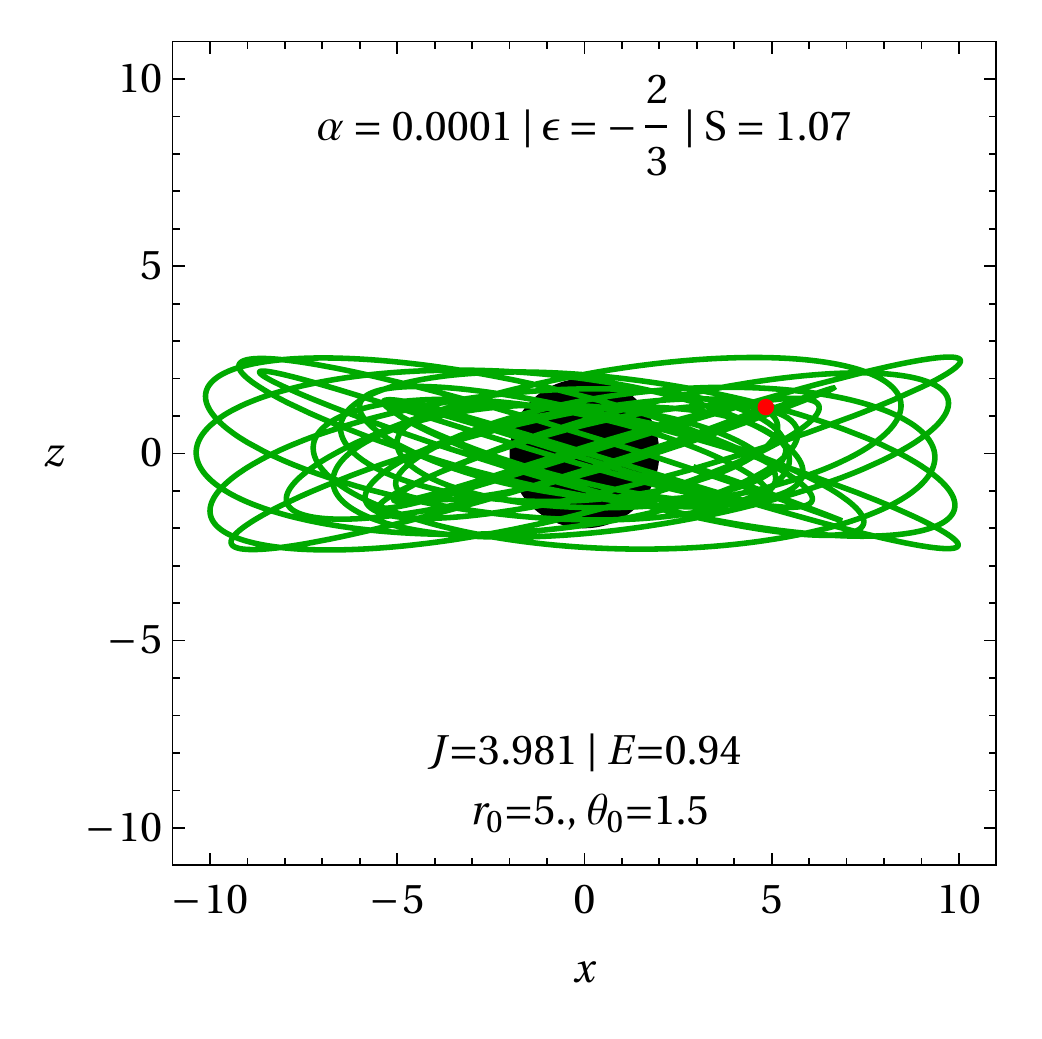}
\includegraphics[width=5cm, height=4cm]{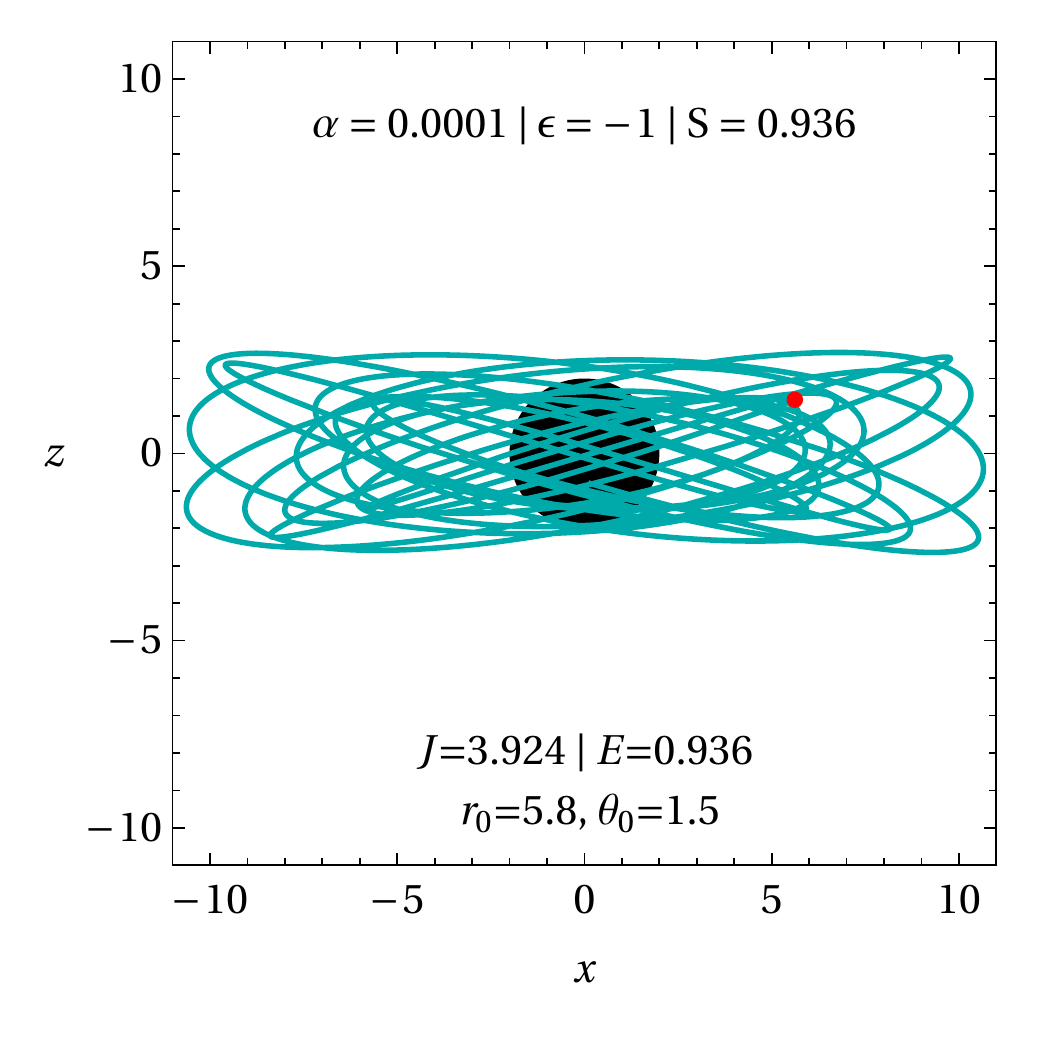}\\
\includegraphics[width=5cm, height=4cm]{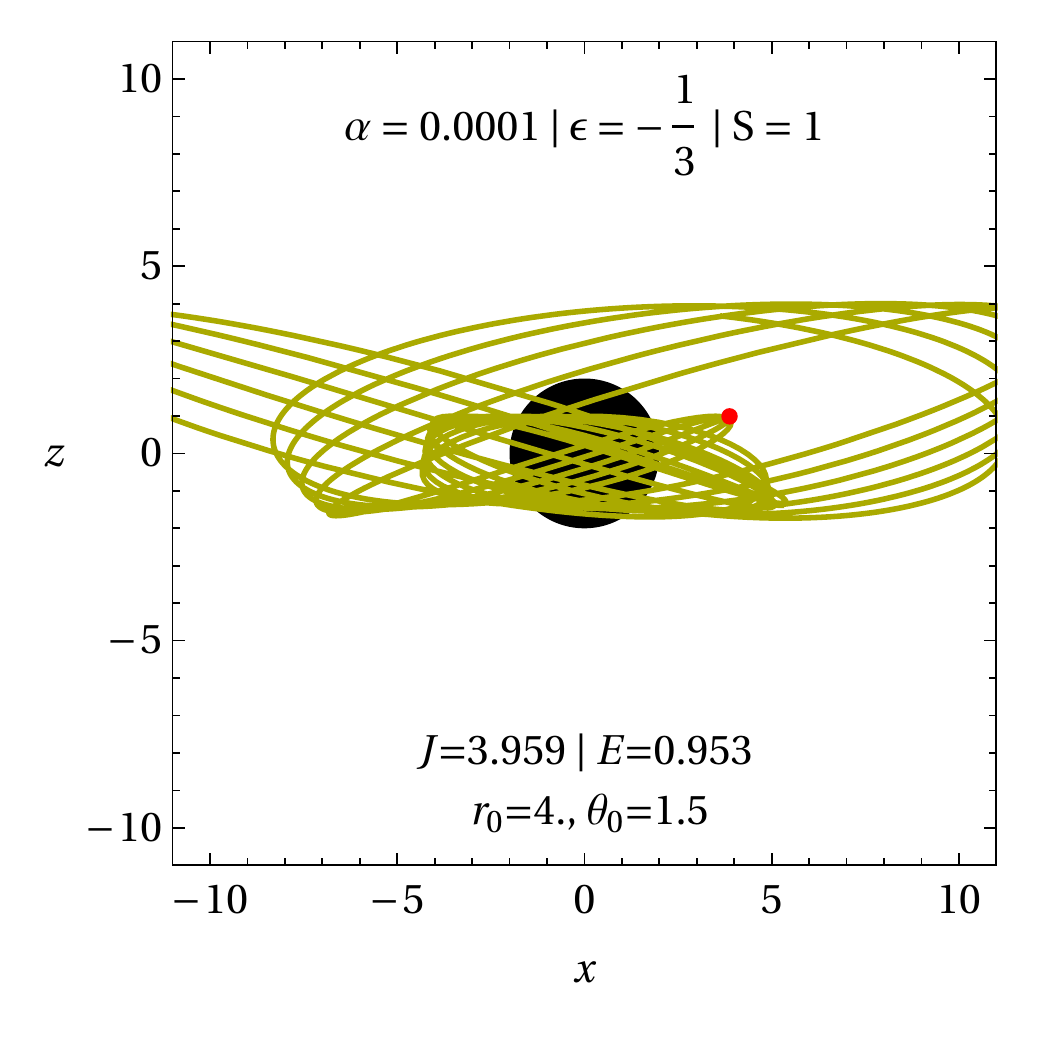}
\includegraphics[width=5cm, height=4cm]{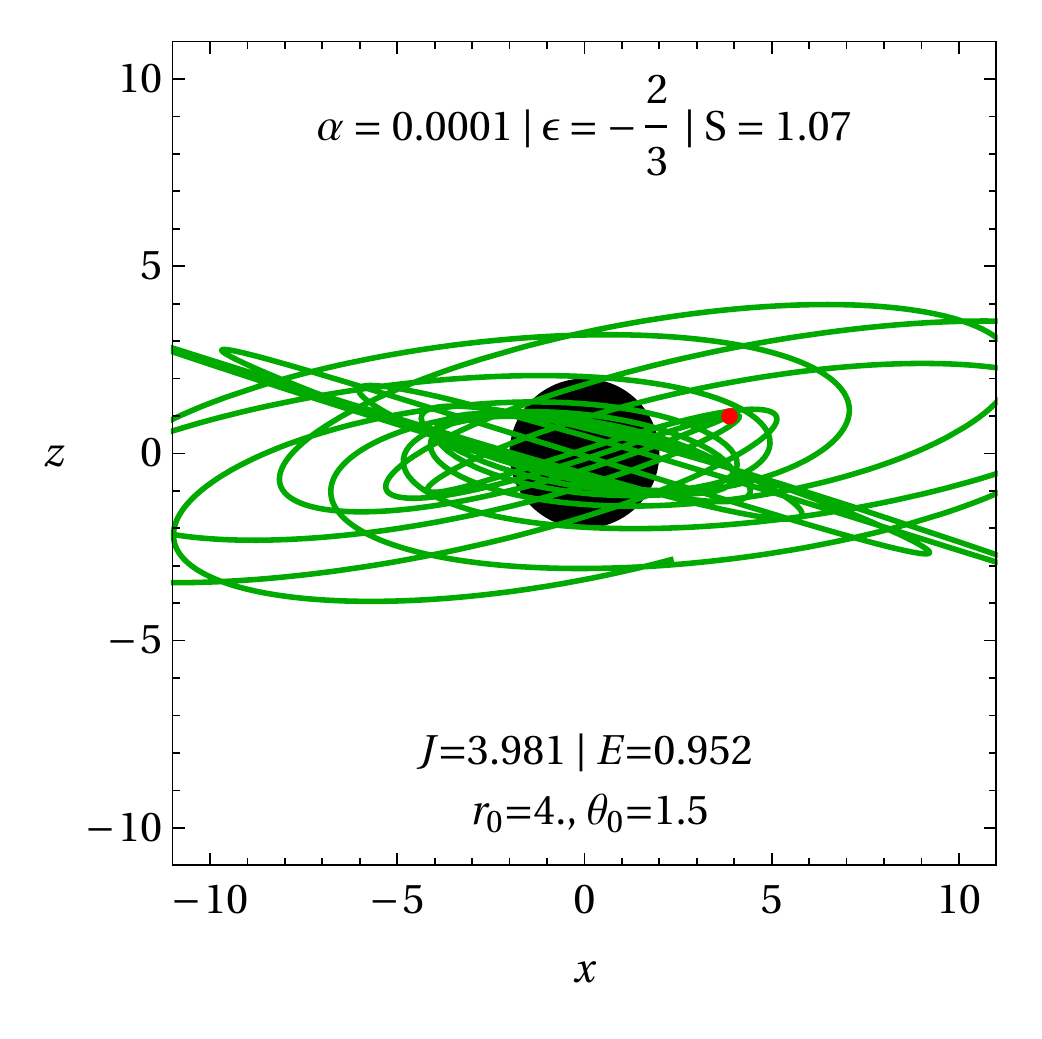}
\includegraphics[width=5cm, height=4cm]{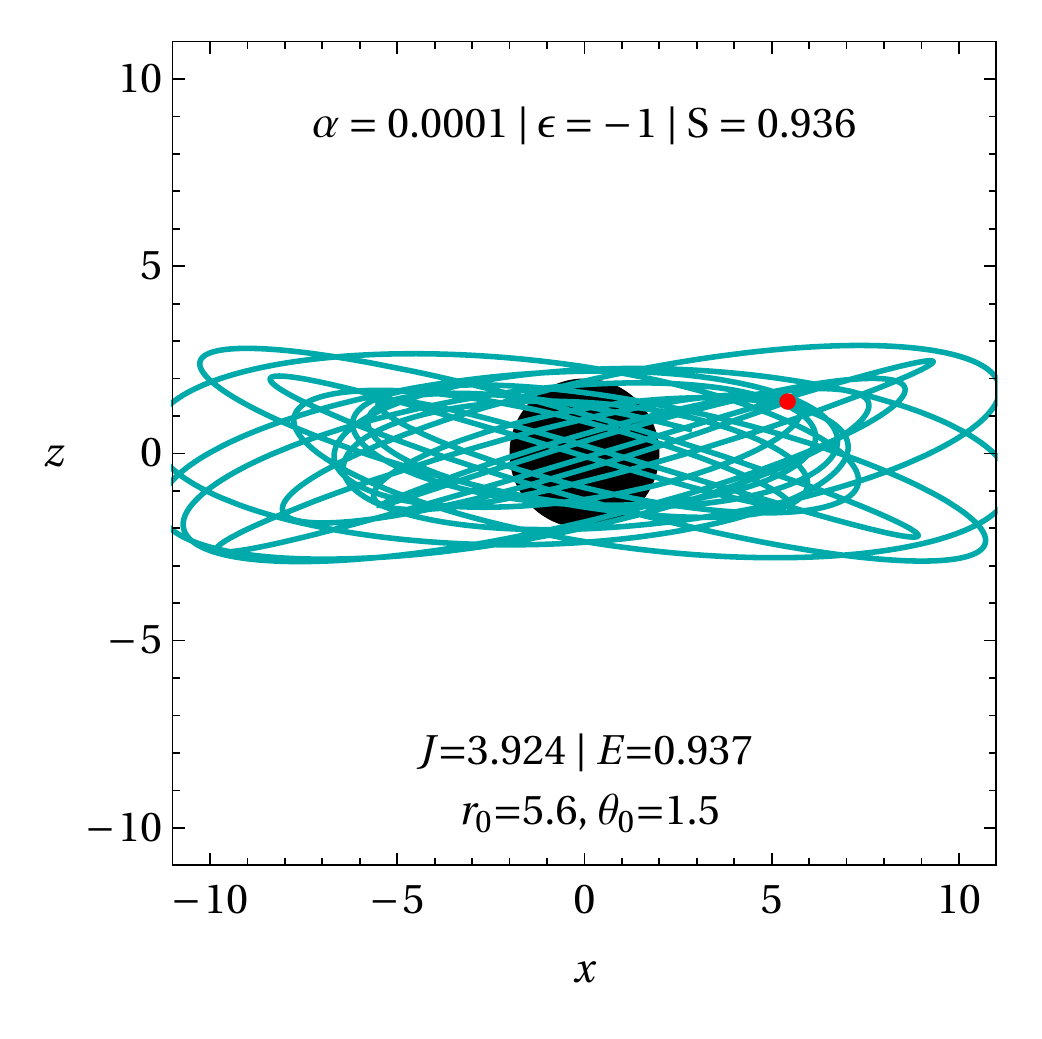}\\
\includegraphics[width=5cm, height=4cm]{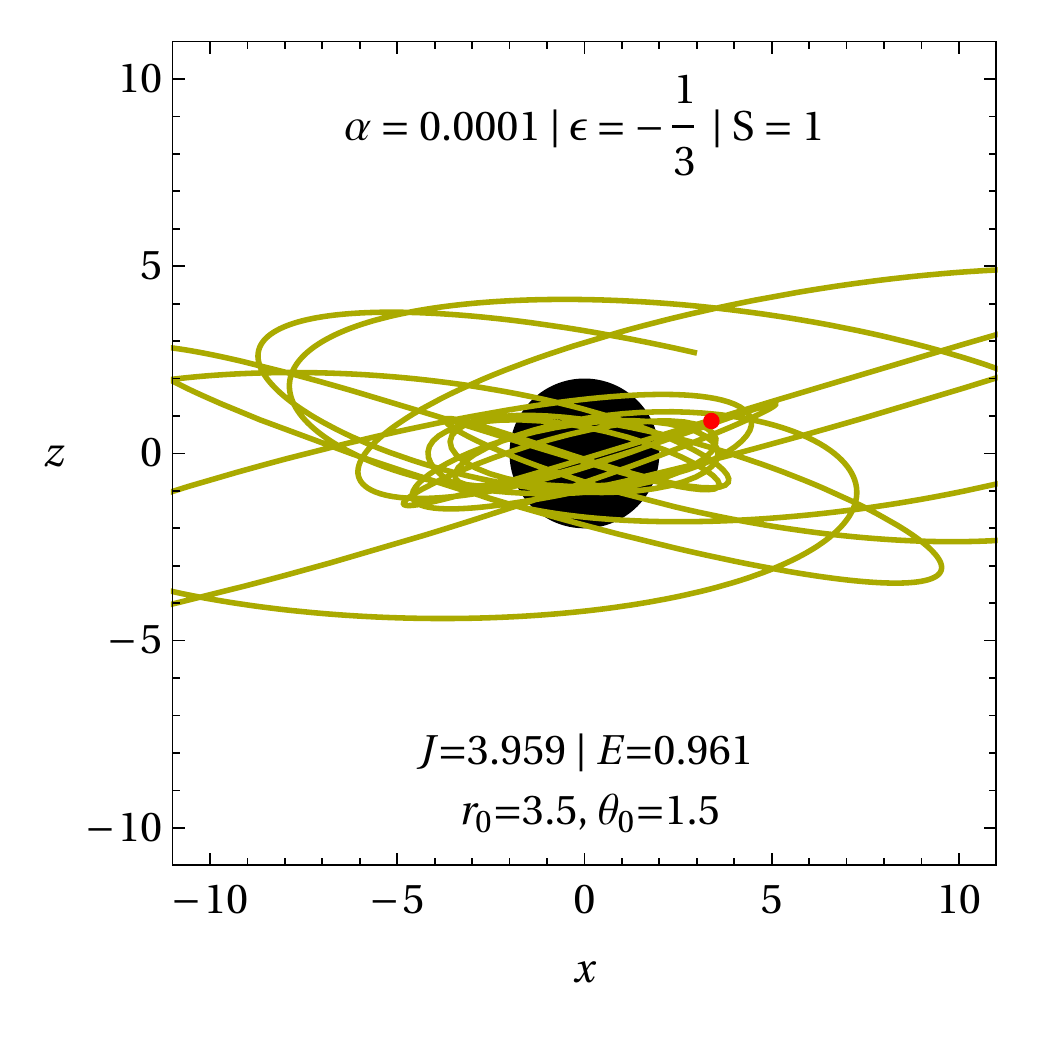}
\includegraphics[width=5cm, height=4cm]{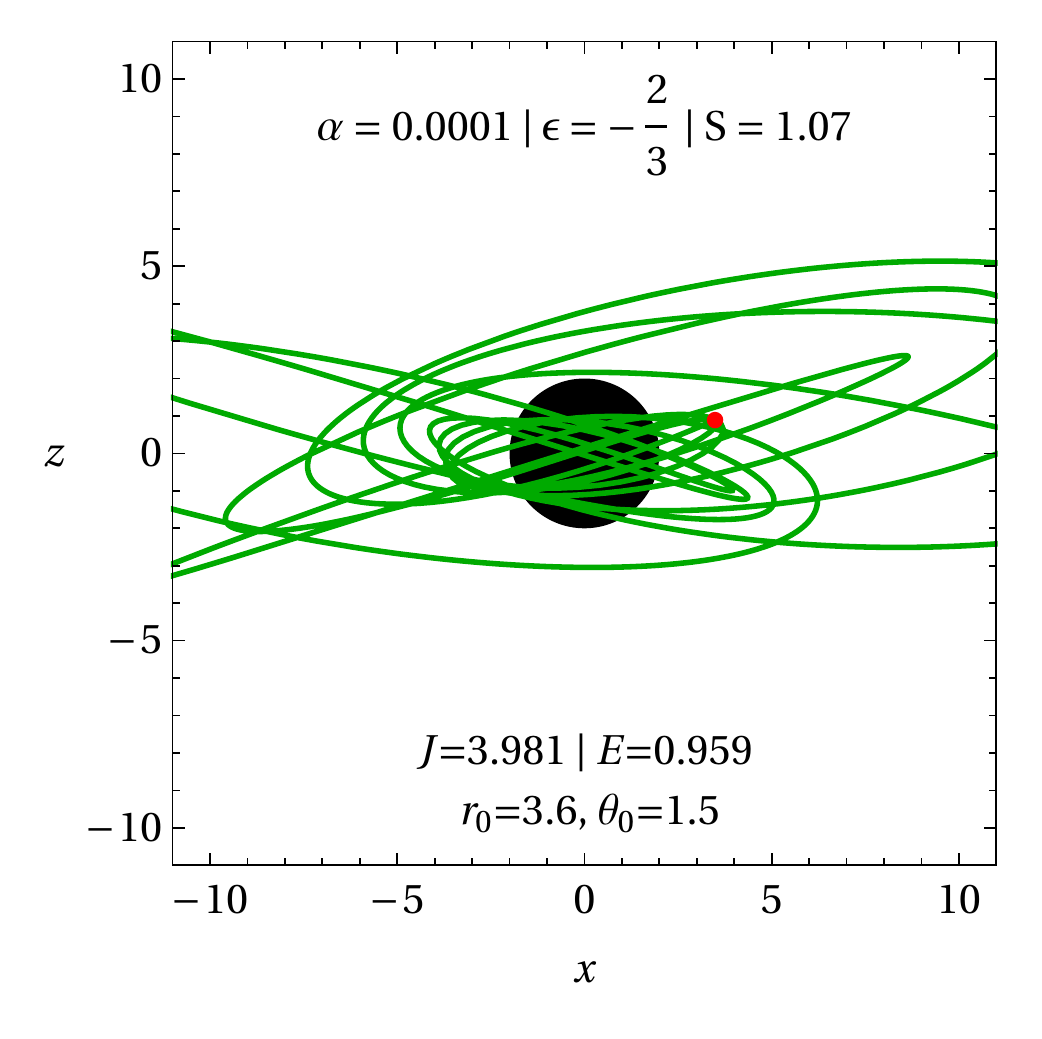}
\includegraphics[width=5cm, height=4cm]{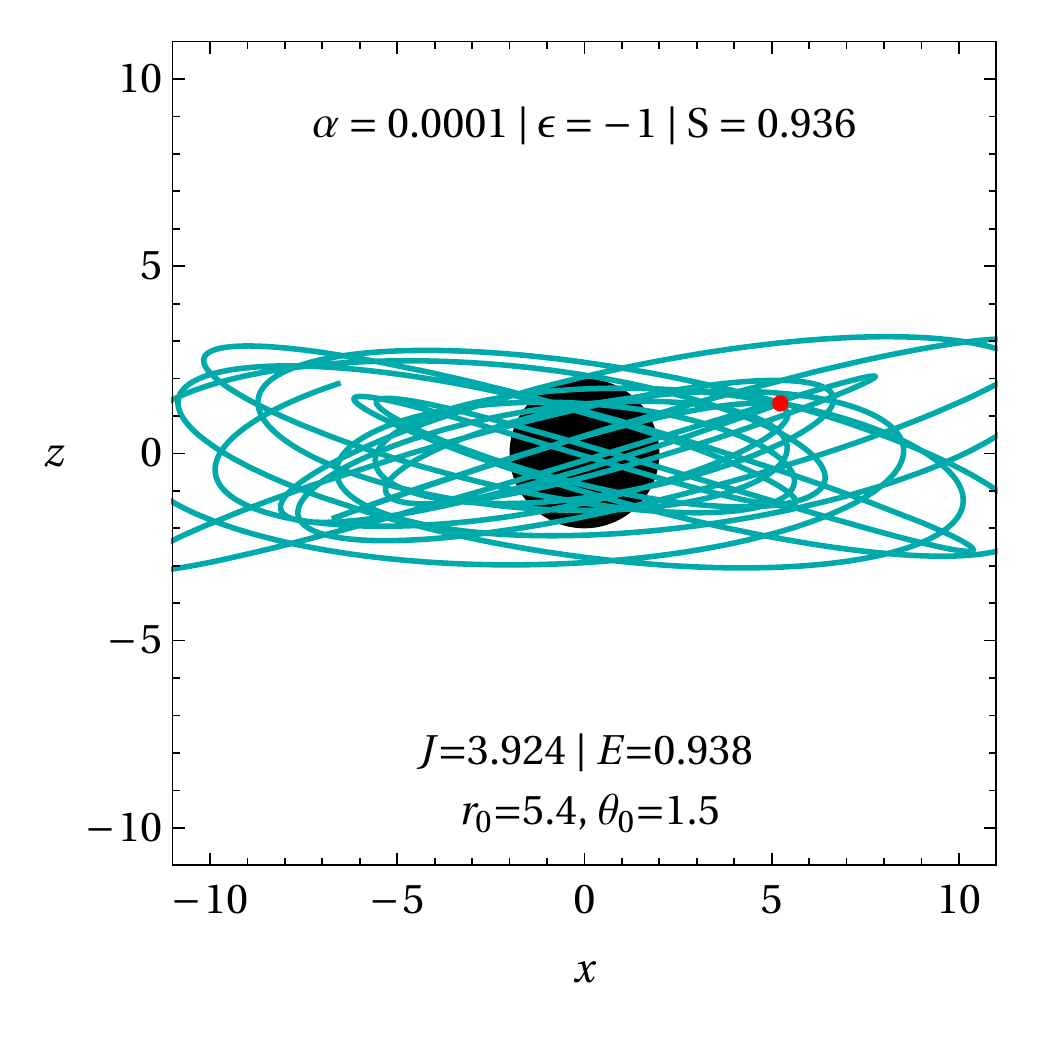}\\
\includegraphics[width=5cm, height=4cm]{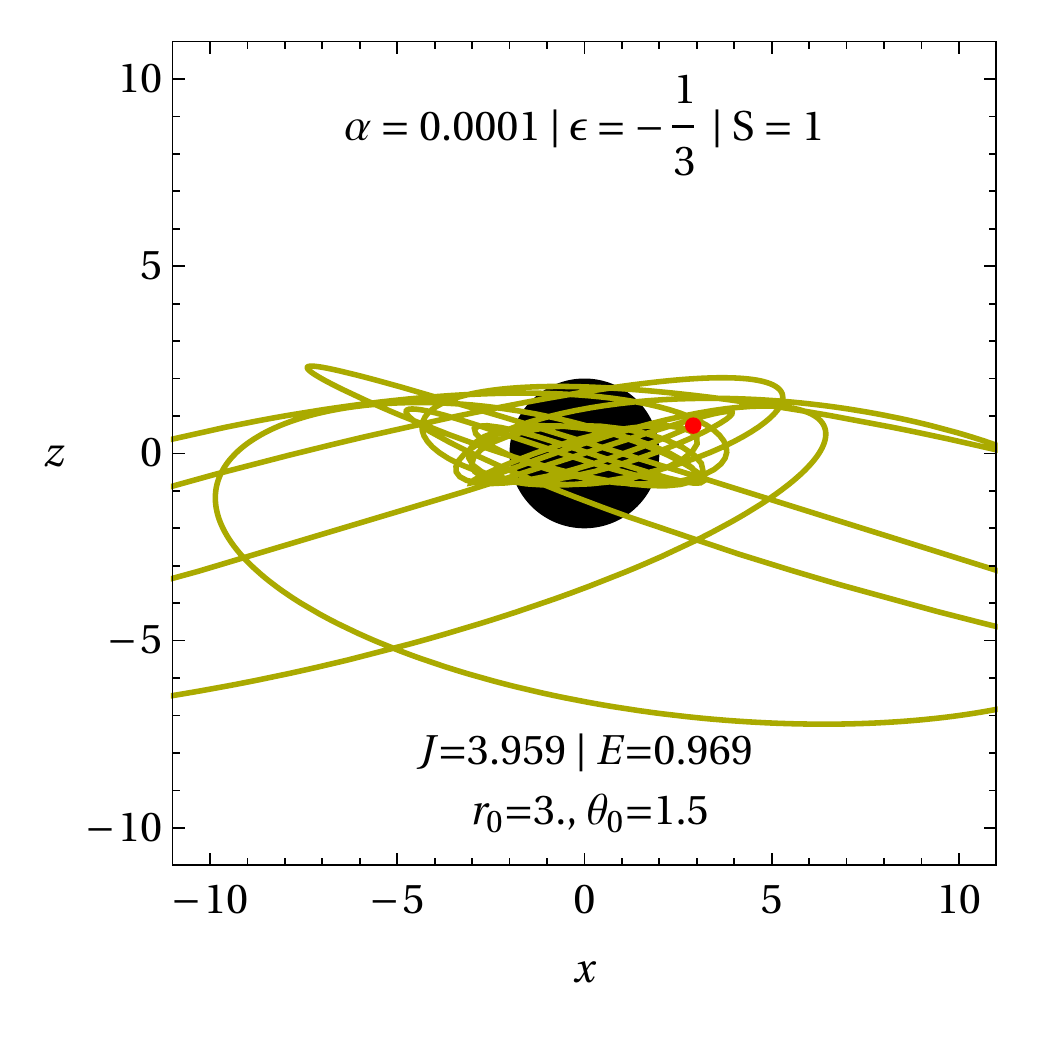}
\includegraphics[width=5cm, height=4cm]{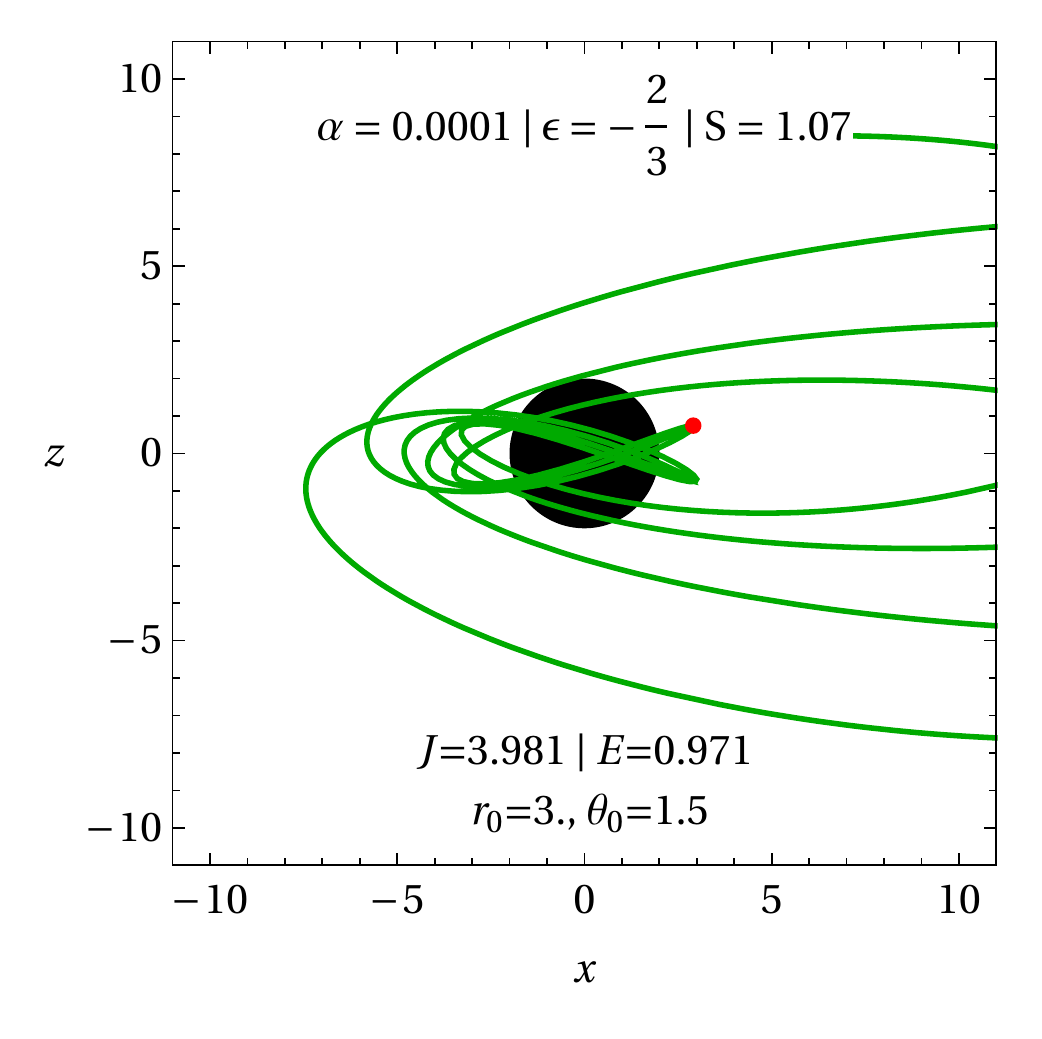}
\includegraphics[width=5cm, height=4cm]{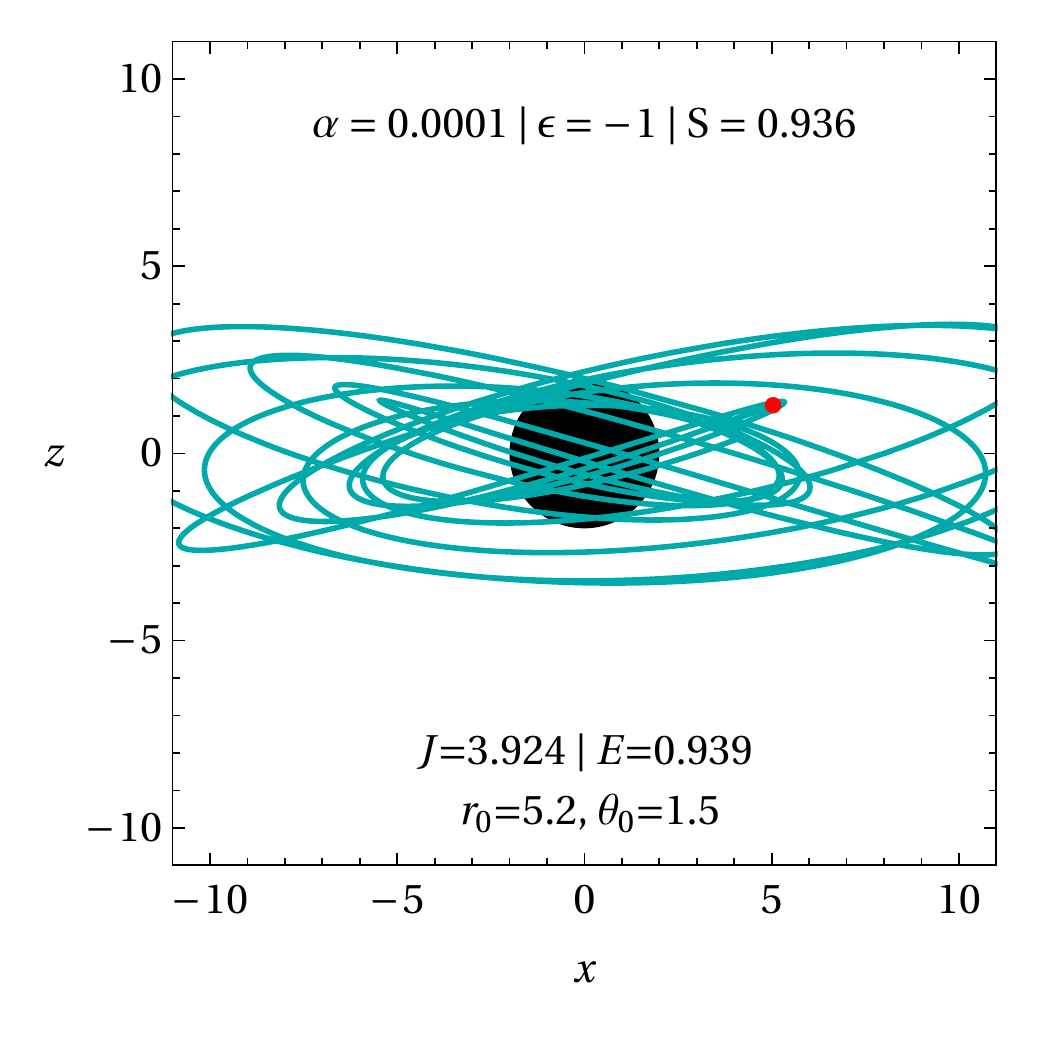}
\end{tabular}
\caption{
The orbits of the spinning particle in the $x-z$ plane for various combinations of parameters $\alpha, \epsilon, S, J, E, \theta, r_{0}$ for the case of $V_{eff}$ that has two saddle points. 
The red dot represents the initial position $r_{0}$ of the spinning particle. Here, the equation of state parameter $\epsilon$ is set to $-1/3, -2/3$ and $-1$ for the {first, second} and {third columns}, respectively.
}\label{fig:Orbits}
\end{figure*}

Further, as the types ($\mathcal{B}$) and ($\mathcal{D}$) of $V_{eff}$ only occur for the case of spinning particle in addition to the types ($\mathcal{A}$) and ($\mathcal{C}$) of $V_{eff}$ which occur even for the non-spinning particles. Therefore, one can easily conclude that these types of $V_{eff}$ arise solely due to spin of the particle. We also numerically study the trajectories of spinning particles for all these four cases of $V_{eff}$ and observe that it is only type ($\mathcal{B}$) $V_{eff}$ for which chaotic motion are observed (see; \cref{fig:Orbits}). This is similar to the observation made in \cite{suzuki1997chaos} for the Schwarzschild BH. 
In \cref{fig:Orbits}, we plot the trajectories of spinning particles moving around the SQBH for various values (i.e, $-1/3, -2/3$ and $-1$) of equation of state parameter $\epsilon$ while keeping the normalization parameter $\lambda=0.0001$ fixed for the type ($\mathcal{B}$) $V_{eff}$. The first, middle and last column are for $\epsilon=-1,3, -2/3$ and $-1$, respectively. In each column, we set the $r_{0}$ near to the minimum of $V_{eff}$ for the top row and it gets closer to saddle points as we move down the each column. However, for the last row we set the $r_{0}$ of the spinning particle between the event horizon and saddle points. It is observed that as we move from top to bottom (till second last row) the chaotic behavior increases and it starts to become non-chaotic as we pass the saddle points. It is important to note that our results are consistent with the results obtained by Suzuki and Maeda for the Schwarzschild BH \cite{suzuki1997chaos}.

\vspace{-0.5cm}
\section{Behaviour of ISCO parameters}\label{sec:ISCO_Parameters}
\begin{figure*}
\begin{tabular}{c c}
\includegraphics[scale=0.45]{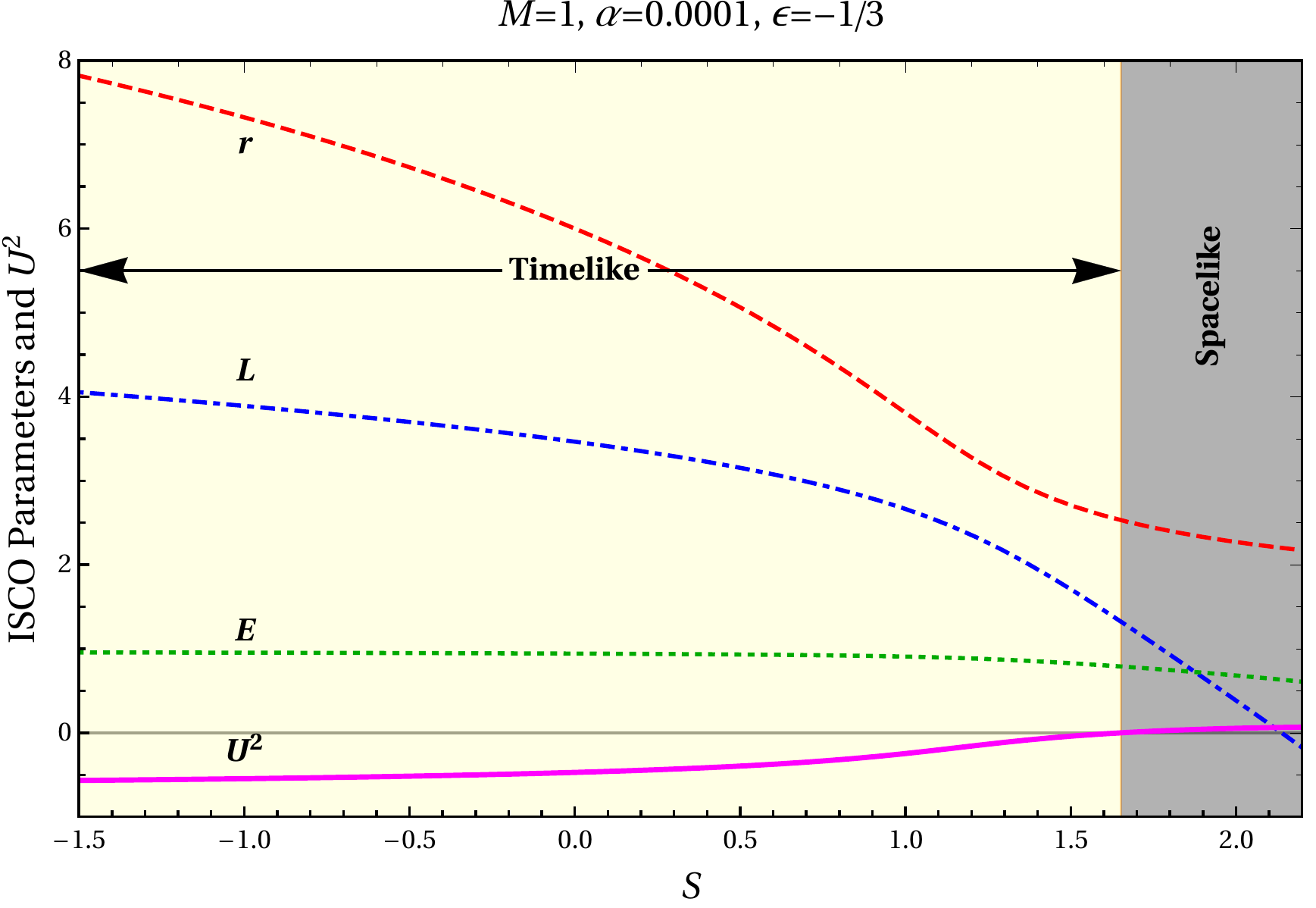}
\includegraphics[scale=0.45]{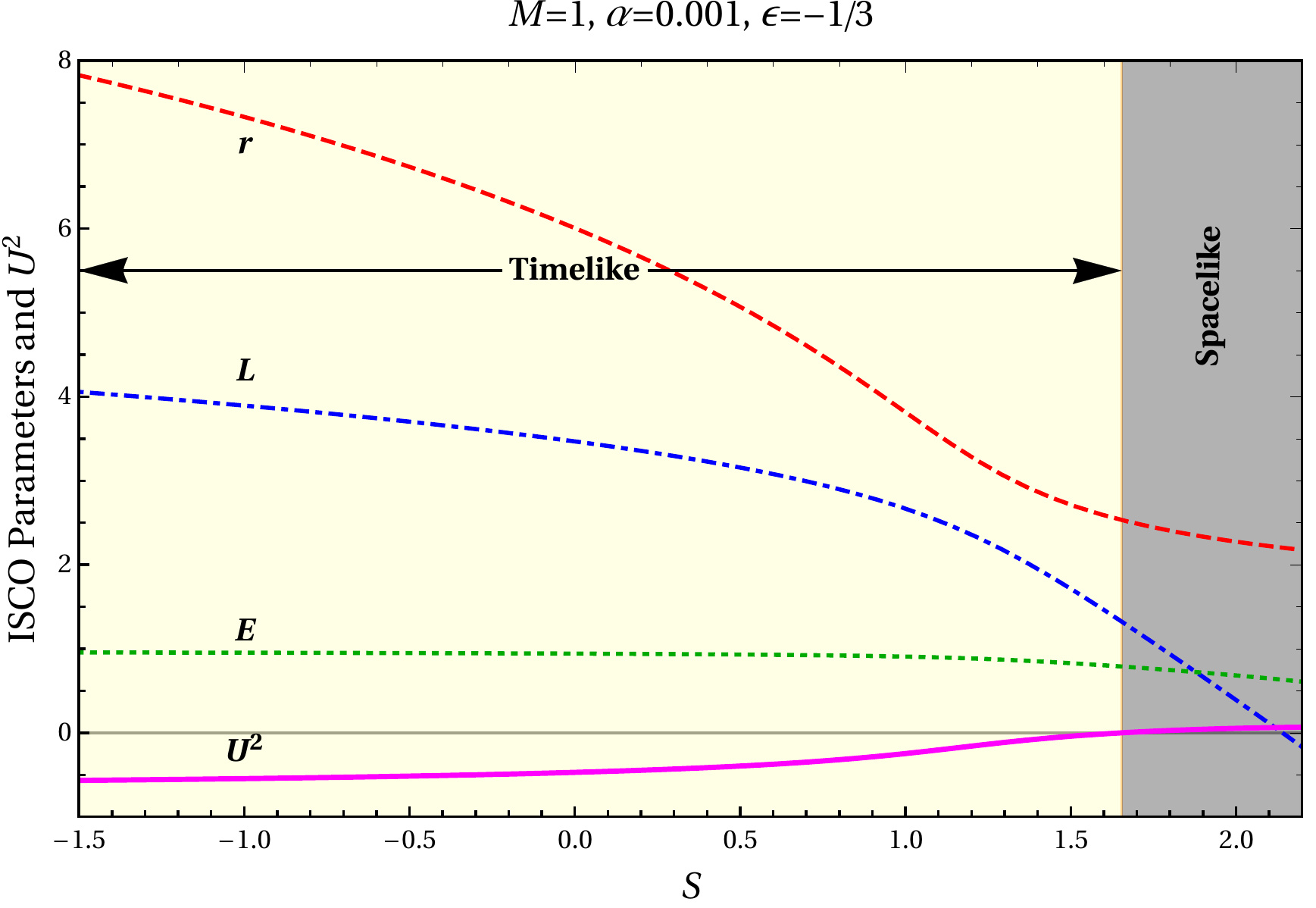}\\
\includegraphics[scale=0.45]{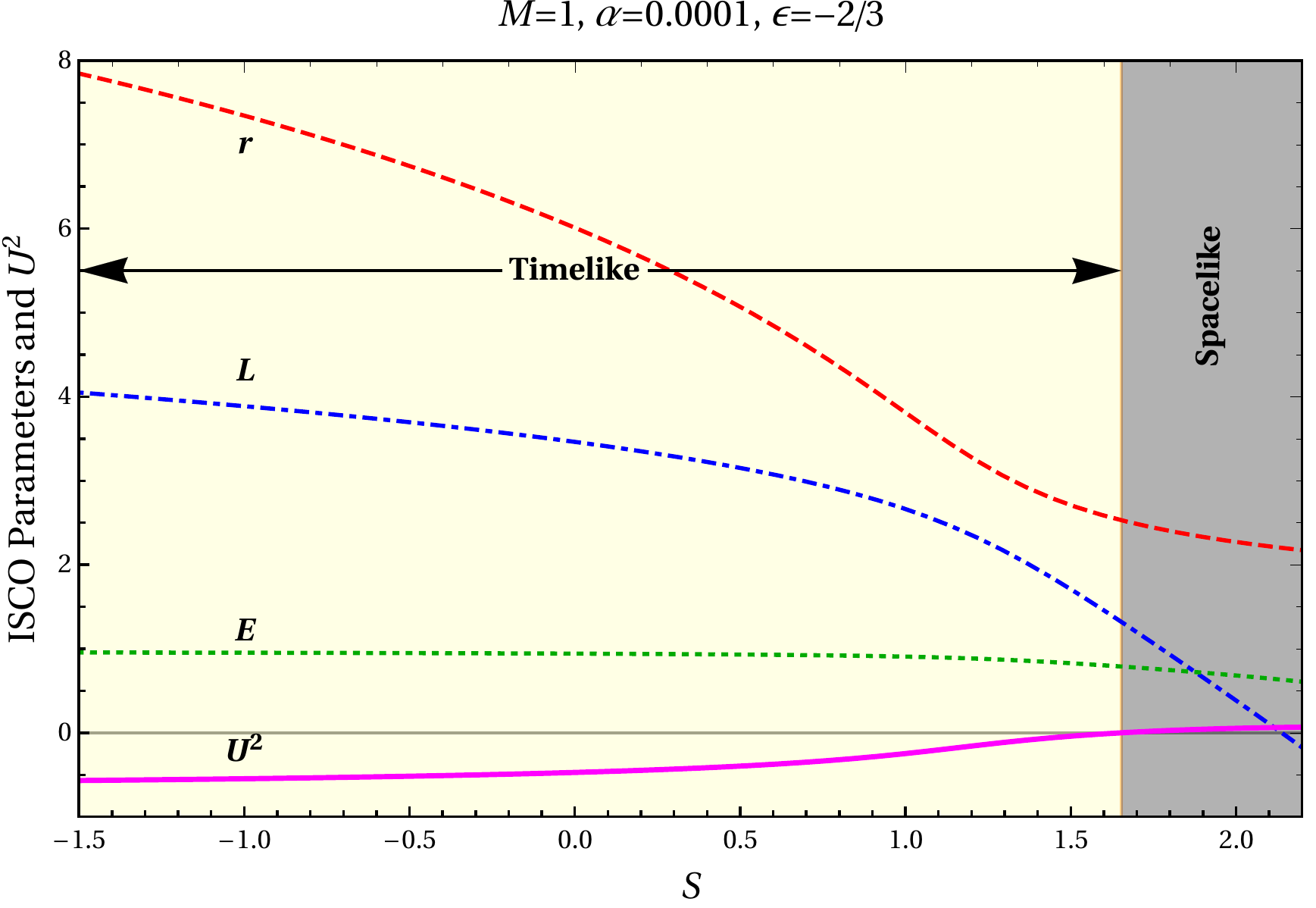}
\includegraphics[scale=0.45]{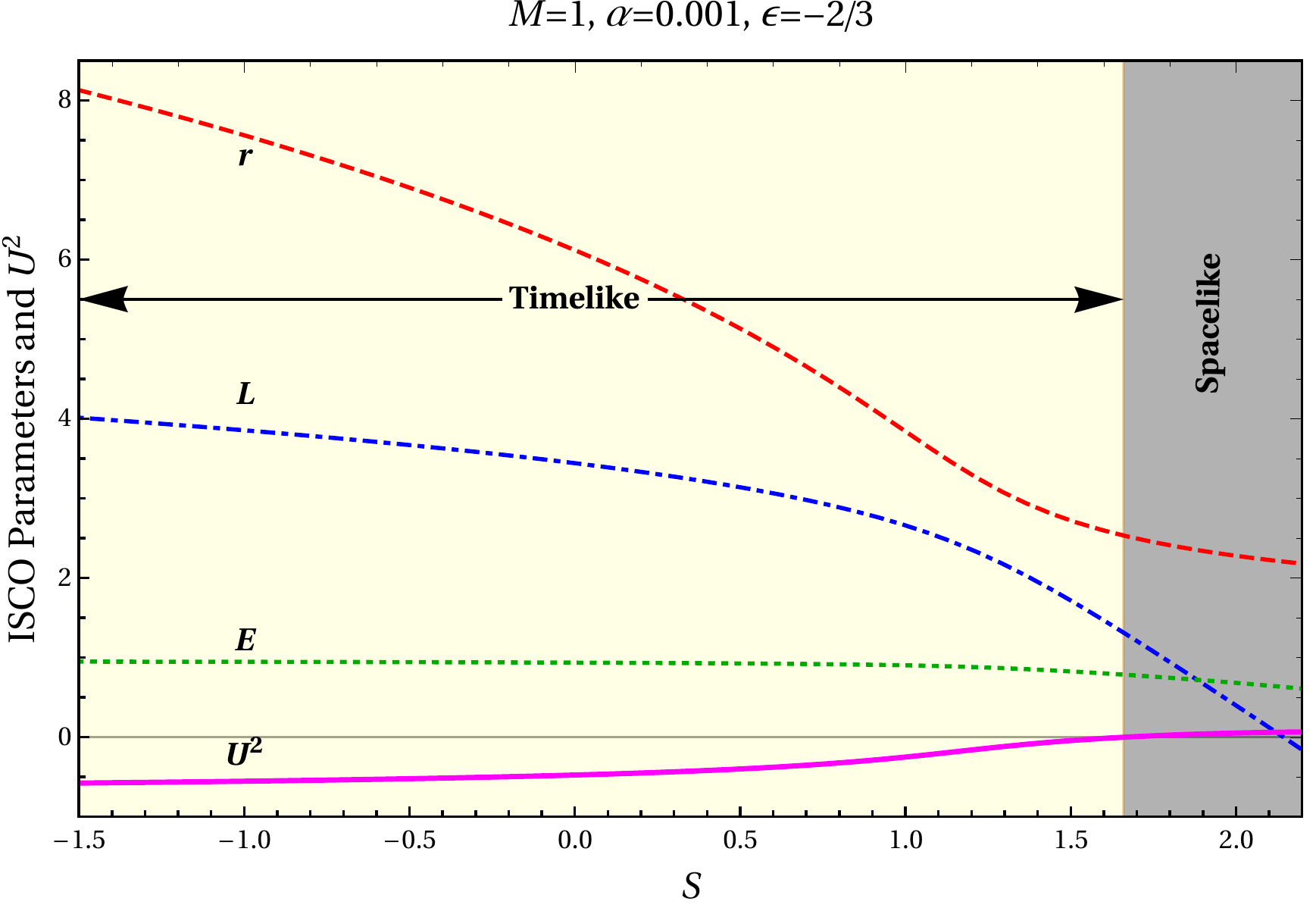}\\
\includegraphics[scale=0.45]{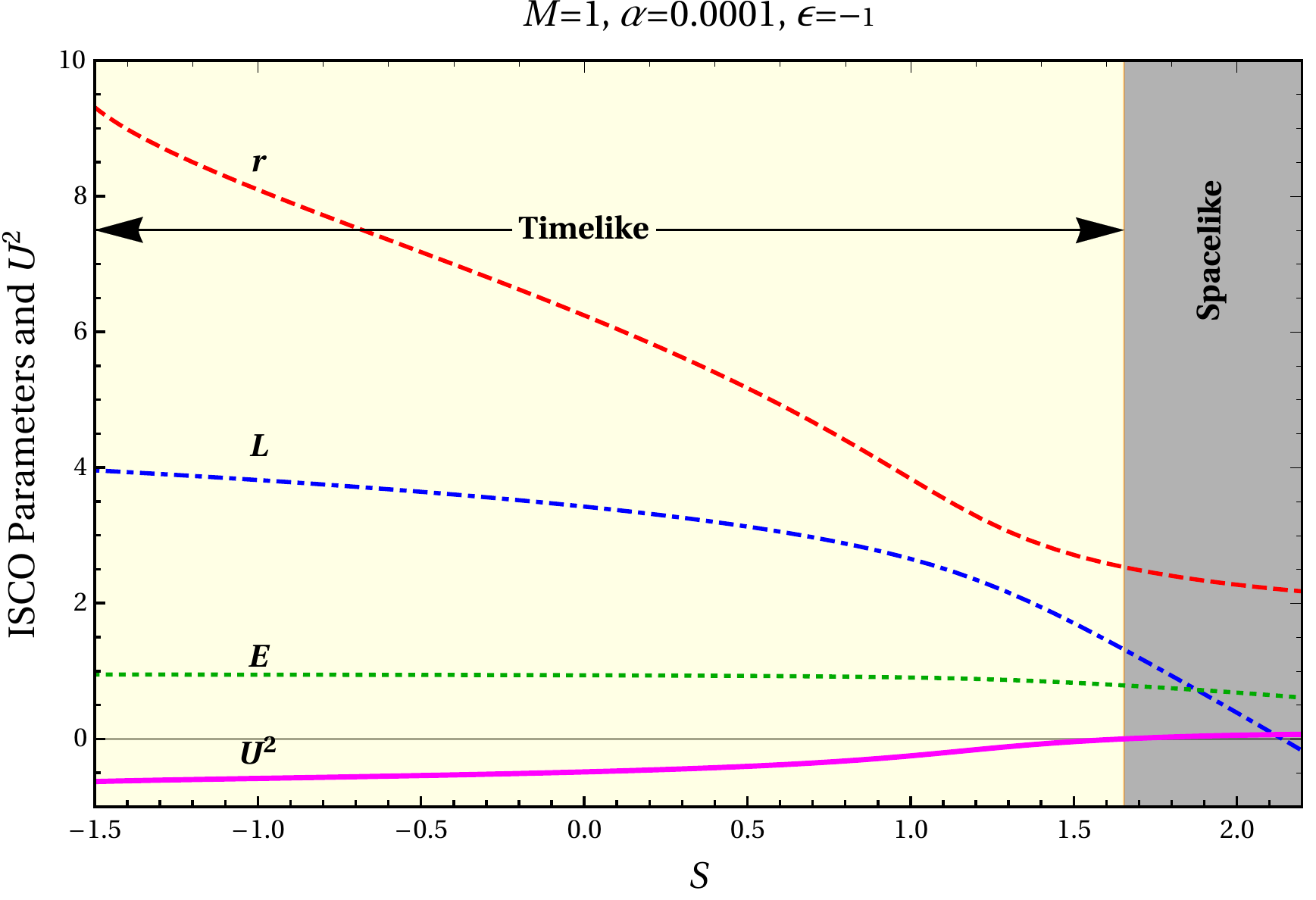}
\includegraphics[scale=0.45]{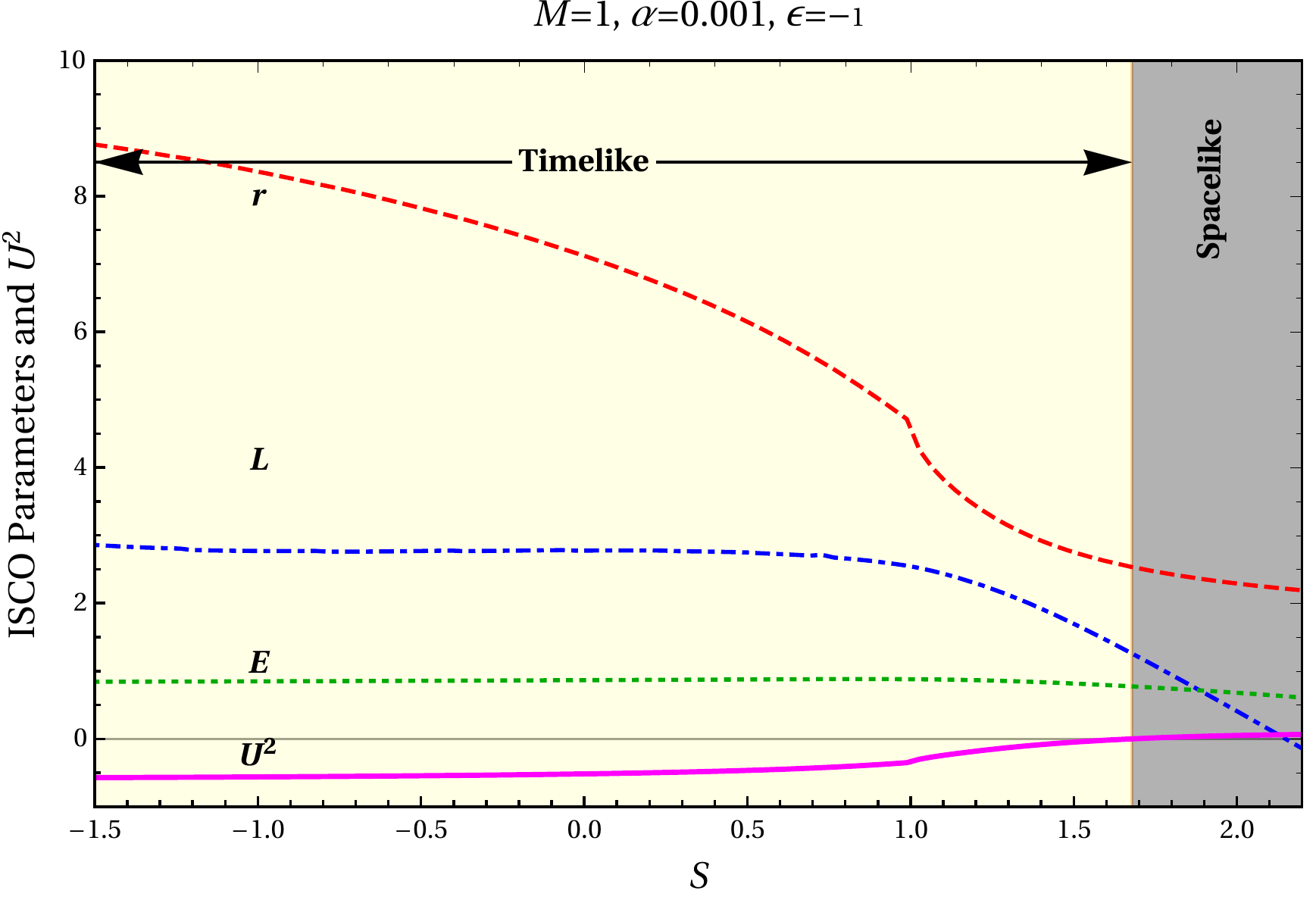}
\end{tabular}
\caption{The variation of the ISCO parameters and $U^{2}$ as a function of spin parameter $S$ for different combinations of parameters $ \epsilon$ and $ \alpha$.}\label{fig:ISCO}
\end{figure*}


In this section, we analyse the properties of ISCOs for a spinning particle moving in the vicinity of SQBH. We begin by defining the radial component of four-momentum $P^{r}$ as function of energy $E$ in addition to the other parameters $J, S, r, M, \alpha$ and $\epsilon$. For this purpose, we use the set of \crefrange{eq:dot_P_t}{eq:dot_S_theta_phi} together with TDSSC and conservation \crefrange{eq:TDSSC1}{eq:TDSSC4}. The square of $P^{r}$ reads

\begin{align}
    \left(\frac{P^{r}}{m}\right)^{2}&=\mathcal{\kappa}\left(\mathcal{A}E^2+\mathcal{B}E+\mathcal{C}\right),\label{eq:Pr_square}
\end{align}
where,
\begin{align}
    \mathcal{\kappa}&= \left[1-f(r)'\frac{S^{2}}{2r}\right]^{-2},\\
    \mathcal{A}&= 1-f(r)\left(\frac{S}{r}\right)^{2},
\end{align}
\begin{align}
    \mathcal{B}&= \frac{JS}{r^{2}}\left(\frac{f(r)'r}{2}-f(r)\right),
  \\
    \mathcal{C}&= f(r)\left(1-\frac{S^{2}f(r)'}{2r}\right)^{2}\left[1-\left(\frac{J}{r}\right)^{2}-f(r)\left(\frac{S}{r}\right)^{2}\right].
\end{align}
As we are intended to analyze the ISCO parameters numerically for the spinning particle, it is easy to define the effective potential as in \cite{PhysRevD.105.104059}:
\begin{align}
    W_{eff}=\left(\frac{P^{r}}{m}\right)^{2}
\end{align}
Now from Newtonian mechanics, we know that to define the circular motion of a particle following two conditions need to be satisfied simultaneously:
\begin{enumerate}[label=(\roman*)]
    \item the radial velocity must vanish, i.e.,
    \begin{align}
    \frac{dr}{d\tau}=0,\label{eq:drbydtau}
    \end{align}
    \item the radial acceleration also vanish,
    \begin{align}
        \frac{d^{2}r}{d\tau^{2}}=0.\label{eq:d2rbydtau}
    \end{align}
 Solving the above system of \cref{eq:drbydtau,eq:d2rbydtau}, one can obtain the expressions for the energy $E$ and orbital angular momentum $L$ in terms of radial coordinate $r$. However, to find the innermost stable circular orbits (ISCOs), one needs to find the value of $r$ for which the maximum and minimum of effective potential merge. Therefore, in order to do so, we need one more condition beside \cref{eq:drbydtau,eq:d2rbydtau} known as the ISCO condition:
 \item \begin{align}
     \frac{d^{2}V_{eff}}{dr^{2}}=0,\label{eq:Isco_Gen}
 \end{align}
 where $V_{eff}$ is effective potential in general.
\end{enumerate}
Therefore, to find the value of parameters (energy $E$, angular momentum $L$ and radius $r$) at the ISCO, one must solve the above three \crefrange{eq:drbydtau}{eq:Isco_Gen} simultaneously .

For the case of spinning particle moving in the curved spacetime (say in the vicinity of SQBH for our case), we employ the above mentioned methodology and find the corresponding ISCO equations, which read as
\begin{align}
      W_{eff}&=0, \label{eq:ISCO1}\\
     \frac{dW_{eff}}{dr}&=0, \label{eq:ISCO2}\\
      \frac{d^{2}W_{eff}}{dr^{2}}&=0. \label{eq:ISCO3}
\end{align}
In \cref{fig:ISCO}, we numerically solve this resulting closed system of \crefrange{eq:ISCO1}{eq:ISCO3} to examine the behaviour of ISCO parameters $E$, $L$ and $r$ as a function of spin parameter $S$ for various combinations of $\alpha$ and $\epsilon$. It is worth to note for the case of spinning particle $L=J-S$, contrary  to the case of non-spinning particle $S=0$ for which both $L$ and $J$ are identical.

Beside \crefrange{eq:ISCO1}{eq:ISCO2}, there is one more quantity that need to take into consideration while discussing the motion of spinning particles. As mentioned earlier for a spinning particle moving in curved spacetime the four momentum $p^{\mu}$ is not always parallel to its four velocity $u^{\mu}$ which means that $u_{\mu}u^{\mu}\equiv u^{2}$ can be negative (timelike) as well as positive (spacelike). Hence,  to ensure that motion of spinning particle is timelike (physical), the superluminal condition in the $\theta=\pi/2$ plane is imposed, i.e,
\begin{align}
    U^{2}<0,
\end{align}
where,
\begin{align}
    U^{2}\equiv \frac{U_{\mu}u^{\mu}}{\left(u^{t}\right)^{2}}=\left(\frac{u}{u^{t}}\right)^{2}=g_{tt}+g_{rr}\left(u^{r}\right)^{2}+g_{\phi\phi}\left(u^{\phi}\right)^{2}.
\end{align}
The behavior of $U^2$ as a function of parameter $S$ is also depicted in \cref{fig:ISCO}, with the highlighted timelike ($U^{2}<0$) and spacelike ($ U^{2}>0$) regions.
\section {Small-Spin corrections and Periastron precession of spinning particles}\label{sec:ssc_and_peri}
In this section, we discuss the periastron precession of a spinning particle moving around the SQBH in an approximately  circular orbit in the equatorial plane $\theta=\pi/2$. Here also, we are mainly interested to bring out the effect of particle spin $S$ and equation of state parameter $\epsilon$ on the periastron precession of the spinning particle.

Since, as we previously explained for arbitrary values of $S$ and $\epsilon$, finding the analytical solution for the system of \crefrange{eq:ISCO1}{eq:ISCO3} is a challenging task. Therefore, we employ small-spin corrections (i.e., $S\ll\mathcal{O}(M)$) and expand the \cref{eq:ISCO1,eq:ISCO2} only up to the linear order in $S$ to find out the analytical expressions for $E$ and $J$.

We thus assume that the solution of \cref{eq:ISCO1,eq:ISCO2} is in the form
\begin{align}
    E&=E_{0}+S E_{1},\label{eq:First_o_s_E}\\
    J&=J_{0}+S J_{1} \label{eq:First_o_s_J}.
\end{align}
Here, $E_{0}$ and $J_{0}$ are the values of $E$ and $J$ for the spinless particle case, whereas $E_{1}$ and $J_{1}$ are the first order spin corrections. It is worth noting here that $S$ is a projection of the particle's spin onto the direction of $J$, and is positive when aligned parallel to $J$ and negative when aligned anti-parallel to $J$, for $J>0$ case in particular. A similar procedure is used for the non-spinning particle case in \cite{hobson2006general} and for the spinning particle case in \cite{jefremov2015innermost,favata2011conservative}. The analytical expression of $E_{0}, E_{1}, J_{0}$ and $J_{1}$ comes out as
\begin{align}
    E_{0}&= \frac{\sqrt{2} f(r)}{\sqrt{2 f(r)-r f'(r)}},\label{eq:E0}\\
    E_{1}&= \frac{\sqrt{r} \left(f(r) f'(r)^{3/2}-r f'(r)^{5/2}+r f(r) \sqrt{f'(r)} f''(r)\right)}{2 \left(2 f(r)-r f'(r)\right)^{3/2}},\label{eq:E1}\\
    J_{0}&= \frac{r^{3/2} \sqrt{f'(r)}}{\sqrt{2 f(r)-r f'(r)}},\label{eq:J0}\\
    J_{1}&= -\frac{-\sqrt{2} r^2 f(r) f''(r)+3 \sqrt{2} r f(r) f'(r)-4 \sqrt{2} f(r)^2}{2 \left(2 f(r)-r f'(r)\right)^{3/2}}\label{eq:J1}.
\end{align}
In the limiting case $\alpha \to 0$ and $r \to 6M$,  \cref{eq:First_o_s_E,eq:First_o_s_J} with help of \crefrange{eq:E0}{eq:J1} read as
\begin{align}
    E &= \frac{2\sqrt{2}}{3}-\frac{S}{36\sqrt{3}M},\label{eq:Schw_E}\\
    J &= 2\sqrt{3}M+\frac{\sqrt{2}S}{3},\label{eq:Schw_J}
\end{align}
which exactly match with the Schwarzschild BH case~\cite{jefremov2015innermost}.

\begin{figure}
\includegraphics[scale=0.65]{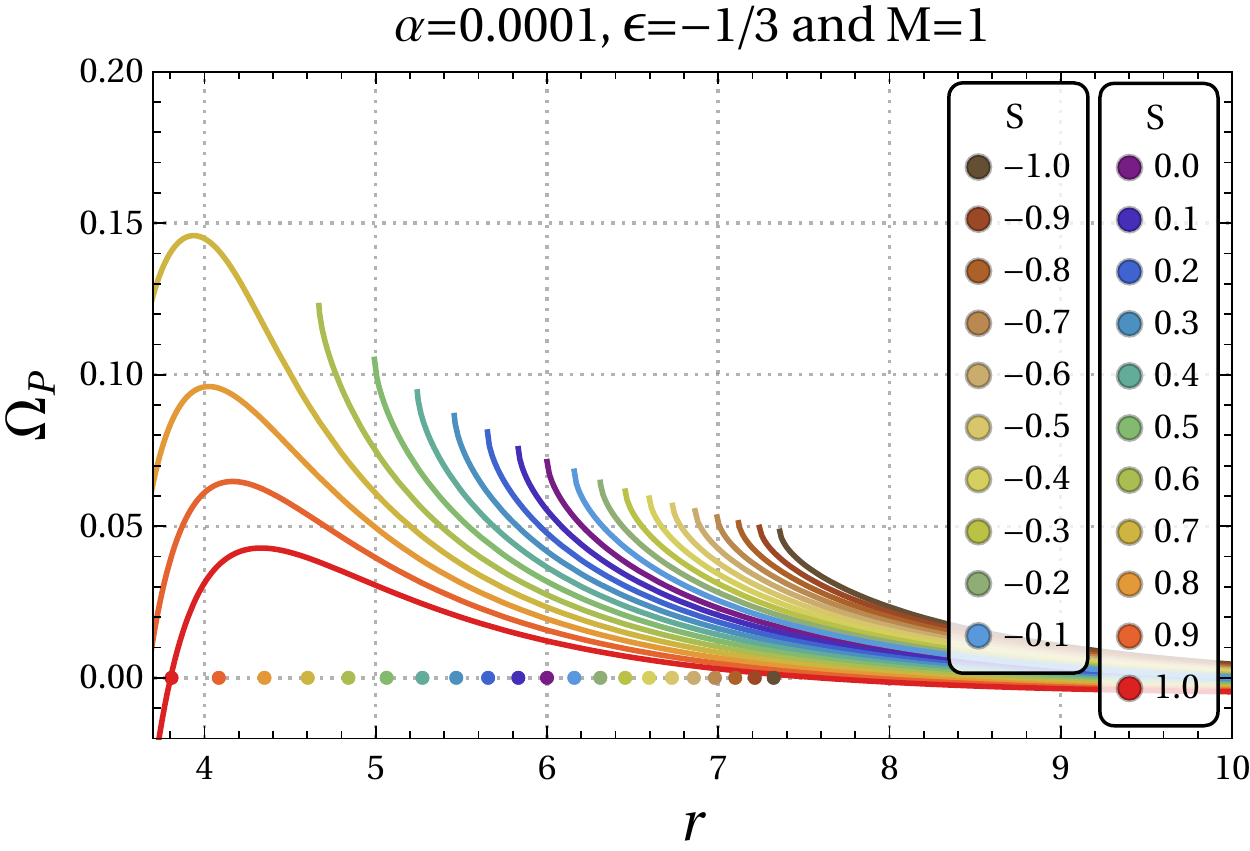}
\includegraphics[scale=0.65]{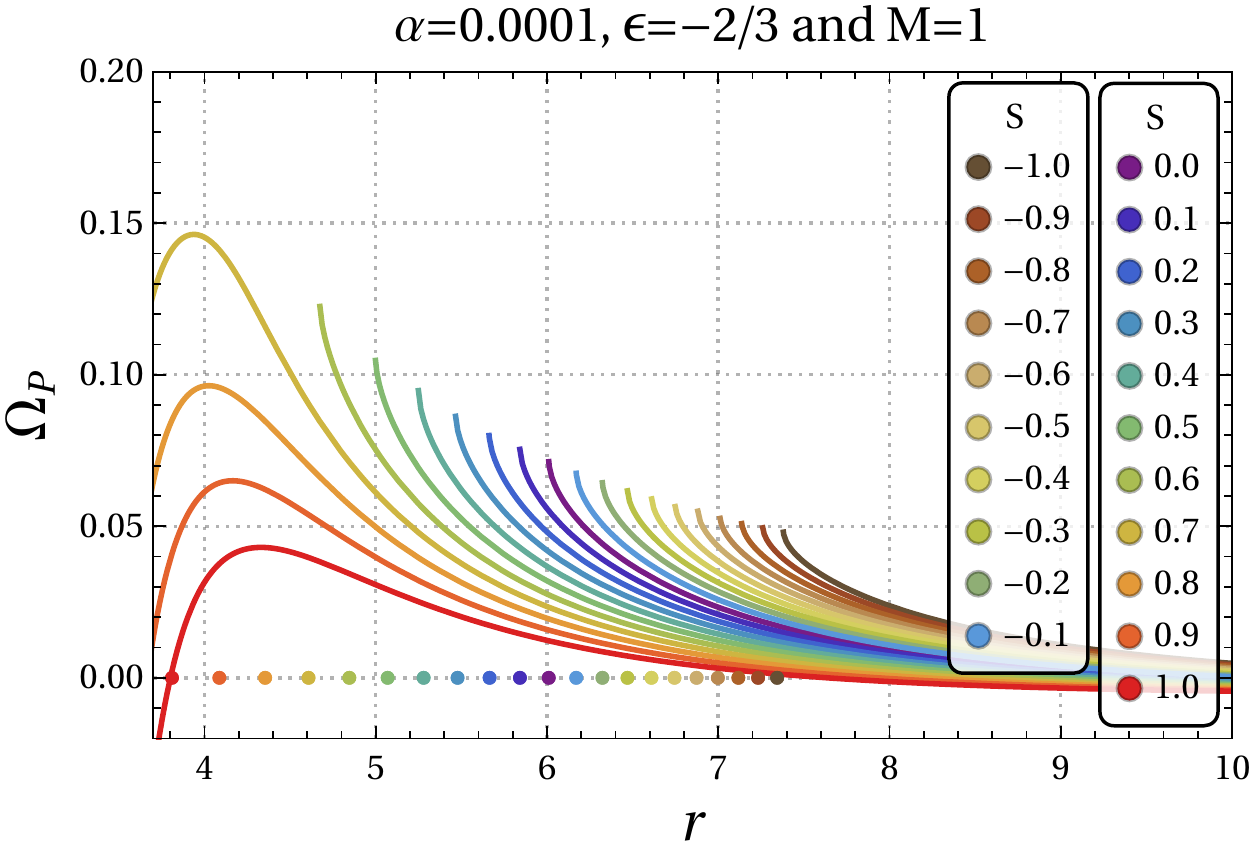}
\includegraphics[scale=0.65]{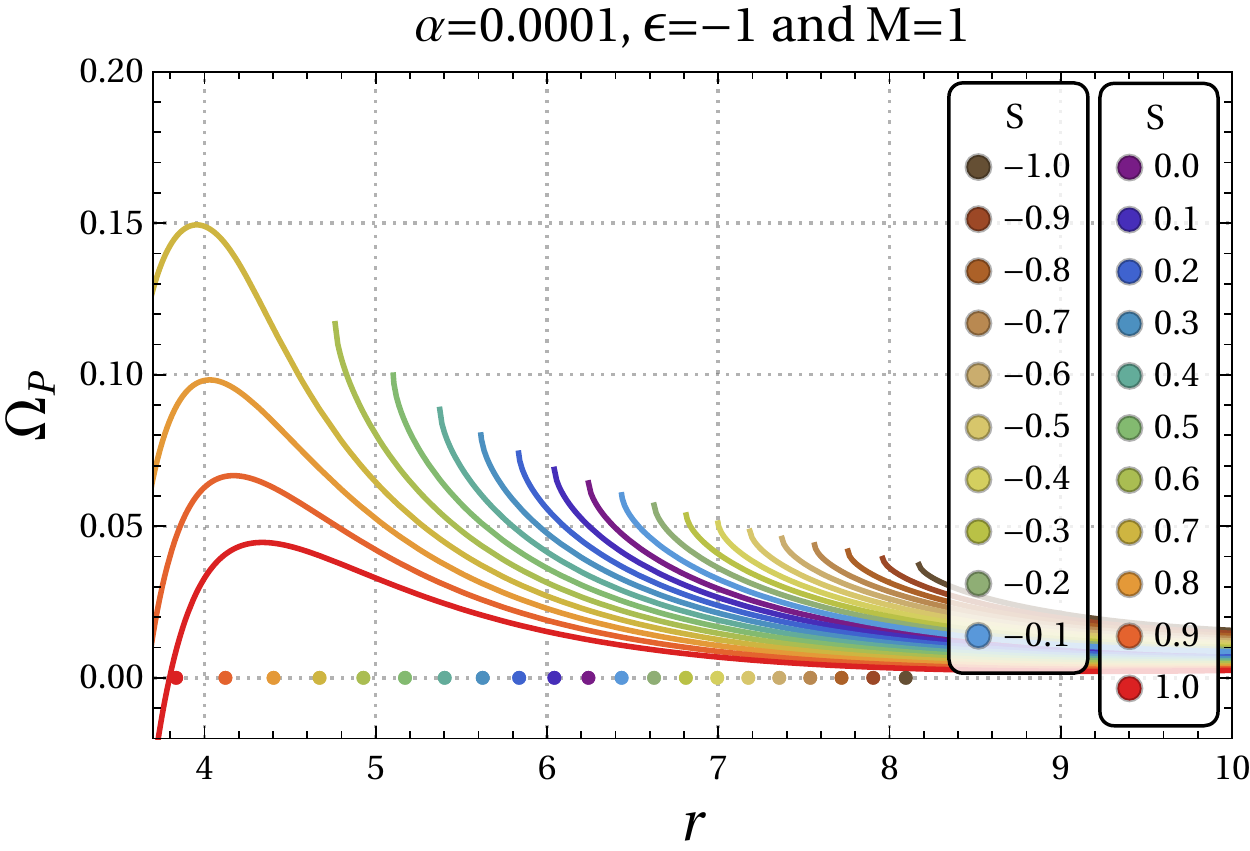}
\caption{For various combinations of the equation of state parameter $\epsilon$ and the particle's spin $S$, plots demonstrate the variation of the periastron precession $\Omega_{P}$ as a function of radial coordinate $r$. The position of ISCOs for the associated values of parameter $S$ is shown by dots on the axis $\Omega_{P}=0$. }\label{fig:Periastron_precession}
\end{figure}

Now, in order to calculate the Periastron precession, we assume that the spinning particle is slightly perturbed from its stable position which in turn results to oscillation about the stable position $r_{s}$ with a radial frequency $\omega^{2}_{r}$ \cite{le2013periastron}, which is defined as
\begin{align}
    \omega^{2}_{r}\equiv -\frac{1}{2}\left(\frac{d^{2}W_{eff}}{dr^{2}}\right)_{r=r_{s}}.\label{eq:Omega_r}
\end{align}
In contrast to Newtonian gravity, when general relativistic effect are taken in consideration (i.e., close to BH), the radial and orbital frequencies are no longer equal (i.e., $\omega_{r}\neq\omega_{\phi}$) \cite{Wald:1984rg}, where orbital frequency $\omega_{\phi}$ is defined as
\begin{align}
    \omega_{\phi}\equiv\frac{d\phi}{d\tau}=\dot{\phi}.\label{eq:Omega_phi}
\end{align}
It is worth noting that both $\omega_{r}$ and $\omega_{\phi}$ are basically contravariant components, and here we have used the subscript instead of the superscript in the definition for the sake of simplicity. In literature, the orbital frequency $\omega_{\phi}$ is also referred to as the azimuthal frequency \cite{LeTiec:2013uey}.

Now, using \cref{eq:Omega_r,eq:Omega_phi}, the periastron precession $\Omega_{P}$ can be defined as
\begin{align}
    \Omega_{P}\equiv \omega_{\phi}-\omega_{r}.\label{eq:Periastron_K}
\end{align}
The behaviour of $\Omega_{P}$ for the spinning particle as a function of radial coordinate $r$ for various combinations of $\epsilon$, $\alpha$ and $S$ is studied numerically and presented in \cref{fig:Periastron_precession}. 

Initially, comparing $\Omega_{P}$ from all three cases of $\epsilon$ reveals that general behaviour of $\Omega_{P}$ is identical in all three cases, whether the spin of the particle is considered or not. Interestingly, we also find that $\Omega_{P}$ for $\epsilon=-1/3$ and $-2/3$ at $S=0$ achieves its maximum value at approximately $r=6$, which is very close to the location of the ISCO of nonspinning particle for Schwarzschild BH case. However, for $\epsilon=-1$ and  $S=0$ the maximum of $\Omega_{P}$ occurs away from $r=6$. Therefore, one can conclude that as the equation of state parameter $\epsilon$ increases the maximum of $\Omega_{P}$ for $S=0$, deviates more from Schwarzschild case.

Further investigations uncover some of $\Omega_{ P}$ most intriguing properties, which are as follows:
\begin{itemize}
 \item   
The maximum occurs for periastron precession $\Omega_{P}$ at $r_{ISCO}$ and matches ISCO values computed using the system of \crefrange{eq:ISCO1}{eq:ISCO3} only when parameter $S$ is in the range $\vert S\rvert<0.5$, i.e., when the parameter $S$ fulfils the M{\o}ller limit $S\ll\mathcal{O}(M)$ (see; \cref{fig:Periastron_precession}) for all three cases of $\epsilon=-1/3, -2/3$, and $-1$, respectively, while keeping the normalization factor minimal (i.e., $\alpha=0.0001$). It has to be so because $\Omega_{P}$ calculations are done with tiny spin corrections (i.e., upto linear order in $S$), whereas ISCO calculations are done without any limit on $S$. It is also observed that in the asymptotic limit $r\to \infty$ all $\Omega_{P}\to 0$ always. 
\item In between the maximum and minimum values, $\Omega_{P}$ decreases monotonically as the radial coordinate advances for both the spinning as well as the non-spinning particles, respectively. It is worth noting here that the characteristic behaviour of $\Omega_{P}$ obtained for SQBH is analogous to $\Omega_{P}$ behaviour seen in the Reissner-Nordstr\"{o}om and Kerr BHs \cite{PhysRevD.97.124006} and the accelerating Kerr BH case \cite{PhysRevD.101.104012}. 
\item
In addition to the above, we also observed that $\Omega_{P}$ is boosted by spin of the particle and increases with rise in parameter $S$ till $S\approx 0.7$, for all three cases of $\epsilon=-1/3, -2/3$, and $-1$. However, the maximum of $\Omega_{P}$ decreases with increasing the spin when $1\geq S\geq 0.7$.
\end{itemize}

It is worth noting that the results presented in \cref{fig:Periastron_precession}, using linear order spin corrections are advantageous from an astrophysical viewpoint because they satisfy the M{\o}ller limit $S\ll\mathcal{O}(M)$; however, if one wishes to investigate large spin values, higher order spin corrections (i.e., square order spin correction and beyond) need to be used.

\section{Concluding remarks} \label{sec:concluding_remarks}
In this paper, we investigated the embedded geometry, analysed the equation of motion of the spinning particle, and explored the characteristics of the ISCO parameters $r$, $L$, and $E$ as a function of particle spin $S$ in detail for the SQBH. As illustrated in \cref{fig:2D_Embedded,fig:3D_Embedded}, the normalization factor $\alpha$ and the equation of state parameter $\epsilon$ have a significant influence on the embedded geometry of the SQBH. It is observed that the embedded geometry respective to the parameter $ \epsilon$ gets separated more from one another as parameter $\alpha$ increases (see; \cref{fig:2D_Embedded,fig:3D_Embedded}).\\
Further, to investigate the possibility of chaotic motion of the spinning particle, we classified the effective potential $V_{eff}$ for the SQBH into four categories based on the existence of saddle points and the minimum of $V_{eff}$: type ($\mathcal{A}$), ($\mathcal{B}$), ($\mathcal{C}$), and  ($\mathcal{D}$).

 It is observed that the orbit of the spinning particle can never be chaotic for the kind ($\mathcal{A}$) $V_{eff}$ that occurs for $S<1<<J$. 
 The particle orbit can be chaotic, only for the type ($\mathcal{B}$) $V_{eff}$ which arises when spin parameter $S$ is very close to unity and $J$ is approximately the same as for type ($\mathcal{A}$).
 The type ($\mathcal{C}$) $V_{eff}$ occurs for parameter $S <1$ while $J$ is smaller than that for the type ($\mathcal{A}$) and ($\mathcal{B}$). The chaotic orbits are not discovered for the type ($\mathcal{C}$) $V_{eff}$ since a comparable type of $V_{eff}$ is also obtained for the case of nonspinning particles for which orbits are non-chaotic.
  The type ($\mathcal{D}$) effective potential is found for parameter $S >1$ while $J$ is comparatively smaller than that for the type ($\mathcal{A}$) and ($\mathcal{B}$). However, like with the type ($\mathcal{C}$) $V_{eff}$, no chaotic orbits are found for the type ($\mathcal{D}$) $V_{eff}$ as well. 
  
  As a result, the orbits in \cref{fig:Orbits} are drawn for the type ($\mathcal{B}$) $V_{eff}$ only for various cases of $\epsilon$. It is interestingly noticed that when the initial position $r_{0}0$ is moved closer to the saddle point (i.e., as we travel down the columns in \cref{fig:Orbits}), the orbits become increasingly chaotic, with the chaos being greatest when the $r_{0}$ is set at the saddle point. Furthermore, when we increased $r_{0}$ distance between the SQBH event horizon and the saddle point, the orbits of the spinning particles became less chaotic (see; the bottom row of \cref{fig:Orbits}).
  
In addition, we numerically analyzed the behavior of the ISCO parameters $E$, $L$, and $r$ as a function of spin parameter $S$ for various combinations of $\epsilon$ and $\alpha$, as well as the behavior of $U^{2}$, which separates the timelike region (for which $U^{2}<0$) of the spinning particles from the spacelike region (for which $U^{2}>0$) in \cref{fig:ISCO}.
The values of the ISCO parameters $E$, $L$, and $r$ drop as the value of parameter $S$ rises, regardless of the value of $\alpha$ and $\epsilon$. It is observed that the ISCO parameter $r$ for the cases of $\epsilon=-1/3$ and $-2/3$ are almost equal but smaller than that for the case of $\epsilon=-1$ for given value of $\alpha$. Further investigations reveal that as $\alpha$ increases (i.e., when moving along the row from left to right in \cref{fig:ISCO}), the ISCO parameters $E$, $L$, and $r$ slightly decrease for $\epsilon=-1/3$ and $-2/3$, whereas the ISCO parameters $E$, $L$, and $r$ decrease significantly in the case when $\epsilon=-1$. 
 
Moreover, the small-spin corrections ($S\ll\mathcal{O}(M)$) of energy $E$ and total angular momentum $J$ of the spinning particle are derived and in prescribed limits of Schwarzschild BH (i.e., $\alpha \to \infty$ and $r \to 6M$), reduce to the expression being the same as for the Schwarzschild BH found in \cite{jefremov2015innermost}. Moreover, the periastron precession $\Omega_{P}$ upto first-order spin correction of the spinning particle moving around the SQBH in the equatorial circular orbit has been investigated in detail. The variation of periastron precession with radial coordinate $r$ for different combinations of parameters $\epsilon$ and $S$ while $\alpha$ is fixed to $0.0001$ has been shown in \cref{fig:Periastron_precession}. It is concluded that the maximum of $\Omega_{P}$ for $S=0$ deviates more from Schwarzschild case as parameter $\epsilon$ increases (i.e., as one moves top to bottom in \cref{fig:Periastron_precession}). On the other hand when $S\neq0$, $\Omega_{P}$ increases with the parameter $S$ for all three cases of $\epsilon$ till $S\leq 0.7$ and decrease when $S$ increases further till $S=1$. 

In the future, we would intend to examine the chaos of spinning particles using processes such as the Fast Lyapunov Indicator \cite{Han:2008zzf}, Poincar\'{e} map \cite{suzuki1997chaos}, and Milnikov technique \cite{Bombelli:1991eg,Polcar:2019kwu} to gain a deeper understanding of the choatic orbits around the SQBH. 

\appendix
  \section {The MPD Equations of Spinning Test Particles Around SQBH}
  \label{sec:MPDeqns}
\noindent In case of spinning particle, spin couples to the curvature of the background spacetime so that the spin force pushes the particle away from the geodesic. Then the deviation from geodesic motion should be very small compared with the curvature tensor of the spacetime. The explicit form of the momentum equations \eqref{eq:MPD1} of MPD equations in case of SQBH are obtained as first order derivative of 4-momentum ($p^\mu$) as follows
\begin{multline}
    \dot{p^{t}}+\frac{f^{'}(r)}{2f(r)} p^{r} +\frac{f^{'}(r)}{2f(r)} \dot{r} p^{t} = \frac{f^{''}}{2f(r)} \dot{r}S^{tr}+ \frac{r}{2}f^{'}(r)\dot{\theta}S^{t\theta}\\ +\frac{r}{2}\sin^{2}\theta f^{'}(r)\dot{\phi}S^{t\phi},\label{eq:dot_P_t}
\end{multline}

\begin{multline}
\dot{p^{r}}+\frac{f(r)}{2}f^{'}(r) p^{t}-\frac{f^{'}(r)}{2f(r)}\dot{r}p^{r}-r f(r) \sin^{2}\theta \dot{\phi}p^{\phi} -r f(r) \dot{\theta} p^{\theta}\\ = \frac{f(r)}{2} f^{''}(r)S^{tr}+\frac{r}{2} f^{'}(r)\dot{\phi}S^{r\phi}+\frac{r}{2}  f^{'}(r) \dot{\theta}S^{r\theta},\label{eq:dot_P_r}
\end{multline}

\begin{multline}
\dot{p^{\theta}}+\frac{\dot{r}}{r}p^{\theta}+\frac{\dot{\theta}}{r}p^{r}-\dot{\phi} \sin\theta\cos \theta p^{\phi} =\frac{f(r)f^{'}(r)}{2 r}S^{t\theta} \\ +\frac{f^{'}(r)}{2rf(r)}\dot{r}S^{r\theta}-2\sin^{2}\theta\left(f(r)-1\right)\dot{\phi}S^{\theta\phi},\label{eq:dot_P_theta}
\end{multline}

\begin{multline}
\dot{p^{\phi}}+\frac{\dot{r}}{r}p^{\phi}+\frac{\cos\theta}{\sin\theta}\dot{\theta}p^{\phi}+\dot{\phi}\left(\frac{p^{r}}{r}+\frac{\cos\theta}{\sin\theta}p^{\theta}\right)=\\ \frac{f(r)f^{'}(r)}{2 r}S^{t\phi}-\frac{f^{'}(r)}{2rf(r)}\dot{r} S^{r\phi}+\left(f(r)-1\right)\dot{\theta}S^{\theta\phi}.\label{eq:dot_P_phi}
\end{multline}

\noindent However, the spin equations \eqref{eq:MPD2} of MPD equations  for SQBH have explicitly took the following form
\begin{multline}
\dot{S^{tr}}-f(r)r \sin^{2}\theta\dot{\phi}S^{t\phi} -r f(r)\dot{\theta}S^{t\theta}= p^{t}\dot{r}-p^{r},\label{eq:dot_S_tr}
\end{multline}
\begin{multline}
\dot{S^{t\phi}}+ \frac{f^{'}(r)}{2 f(r)}S^{r\phi} + \dot{r} \left(\frac{f^{'}(r)}{2 f(r)}+\frac{1}{r}\right)S^{t\phi}+\dot{\theta}\left(\frac{\cos\theta}{\sin\theta}\right)S^{t\phi}\\ +\dot{\phi}\left(\frac{S^{tr}}{r}+\frac{\cos\theta}{\sin\theta}S^{t\theta}\right)= p^{t}\dot{\phi}-p^{\phi},\label{eq:dot_S_t_phi}
\end{multline}
\begin{multline}
\dot{S^{t\theta}}+ \frac{f^{'}(r)}{2 f(r)}S^{r\theta} + \dot{r} \left(\frac{f^{'}(r)}{2 f(r)}+\frac{1}{r}\right)S^{t\theta}+\dot{\theta}\frac{S^{tr}}{r} \\ -\dot{\phi} \sin\theta\cos\theta S^{t\phi} = p^{t}\dot{\theta}-p^{\theta},\label{eq:dot_S_t_theta}
\end{multline}
\begin{multline}
\dot{S^{r\phi}}+ \frac{f^{'}(r) f(r)}{2}S^{t\phi} + \dot{r} \left(-\frac{f^{'}(r)}{2 f(r)}+\frac{1}{r}\right)S^{r\phi}+\\ \dot{\theta}\left(-r f(r)S^{\theta\phi}+\frac{\cos\theta}{\sin\theta}S^{r\phi}\right)+\dot{\phi} \frac{\cos\theta}{\sin\theta} S^{r\theta}= p^{r}\dot{\phi}-\dot{r}p^{\phi},\label{eq:dot_S_r_phi}
\end{multline}
\begin{multline}
\dot{S^{r\theta}}+ \frac{f^{'}(r) f(r)}{2}S^{t\theta} + \dot{r} \left(-\frac{f^{'}(r)}{2 f(r)}+\frac{1}{r}\right)S^{r\theta}\\ +\dot{\phi}\left(r f(r)\sin^{2}\theta S^{\theta \phi}-\sin\theta \cos\theta S^{r\phi}\right) = p^{r}\dot{\theta}-\dot{r}p^{\theta},\label{eq:dot_S_r_theta}
\end{multline}

\begin{multline}
\dot{S^{\theta\phi}}+ \frac{2\dot{r}}{r}S^{\theta\phi}+\dot{\theta}\left(\frac{S^{r\phi}}{r}+\frac{\cos\theta}{\sin\theta}S^{\theta\phi}\right)+\dot{\phi}\frac{S^{\theta r}}{r} \\ =p^{\theta}\dot{\phi}-\dot{\theta}p^{\phi}.\label{eq:dot_S_theta_phi}
\end{multline}
The TDSSC \eqref{eq:SSS} in this case are obtained in the following explicit form 
 \begin{equation}
\frac{p^{r}}{f(r)}S^{tr}+r^{2}p^{\theta}S^{t\theta}+r^{2}\sin^{2}\theta p^{\phi}S^{t\phi}=0,\label{eq:TDSSC1}
\end{equation}
\begin{equation}
f(r)p^{t}S^{tr}+r^{2}p^{\theta}S^{r\theta}+r^{2}\sin^{2}\theta p^{\phi}S^{r\phi}=0,\label{eq:TDSSC2}
\end{equation}
\begin{equation}
f(r)p^{t}S^{t\theta}+\frac{p^{r}}{f(r)}S^{\theta r}+r^{2}\sin^{2}\theta p^{\phi}S^{\theta\phi}=0,\label{eq:TDSSC3}
\end{equation}
\begin{equation}
f(r)p^{t}S^{t\phi}+\frac{p^{r}}{f(r)}S^{\phi r}+r^{2}p^{\theta}S^{\phi\theta}=0.\label{eq:TDSSC4}
\end{equation}

In equatorial or $\theta=\pi/2$ plane, the components $(u^{t}, u^{r}, u^{\theta}, u^{\phi})$ of \cref{eq:u_mu} with the help of \cref{eq:Y} reduce to
\begin{widetext}
\begin{align}
    {u}^{t}&=-\frac {p_{{t}} \left[2f(r)m^{2}r^{3}+s^{2}f'(r)\left(r^{2}f(r)^{2}p_{r}^{2}-r^{2}p_{t}^{2}+f(r)p_{\phi}^{2}\right)\right]}{r\;f(r)\Phi},\label{eq:ut}
    \end{align}
\begin{align}
u^{r}&=\frac {f(r)p_{{r}} \left[2f(r)m^{2}r^{3}+s^{2}f'(r)\left(r^{2}f(r)^{2}p_{r}^{2}-r^{2}p_{t}^{2}+f(r)p_{\phi}^{2}\right)\right]}{r\;\Phi},\label{eq:ur}\\
u^{\theta}&=0,\label{eq:utheta}\\
 u^{\phi}&=\frac {p_{{\phi}} \left[2f(r)m^{2}r^{2}+s^{2}f''(r)\left(r^{2}f(r)^{2}p_{r}^{2}-r^{2}p_{t}^{2}+f(r)p_{\phi}^{2}\right)\right]}{r^{2}\;\Phi},\label{eq:uphi}\\
\text{where,}\hspace{1cm}&\nonumber\\
    \Phi&=2m^{2}r^{2}f(r)+s^{2}rf'(r)\left(f(r)^{2}p_{r}^{2}-p_{t}^{2}\right)+s^{2}f(r)f''(r)p_{\phi}^{2}.\label{eq:Phi}
\end{align}
\end{widetext}


\begin{acknowledgments}
S.G. acknowledges the financial support provided from University Grants Commission (UGC), New Delhi, India as a Senior Research Fellow through UGC-Ref.No. 1479/CSIR-UGC NET JUNE 2017.
P.S. acknowledges support under University Grant Commission (UGC)-DSKPDF scheme (Govt. of India) through grant No. F.4-2/2006(BSR)/PH/20-21/0053. The authors would also  like to acknowledge the facilities used at IUCAA Centre for Astronomy Research \& Development (ICARD), Gurukula Kangri (Deemed to be University), Haridwar, India. S.S. acknowledges the support from Research F-FA-2021-432 of the Uzbekistan Ministry for Innovative Development.    
\end{acknowledgments}



\bibliography{shobitref}

\end{document}